\documentclass[11pt]{article}
\usepackage{fullpage} 
\usepackage{graphicx}
\usepackage{amsmath}
\usepackage{graphics}
\usepackage{amssymb}
\usepackage{hyperref}
\usepackage{natbib}
\usepackage{color}
\usepackage{amsthm}
\usepackage{amsfonts}
\usepackage{subfig}
\usepackage{float}
\usepackage{comment}

\usepackage{srctex}

\def\bs{\boldsymbol}

\def\ol{\overline}

\definecolor{JP}{RGB}{255,0,0}
\newcommand{\JP}[1]{\textcolor{JP}{[J: #1]}}

\definecolor{DP}{RGB}{255,0,255}
\newcommand{\DP}[1]{\textcolor{DP}{[D: #1]}}

\begin{document}

\title{\Large\bf Estimating fiber orientation distribution from diffusion MRI with spherical needlets}
\author{Hao Yan$^1$, Owen Carmichael$^2$, Debashis Paul$^1$, Jie Peng$^{1,  \footnote{Correpondence author: Email: jiepeng@ucdavis.edu}}$ \\
for the Alzheimer's Disease Neuroimaging Initiative$^{\footnote{Data used in preparation of this article were obtained from the Alzheimer’s Disease Neuroimaging Initiative
(ADNI) database (\url{adni.loni.usc.edu}). As such, the investigators within the ADNI contributed to the design
and implementation of ADNI and/or provided data but did not participate in analysis or writing of this report.
A complete listing of ADNI investigators can be found at:
\url{http://adni.loni.usc.edu/wp-content/uploads/how_to_apply/ADNI_Acknowledgement_List.pdf}}}$\\\\
\textit{$1$: Department of Statistics, University of California,  Davis }\\
\textit{One Shields Ave., Davis, CA 95616} \\\\
\textit{$2$ : Pennington Biomedical Research Center, Louisiana State University} \\
\textit{ 6400 Perkins Road, Baton Rouge, LA 70808}
}

\date{}

\maketitle

\clearpage
\newpage
\begin{abstract}
We present a novel method for estimation of the fiber orientation distribution (FOD) function based on diffusion-weighted
Magnetic Resonance Imaging (D-MRI) data.  We formulate the problem of FOD estimation as a regression problem through spherical deconvolution and a sparse representation of
the FOD by a \textit{spherical needlets} basis that form a multi-resolution tight frame for spherical functions.  This sparse representation allows us to estimate FOD by an $l_1$-penalized regression under a non-negativity constraint.
The resulting convex optimization problem is solved by an alternating direction method of multipliers (ADMM) algorithm. The proposed method leads to a reconstruction of the FODs that is accurate, has low variability and preserves sharp features. Through extensive experiments, we demonstrate the effectiveness and  favorable  performance of the proposed method compared with two existing methods. Particularly, we show the ability of the proposed method in successfully resolving fiber crossing at small angles and in automatically identifying isotropic diffusion. We also apply the proposed method to real 3T D-MRI data sets of healthy elderly individuals. The results show realistic descriptions of crossing fibers that are more accurate and less noisy than competing methods even with a relatively small number of gradient directions.
\end{abstract}


\section{Introduction}\label{sec:intro}

Diffusion-weighted MRI (D-MRI) \citep{LeBihanEtal2001,Mori2007} has become a widely used,
non-invasive tool for clinical and experimental neuroscience due to its capability of characterizing
tissue microstructure \textit{in vivo} by making use of  diffusion displacement measures of water molecules.  Specifically, high angular resolution diffusion imaging (HARDI) enables extraction of accurate and detailed information about fiber tract directions through diffusion measurements made along a large number of spatial directions \citep{TuchEtal2002}. Such information is then used in fiber reconstruction algorithms  (tractography)  to facilitate better mapping of neuronal connections.

Following \cite{AssemlalTBS2011}, \cite{DescoteauxEtAl2011}, \cite{JonesKT2013}, \cite{ParkerEtAl2013}, \cite{TournierML2011},
\cite{YehWT2010} and others, HARDI techniques used for describing local diffusion characteristics may be categorized by the different sampling schemes in the q-space, namely, single $q$-shell techniques and multiple $q$-shell techniques. Single $q$-shell techniques  sample  gradient vectors with a single \textit{bvalue}; whereas multiple $q$-shell techniques sample  gradient vectors with multiple \textit{bvalues} \citep{JensenEtAl2005,LuJRH2006,WeedenEtAl2005,WeedenEtAl2008, WuA2007,WuFA2008, LiuBAM2004, DescoteauxEtAl2011,  AssemlalTB2009,ChengGJD2010}.  
Single $q$-shell techniques include angular reconstruction models such as 
fiber orientation distribution (FOD or fFOD) estimation
\citep{TournierEtal2004,TournierEtal2007,JianV2007,YehT2013},  diffusion orientation distribution function (ODF or dODF) estimation \citep{Tuch2004,DescoteauxAFD2007,DescoteauxDKA2009,AganjEtAl2010},
diffusion orientation transform (DOT) \citep{OzarslanEtAl2006} and fiber ball imaging \citep{JensenRH2016}.
Different $q$-space sampling schemes and data representations are optimal for different analytical goals; For example multiple $q$-shell techniques enable explicit modeling of separate cellular contributors to the D-MRI signal.  In this paper we focus on one single $q$-shell data representation, the FOD, because it is designed to describe the local spatial arrangement of axonal fiber bundles in such a way that sharply-defined geometric features are preserved \citep{JensenRH2016}. This property makes the FOD an attractive representation if the ultimate downstream analytic goal is fiber reconstruction through tractography.

In this paper, we propose a novel FOD estimation method
that makes use of sparse representation of the FOD in a multiresolution basis on the unit sphere,
named needlets \citep{NarcowichPW2006b,MarinucciP2011}. To estimate the FOD from observed D-MRI data, we adopt the spherical deconvolution framework described in \cite{TournierEtal2004} and \cite{TournierEtal2007}, and formulate
the FOD estimation problem as a regression problem. Our method has the 
advantages of explicitly making use of the excellent function representation
characteristics of needlets, which we exploit through the imposition of a
sparsity inducing $\ell_1$ penalty on the needlet coefficients of the 
FOD. 
In addition, we impose a non-negativity constraint on 
the estimated FOD and design a computationally efficient
estimation procedure. Furthermore, we propose an effective and statistically
reliable method for selecting the penalty parameters. 
These methodological innovations lead to improved statistical efficiency and better detection of fiber directions, as is demonstrated through extensive experiments. To the best of our knowledge, none of the existing methods for FOD estimation incorporates all these salient features altogether.

A key condition for successful spherical deconvolution-based estimation of FOD is the availability of a parsimonious
representation of the FOD. This is due to the ill-conditioned nature of deconvolution problems, which manifests itself in the form of noise amplification  when the effective number of coefficients  in function representation is large  \citep{KerkyacharianPPW2007,JohnstoneP2014}.
Several popular FOD estimation methods rely on spherical harmonics (SH) representation of the FOD \citep{TournierEtal2004,TournierEtal2007} due to the nice analytical properties of SH basis, including
the fact that they have closed form expressions in terms of Legendre   polynomials, and they are eigenfunctions
of the Laplace-Beltrami operator on $\mathbb{S}^2$ \citep{AtkinsonH2012}. 
However, when the function being represented has localized sharp peaks, as  is expected of FODs,  SH basis
does not provide an efficient representation due to the global support of its basis functions.

On the other hand, a collection of spherical functions that are localized in space and scale/frequency will yield a parsimonious representation for a spiky but otherwise regular function  \citep{NarcowichPW2006a,NarcowichPW2006b}. This is highly desirable for FOD estimation since it leads to an increase of effective signal-to-noise ratio and allows for utilizing 
sparsity-inducing regularization schemes.
A class of such functions is the \textit{needlets} 
\citep{MarinucciP2011,NarcowichPW2006b,Fan2015},  described in 
detail in Section \ref{subsec:needlet}, which forms a special class of spherical wavelet frames. Moreover, needlets are not only localized in both space and frequency, they are also smooth functions, which together ensure  stable reconstruction of a spherical function in a needlet frame.
In particularly, the tightness of the needlet frame, which translates into low mutual coherence and a simpler representation of spherical functions, is crucial for statistical efficiency when solving an inverse problem  based on noisy data such as FOD estimation from D-MRI data studied here \citep{KerkyacharianPPW2007,JohnstoneP2014}.
Due to these advantages, we propose to use needlets as a
representer of FODs.

Relying on the sparse representation enabled by the needlets, we propose a  sparsity-inducing $\ell_1$-norm
penalty on the needlet coefficients of the FOD. 
The use  of $\ell_1$ penalization enables a good estimation with sharp localized peaks even when the number of gradient directions and signal level are both moderate. In addition, since FOD has a density interpretation, we impose a nonnegativity constraint on the estimated FOD. Importance of the nonnegativity restriction has been studied by \cite{ChengEtAl2014}, among others.

The sparse inducing penalty ($\ell_1$ and/or $\ell_0$ ) has been used for FOD estimation by \cite{YehT2013} and \cite{DaducciVTW2014}. In \cite{YehT2013}, FODs are represented in a mono-resolution basis consisting of a large class of putative dODF functions, rather than a stable multi-resolution basis designed to enable a sparse representation. 
In \cite{DaducciVTW2014}, an ad-hoc dictionary of functions is used to represent  
 FODs under $\ell_1$ and $\ell_0$ penalization schemes. While these methods aim to give sparse solutions, their choice of representer may not lead to sparse representation of FODs that are inhomogeneous spherical functions with sharp peaks. 
Thus these estimators could suffer from bias and inefficiency in FOD estimation.


Another class of methods relies on  multi-tensor models, whereby the 
FOD itself is modeled as a discrete mixture of unit direction vectors
representing the orientation of different 
fiber bundles. A prominent work under this framework is by \cite{LandmanEtAl2012} who use an $\ell_1$ penalty with   a nonnegativity constraint for estimating
the mixture fractions. 
For FOD estimation using a multi-tensor model, the number of mixture components needs to be reasonably small to ensure identifiability of the parameters \citep{scherrer2010multiple,WongLPP2016} and low variability of the estimator. 
In contrast, needlets not only yield stable and sparse representation for smooth inhomogeneous functions, it also leads to an accurate and  efficient approximation  of  finite mixtures of distinct directions since these can be well approximated by smooth inhomogeneous functions. Therefore, needlet representation can lead to good estimation for a broader class of FODs. 
One advantage of the approach in \cite{LandmanEtAl2012} is that it naturally
incorporates a multi-compartment framework with different response functions.
It is possible to  extend our approach to  accommodate multiple compartments 
as is discussed in Section
\ref{sec:conclusion}.

The proposed needlets based procedure also 
enables developments of model selection and inferential schemes
within the well-established framework of nonparametric function estimation.  
Specifically, we propose a data-driven scheme for the selection of the penalty parameter that aims to 
balance bias and variance in FOD estimation. This is in contrast to the requirement of making ad-hoc choices 
of the regularization parameters in many existing methods in the literature. 

Finally, we also develop an efficient computational scheme  based on  a fast implementation
of spherical needlets transform; and  an efficient implementation of the nonnegativity constrained
$\ell_1$ penalization problem through an \textit{Alternating Direction  Method of Multipliers}  algorithm \citep{BoydEtAl2011}.
We conduct extensive experiments  based on synthetic data to demonstrate the effectiveness of the proposed method and to
compare it with two existing methods, namely, an SH based deconvolution  with Tikhonov regularization proposed by
\cite{TournierEtal2004} and the SuperCSD sharpening of the aforementioned SH-based estimator  \citep{TournierEtal2007}.  We specifically look into small crossing angles since this is a challenging yet very important problem in resolving crossing fibers and we  focus on high \textit{bvalue} and/or high signal-to-noise ratio for such cases since this is the direction where the field is moving \citep{SetsompopEtAl2013, VanEssenEtAl2013}.
	The proposed method not only shows favorable performance compared to the competing methods,  it is also able to recover FOD when fiber crossing angle is as small as $30^{\circ}$ and  automatically identify isotropic diffusion.
We also apply these methods to two real 3T 
D-MRI data sets of healthy elderly individuals from the Alzheimer's
Disease Neuroimaging Initiative (ADNI) database.

The rest of the paper is organized as follows. We describe the proposed method and  experiments based on both synthetic data and real D-MRI data in
Section \ref{sec:method}. We report the results of the experiments in Section \ref{sec:results}.
We make conclusions and  discuss future directions in Section \ref{sec:conclusion}. Much of the technical details and additional
experimental results are reported in a Supplementary Material.

\section{Material and Methods}\label{sec:method}

\subsection{Needlets representation of spherical functions}\label{subsec:needlet}

We give a brief introduction
to needlets and some of their relevant properties. Detailed treatments,
including their localization characteristics and approximation properties,
are available in \citep{MarinucciP2011,NarcowichPW2006a,NarcowichPW2006b}.

Needlets are constructed by complex-valued SH functions $\{\Phi_{lm}:m=-l,\ldots,l\}_{l=0,1,2,\ldots}$, which
form an orthonormal basis for $L^2(\mathbb{S}^2)$, the space of square integrable functions defined on unit sphere $\mathbb{S}^2$.
The indices $l$ and $m$, referred to as \textit{level index} (larger $l$ corresponds to higher angular frequency) and \textit{phase index}, respectively,
together determine the wavy pattern of the function $\Phi_{lm}$.
Needlets construction is based  on two key ideas \citep{MarinucciP2011}:
(a) discretization of $\mathbb{S}^2$:
If ${\cal H}_l$ denotes the space spanned by $\{\Phi_{lm}:m=-l,\ldots,l\}$, and
${\cal K}_j := \bigoplus_{l=0}^j {\cal H}_l$, then for every $j \in \mathbb{N}$, there exists
a finite subset $\chi_j = \{\zeta_{jk}\}_{k=1}^{p_j} \subset \mathbb{S}^2$ (\textit{quadrature  points}) and positive weights
$\{\lambda_{jk}\}_{k=1}^{p_j}$ (\textit{quadrature  weights}) such that, for any $f \in {\cal K}_j$,
\begin{equation}\label{eq:quadrature_formula}
\int_{\mathbb{S}^2}  f(\mathbf{x}) d\omega(\mathbf{x}) = \sum_{k=1}^{p_j} \lambda_{jk} f(\zeta_{jk}),
\end{equation}
where $d\omega(\mathbf{x})$ denotes the surface element of $\mathbb{S}^2$ at $\mathbf{x}$;
and (b)  Littlewood-Paley decomposition:  This is through a function $b$ defined on
$\mathbb{R}^+$ satisfying:
(i) $b(\cdot) > 0$ on $(B^{-1},B)$ for some $B > 1$, and equal to zero on $(B^{-1},B)^c$; (ii)
$\sum_{j=0}^\infty b^2(y/B^j) = 1$ for all $y \in \mathbb{R}^+$; and (iii) $b(\cdot) \in C^M(\mathbb{R}^+)$. ($B=2$ is used in our construction).

With the above, a class of needlets  $\{\psi_{jk}: 1\leq k \leq p_j\}_{j\geq 1}$ can be defined as follows: for $\mathbf{x}\in
\mathbb{S}^2$,
\begin{equation}\label{eq:needlet_construct}
\psi_{jk}(\mathbf{x}) = \sqrt{\lambda_{jk}} \sum_{l=\lceil B^{j-1}\rceil}^{\lfloor B^{j+1}\rfloor}
b\left(\frac{l}{B^j}\right)\sum_{m=-l}^{l}  \Phi_{lm}(\zeta_{jk}) \ol{\Phi}_{lm}(\mathbf{x})
=  \sqrt{\lambda_{jk}} \sum_{l=\lceil B^{j-1}\rceil}^{\lfloor B^{j+1}\rfloor}
b\left(\frac{l}{B^j}\right) \frac{2l+1}{4\pi} P_l(\langle \zeta_{jk}, \mathbf{x}\rangle),
\end{equation}
where  $j\in \mathbb{N}$ encodes \textit{scale/frequency} and $k \in \{1,\ldots,p_j\}$
encodes \textit{location} (determined by $\zeta_{jk}$) and $P_l$ is the $l$-th Legendre polynomial.
Note that, the needlets  are real valued spherical functions.
 
It can be shown that needlets thus constructed are localized
in both scale and space, with exponentially increasing concentration around the quadrature point $\zeta_{jk}$ 
as the scale  index $j$ increases.  The quadrature formula (\ref{eq:quadrature_formula}) satisfied by 
needlets is an important factor behind the spatial concentration of needlets and the consequent advantages in terms of function approximation.

The collection of needlets together with 
$\Phi_{00}$ (i.e., the constant function on sphere) also form a \textit{tight frame} (referred to as the needlets frame), i.e., for $f \in L^2(\mathbb{S}^2)$,
$\int_{\mathbb{S}^2} (f(\mathbf{x}))^2 d\omega(\mathbf{x}) = |\langle f,\Phi_{00}\rangle|^2 + \sum_{j=1}^\infty \sum_{k=1}^{p_j} |\langle f,\psi_{jk}\rangle_{L^2}|^2$,
where 
\begin{equation}
\label{eq:needlet_coef}
\beta_{jk}=\langle f,\psi_{jk}\rangle_{L^2} = \int_{\mathbb{S}^2} f(\mathbf{x}) \psi_{jk}(\mathbf{x})d\omega(\mathbf{x})
\end{equation}
is  called the
\textit{needlet coefficient} of $f$ corresponding to index pair $(j,k)$.
The tight frame property, together with the localization in space and scale,
imply that needlets can be used to perform \textit{multiresolution analysis} of functions in $L^2(\mathbb{S}^2)$.

Moreover, by (\ref{eq:needlet_construct}), the needlet coefficients of $f$ have the linear representation in terms of SH coefficients:
\begin{equation}\label{eq:representation_SH_coeff}
\beta_{jk}=\langle f,\psi_{jk}\rangle_{L^2} = \sqrt{\lambda_{jk}} \sum_{l=\lceil B^{j-1}\rceil}^{\lfloor B^{j+1}\rfloor}
b\left(\frac{l}{B^j}\right)\sum_{m=-l}^{l} \langle f, \Phi_{lm}\rangle_{L^2} \Phi_{lm}(\zeta_{jk}).
\end{equation}
(\ref{eq:representation_SH_coeff}) provides a very useful computational tool since fast
computational algorithms are available for SH transform \citep{DriscollH1994,Fan2015}.

In order to obtain the quadrature points $\{\zeta_{jk}\}$ and corresponding quadrature weights $\{\lambda_{jk}\}$, we make use of the HEALPix construction due to \cite{GorskiEtAl2005} that partitions $\mathbb{S}^2$
into $N_{j,pix} = 12 N_{j,side}^2=12\times 2^{j-1}$ spherical triangles of equal area, where $N_{j,side}$ is a power of
two determining the resolution. Then, $\zeta_{jk}$'s are chosen as the centroid
of the triangles, while $\lambda_{jk}$'s are all equal to $4\pi/N_{j,pix}$.

Since the FODs are symmetric and real-valued, we construct symmetrized needlets which can be easily derived from the needlets. 
Henceforth (with slight abuse of notations), $\{\Phi_{lm}\}$ denote the real symmetric
SH functions where $l=0,2,4,\cdots$ \citep{AtkinsonH2012,DescoteauxAFD2007} and $\{\psi_{jk}\}$ denote the symmetrized needlets functions. See the Section \ref{sec:design} in  Supplementary Material for details.

\subsection{Regression model for D-MRI measurements}\label{subsec:convolution_model}

In this section, we first describe the spherical convolution model  that relates FOD,
denoted by $F(\cdot)$, with the diffusion signal function, denoted by $S(\cdot)$.
We view the observed diffusion weighted measurements $\mathbf{y}=\{y_i\}_{i=1}^n$ corresponding to $n$ gradient
directions as noise corrupted samples
from the diffusion signal function $S(\cdot)$ evaluated at the respective gradient directions.
After representing the FOD  in the SH basis, we can then model the observed  measurements by a linear
regression model where the needlet coefficients of the FOD are the  regression coefficients.

Following, \cite{TournierEtal2004}, \cite{TournierEtal2007},
\cite{SakaieL2007} and \cite{LengletEtAl2009},
it is assumed that the diffusion signal function $S(\cdot)$ is a spherical convolution
of the FOD $F : \mathbb{S}^2 \to \mathbb{R}^+$, a symmetric spherical
distribution; with an azimuthal symmetric kernel $R : L^1([-1,1])
\to \mathbb{R}$. The kernel $R$,  referred to as the \textit{response function},
represents the local diffusion characteristics of water molecules along neuronal fibers,
which are assumed to be the same across different fiber bundles and voxels.  In the following,
we assume that the response function $R(\cdot)$, and hence its  SH coefficients, are known.
In practice, we may assume that $R(\cdot)$ belongs to a parametric
family of nonnegative functions on $[-1,1]$ (e.g., specified by a tensor model) and then estimate the parameters based on voxels with a single dominant fiber bundle, typically characterized by high
fractional anisotropy (FA) values. Experimental results show that our method is robust to the specification of the response function (results not reported).

Denote $S = R \star F$, and it is defined as
\begin{equation}\label{eq:FOD_convolution}
S(\mathbf{x}) = \int_{\mathbb{S}^2} R(\mathbf{x}^T \mathbf{y}) F(\mathbf{y})
d\omega(\mathbf{y}), \qquad \mathbf{x} \in \mathbb{S}^2.
\end{equation}


Diagonalizing in the SH basis, the SH coefficients of the diffusion signal function $S(\cdot)$ follow:
\begin{equation}\label{eq:convolution_SH_diagonal}
s_{lm} := \langle S, \Phi_{l,m}\rangle = \sqrt{\frac{4\pi}{2l+1}} r_l f_{lm}, ~~~l=0,2,\cdots, m=-l,\cdots, 0, \cdots, l.
\end{equation}
where $r_l := \langle R, \Phi_{l,0}\rangle$ and $f_{lm} := \langle F,
\Phi_{l,m}\rangle$, are the rotational harmonics and spherical harmonics coefficients of the response function and FOD, respectively.  Moreover,  by the orthonormality of the SH basis,
we can express the diffusion signal function as:
\begin{equation*}
S(\mathbf{x}) = \sum_{l}\sum_{m=-l}^l s_{lm} \Phi_{l,m}(\mathbf{x}) = \sum_{l}
\sqrt{\frac{4\pi}{2l+1}} r_l  \sum_{m=-l}^l f_{lm} \Phi_{l,m}(\mathbf{x}).
\end{equation*}

In the following, we assume that both  $S(\cdot)$ and $F(\cdot)$ can be well approximated in a finite order real symmetric SH basis
$\{\Phi_{l,m} :  -l \leq m\leq l\}_{l=0,2,\cdots, l_{\max}}$ consisting of $L =
(l_{\max} + 1)(l_{\max}+2)/2$ basis functions. The observed DWI measurements thus can be modeled as:
\begin{equation}
\label{eq:dwi}
\mathbf{y} = \bs\Phi \mathbf{R}
\mathbf{f} + \bs\varepsilon,
\end{equation}
where $\mathbf{R}$ is an $L \times L$ diagonal
matrix with diagonal elements $\sqrt{4\pi/(2l+1)} r_l$ (in blocks of length
$2l+1$); $\mathbf{f}= (f_{lm})$ is the $L \times 1$ vector of SH
coefficients of the FOD $F$; $\bs\Phi$ is the $n \times L$ matrix with the $i$-th row
being $\{\Phi_{l,m}(\theta_i,\phi_i)\}$,
where $\theta_i$ and $\phi_i$ are elevation angle and azimuthal angle, respectively, of
the $i$-th  gradient direction. The $i$-th coordinate of $\mathbf{y}=(y_i)_{i=1}^n$ corresponds to the observed diffusion measurement along the $i$-th gradient direction; 
 and $\bs\varepsilon=(\epsilon_i)_{i=1}^n$ is an $n \times 1$ vector
representing observational noise and possible approximation error.

\subsection{$\ell_1$-penalized estimation of FOD under needlets representation}\label{subsec:constrained_FOD}


Since 
	the FOD $F(\cdot)$ is expected to 
be either a constant function on the sphere (when the diffusion is isotropic) or a spherical function with a few sharp peaks (each corresponding to a distinct major fiber bundle),
 the localization and tight frame properties of the needlets would imply that the
needlet coefficients of $F(\cdot)$ form a sparse vector \citep{NarcowichPW2006b}, i.e., $F(\cdot)$  can be well approximated by a small fraction of needlet functions.

Consider a symmetrized needlets frame with the first $N= 1 + (\sum_{1 \leq j \leq j_{max}} N_{j,pix})/2= 2^{2 j_{\max}+1}-1$
frame elements, where  $j_{max} \geq 1$ is the maximum level of needlets being used. $j_{\max}$ is set to be $\lceil\log_2(l_{\max})+1\rceil$ such that SH functions  up to level $l_{\max}$  can be linearly 
represented in the first $N$ needlets.  Denote the $N \times 1$ needlet
coefficients vector of $F(\cdot)$ by $\bs\beta$.
Then the SH coefficients $\mathbf{f}$ of $F(\cdot)$ can be expressed as $\mathbf{f} = \mathbf{C} \bs\beta$, where $\mathbf{C}$ is an $L \times N$ matrix 
(see Section \ref{sec:design} in Supplementary Material).
This allows us to rewrite equation (\ref{eq:dwi}) as
\begin{equation}\label{eq:dwi_representation}
\mathbf{y} = \bs\Phi \mathbf{R} \mathbf{C} \bs\beta + \bs\varepsilon.
\end{equation}
A key point here is that $\mathbf{C}$ can be
easily computed, and so is the design matrix $\bs\Phi \mathbf{R} \mathbf{C}$.
Furthermore, since the response function $R$ is the same across voxels, the design matrix needs to be computed only once. 

The sparseness of the FOD needlet coefficients $\bs\beta$  
motivates us to
propose a  penalized regression estimate:
\begin{equation}\label{eq:wavelet_penalized_regression}
\widehat{\bs\beta}_\lambda = \arg\min_{\bs\beta : \tilde{\bs\Phi} \mathbf{C} \bs\beta
\succeq \mathbf{0}} ~\parallel \mathbf{y} - \bs\Phi \mathbf{R} \mathbf{C}
\bs\beta\parallel^2 + P_\lambda(\bs\beta)
\end{equation}
where $P_\lambda(\bs\beta)$ denotes a sparsity-inducing penalty, with the tuning parameter $\lambda \geq 0$ controlling the degree of regularization, and $\tilde{\bs\Phi}$ 
is the matrix of SH basis functions (up to level $l_{\max}$) evaluated on a pre-specified dense evaluation grid.
The constraint $\tilde{\bs\Phi}
\mathbf{C} \bs\beta \succeq \mathbf{0}$ ensures that the estimated $F(\cdot)$
evaluated on this grid, i.e., $\widehat{F}:=\tilde{\bs\Phi} \mathbf{C} \hat{\bs\beta}$, is nonnegative.   
In the subsequent experiments, we use an equiangular grid with $2562$ grid points. 

Following \cite{Tibshirani1996} and subsequent developments in the statistical literature, we propose to use  an $\ell_1$ penalty
$P_\lambda(\bs\beta) = \lambda \sum_{j,k} |\beta_{jk}|$.
The estimation problem is then a  convex
optimization problem with a non-negativity constraint.
We develop  a computationally efficient algorithm based on the
\textit{Alternating Direction  Method of Multipliers (ADMM)} \citep{BoydEtAl2011,SraNW2012}  to solve (\ref{eq:wavelet_penalized_regression}).  ADMM  is a general-purpose algorithm for solving convex optimization problems with constraints.
Finally, we rescale the estimated $F(\cdot)$ such that it integrates to one on the unit sphere.
The details of the ADMM algorithm is given in Section \ref{subsec:ADMM} in Supplementary Material.


Since our goal here is to get a good estimate of FOD, particularly one that is useful for subsequent analyses such as tractography,  
thus slight overfitting 
is less detrimental than underfitting. 
Thus we propose a criterion which chooses  the largest $\lambda$ such that the penalty parameter values  smaller than this value will lead to essentially the same residual sum of squares (RSS)  (See Section \ref{subsec:penalty} in Supplementary Material for details). 
The commonly used model selection criteria such as BIC 
\citep{Schwarz1978} and AIC \citep{Akaike1974} require specification of  the degrees of freedom for the model  which is difficult when the design matrix 
is ill-conditioned and non-smooth penalties such as the $\ell_1$ norm are used.
Experiments based on synthetic data show that the proposed strategy is able to strike a good balance between bias and variance in FOD estimation and leads to better results than BIC or AIC (results not reported).


After obtaining an estimated FOD, we may want to identify major fiber bundle orientation(s),
which could be used for subsequent analyses.
Ideally, this can be done through peak detection, i.e., locating the local maxima of the (estimated)
FOD. However, since the estimated FOD may have spurious peaks due to noise, we need to eliminate peaks
that are likely to be false. We propose a simple yet effective peak detection algorithm based on grid search, followed 
by a pruning step and a clustering step to filter out potential false peaks. Details are given in  Section \ref{subsec:peak} in  Supplementary Material.


\subsection{Synthetic Data Experiments}\label{subsec:synthetic_data}

In this section, we describe  experiments based on synthetic data  to study the performance of the proposed estimator,  referred to as \texttt{SN-lasso}, and to compare it with two competing FOD estimators described below. 

\subsubsection{Competing FOD estimators}

In addition to the proposed \texttt{SN-lasso},  we also consider two existing FOD estimators:

(i) The \texttt{SH-ridge} estimator  \citep{TournierEtal2004} through a ridge type regression by minimizing: 
\begin{equation*}\label{eq:LB_regression}
\parallel
\mathbf{y} - \mathbf{\Phi} \mathbf{R}  \mathbf{f}\parallel_2^2 +
\lambda \mathbb{E}(F), ~~~\mathbb{E}(F): = \int_\Omega(\Delta_b F)^2d\Omega = \mathbf{f}^T \mathbf{P} \mathbf{f}, 
\end{equation*}
where  $\Delta_b$ is the spherical Laplacian operator and $\mathbf{P}$ is a diagonal matrix with entries $l^2(l+1)^2$ in blocks of size $2l+1$ ($l=0,2,\cdots, l_{\max}$).  $\mathbb{E}(F)$, referred to as the Laplace-Beltrami penalty, is a measure of roughness of spherical functions. The \texttt{SH-ridge} estimator is solved explicitly by:
$$
\hat{\mathbf{f}}^{LB} = (\mathbf{R}^{T}\mathbf{\Phi}^{T}\mathbf{\Phi} \mathbf{R} + \lambda \mathbf{P})^{-1}\mathbf{R}^T \mathbf{\Phi}^T \mathbf{y},~~~
\hat{F}^{LB}= \sum_{l,m} \hat{f}^{LB}_{l,m} \Phi_{l,m}.
$$
The penalty parameter $\lambda$ can be chosen by the Bayesian Information Criterion \citep{Schwarz1978}.

(ii) \texttt{super-CSD} estimator \citep{TournierEtal2007}. The idea is to suppress  small values of the estimated FOD and consequently sharpen the peak(s) of the FOD estimator using an SH representation of order 
$l^s_{\max}$. In our experiments, we apply \texttt{super-CSD} algorithm to the \texttt{SH-ridge} estimates and consider $l^s_{\max}=8,12, 16$. We refer to the corresponding estimators as \texttt{SCSD8}, \texttt{SCSD12}, and \texttt{SCSD16},  respectively.
For other parameters in the \texttt{super-CSD} algorithm, we follow the recommended values in \cite{TournierEtal2007}. For more details, see Section \ref{subsec:scsd} in Supplementary Material.

\subsubsection{Experimental setting}\label{subsec:simul_separation}

We consider FOD estimation for a voxel with various scenarios of  fiber populations. We use $sep$ to denote the separation angle between a pair of crossing fiber bundles. Our experimental settings include $K=2$ fiber bundles crossing at $sep=90^{\circ}, 75^{\circ}, 60^{\circ}, 45^{\circ}, 30^{\circ}$. We also consider  $K=0$ fiber bundle, i.e., isotropic diffusion; $K=1$ fiber bundle, i.e., no crossing fiber; and 
$K=3$ fiber bundles with pairwise crossing at 
$sep=90^{\circ},75^{\circ}, 60^{\circ}$.

We simulate noiseless diffusion weighted signals according to the convolution model (\ref{eq:FOD_convolution}), where the true FOD $F$:
\begin{equation*}
F (\theta, \phi)= \sum_{k=1}^K  w_k \delta_{\theta_k,\phi_k}(\theta, \phi), ~~~ \theta \in  [0,\pi], ~~ \phi \in [0, 2 \pi),
\end{equation*}
with  $w_k > 0, ~~\sum_{k=1}^K w_k= 1$ being the volume fractions and  $\theta_k$ (elevation angle) and $\phi_k$ (azimuthal angle) being the spherical coordinates  of the orientation of the $k$-th fiber  bundle, respectively. 
The response function  is set as: 
$$
R(\cos(\theta))= S_0\exp^{-b (\lambda_1 \sin^2\theta  + \lambda_3 \cos^2\theta )}, ~~~ \theta \in  [0,\pi],
$$
where throughout we fix $\lambda_3=1\times 10^{-3} mm^2/s$ and set the ratio between $\lambda_3$ and $\lambda_1$ as $ratio=10$. 
The  volume fractions are set as $w_1=w_2=0.5$ for the two-fiber case and $w_1=w_2=0.3, w_3=0.4$ for the three-fiber case.

In terms of the D-MRI experimental parameters, we consider \textit{bvalue} at $b=1000s/mm^2, 3000s/mm^2$ and  $5000s/mm^2$ (for small crossing angles only); and three angular resolutions, namely, $n=41, 81, 321$  gradient directions on an equiangular grid. These settings aim to cover both commonly used values in large-scale D-MRI experiments such as ADNI as well as high \textit{bvalue} and/or high angular resolution D-MRI experiments.

The observed diffusion weighted measurements along the $n$ gradient directions are generated by adding independent Rician noise  \citep{GudbjartssonP1995,HahnPHH2006,PolzehlT2008}
to the respective noiseless diffusion signals.
The signal-to-noise ratio (SNR), defined as the ratio between the $b_0$ image intensity $S_0$ and the Rician noise level $\sigma$ \citep{TournierEtal2007}, is set at $SNR :=S_0/\sigma=20$ and $50$ (for small crossing angles only).


\subsubsection{Evaluation metrics}

We report the statistical characteristics of the estimators (across $100$ independent data sets)  by depicting  the mean estimated FOD  in opaque color and the mean plus two standard deviations in translucent color. We also examine the performance of each estimator using various numerical metrics including: 
(i) The success rate of the peak detection algorithm applied to the $100$ estimated FODs, where `success' means that the algorithm identifies the correct number of fiber bundles.
(ii) Among the successful estimators, the mean angular errors between the identified peaks and the true fiber  directions, as well as the bias of the estimated separation angle between pairs of fiber bundles. 
Throughout,  ``true FOD" refers to the true FOD projected on to an SH basis. 

\subsection{Real D-MRI Data Experiments}\label{sec:application}

Data used in the preparation of this article were obtained from the Alzheimer’s Disease Neuroimaging
Initiative (ADNI) database (\url{adni.loni.usc.edu}). The ADNI was launched in 2003 as a public-private
partnership, led by Principal Investigator Michael W. Weiner, MD. The primary goal of ADNI has been to
test whether serial magnetic resonance imaging (MRI), positron emission tomography (PET), other
biological markers, and clinical and neuropsychological assessment can be combined to measure the
progression of mild cognitive impairment (MCI) and early Alzheimer’s disease (AD).


Specifically, we analyze D-MRI data sets from two participants in the second phase of the ADNI project (ADNI-2).  Both participants were scanned on 3 Tesla MRI machines produced by General Electric.  The scanning protocol was analogous across scanners, and was optimized prior to the study to provide harmonized data across scanners.  One participant was a 57 year old cognitively healthy female scanned at the University of Rochester.  The other participant was a 56 year old cognitively healthy female scanned at the Wein Center for Alzheimer’s Disease and Memory Disorders.

We use eddy-current-corrected diffusion images  with diffusion signals measured on a $256 \times 256 \times 59$ 3D grid  along $41$ distinct gradient directions under $b=1000s/mm^2$. There are also five $b_0$ images based on which we estimate the $b_0$ image intensity  $S_0$  and the Rician noise level $\sigma$. The signal-to-noise ratio $S_0/\sigma$ has median $40$ and $36$ (across voxels) for these two data sets, respectively.  We first fit the single diffusion tensor model \citep{LeBihan1995,Basser2002,Mori2007,CarmichaelCPP2013} to each voxel and calculate the \textit{fractional anisotropy (FA)} value and \textit{mean diffusivity (MD)}.
 We then 
identify voxels with a single dominant fiber bundle characterized by $FA>0.8$ and the ratio between the two minor eigenvalues of the tensor $<1.5$. We use these voxels to estimate the response function $R$ under a Gaussian diffusion model. The (estimated) response functions have the leading eigenvalue $1.5\times 10^{-3} mm^2/s$ and $2\times 10^{-3} mm^2/s$ for the two data sets, respectively. The ratios between the leading and the minor eigenvalue are $6.5$ and $7.3$ for the two data sets, respectively.
See Section \ref{sec:realdatasupp} in Supplementary Material for more details.

We look into a $16 \times 16$ region  of interest (ROI) with $x$-slice coordinates from $108\sim 123$, $y$-slice coordinates from $124 \sim 139$ at $z$-slice $40$ from participant 1 (referred to as ROI I). This region is at the crossing of corona radiata and corpus callosum and we use it as an example for regions with crossing fibers. ROI I is indicated by the white box in Figure \ref{fig:realdata_regionb}(a), left panel (also see Figure \ref{fig:realdata_FOM}(a) for an enlarged plot). 
In the same figure, we also show the fiber orientation colormap, FA map and MD map zoomed into this region.   The fiber orientation colormaps depict the orientation of the principal eigenvector  under a single tensor model, where  red  indicates left-to-right direction, green indicates up-to-down direction and blue indicates into-to-out-of-page direction. Moreover, fiber orientation colormaps are modulated by the FA values such that voxels with small FA values are darker. The FA map and MD map are also calculated under the single tensor model,  where darker colors correspond to smaller FA/MD values.
The second region we consider is a $10\times 10$ region with $x$-slice from $112 \sim 121$, $y$-slice from $97 \sim 106$ and $z$-slice $32$ from participant 2 (referred to as ROI II). This region is shown by the white box on Figure \ref{fig:realdata_regionc}(a), left panel (also see Figure \ref{fig:realdata_FOM}(b) for an enlarged plot).  
We choose this region to illustrate a scenario where there is a mixture of
	voxels with a single direction (here in corpus callosum) and isotropic diffusion (here cerebrospinal fluid (CSF)).

\section{Results}
\label{sec:results}
\subsection{Synthetic Data Experiments} \label{sec:details_synthetic_data}

In the main text, we only report graphical summaries of the results for some of  the two-fiber ($K=2$) crossing cases with 
the number of gradient directions being $n=41$. 
We shall comment on the results under the rest of the settings and defer details to the Supplementary Material.

\begin{figure}[H]
	\centering
	\caption{
		\small{\textbf{Two fiber crossing at $\mathbf{90^{\circ}, 75^{\circ}, 60^{\circ}}$ with $\mathbf{b=1000 s/mm^2,SNR=20}$}.  $n=41$ gradient directions and $ratio=10$. 
			The lines indicate the true fiber directions, the opaque part in the plots corresponds to mean estimated FOD across $100$ replicates, and the translucent part in the plots corresponds to mean plus two standard deviations of the estimated FOD.}
		\label{fig:b1_ratio10_lmax8_N41}    }
	\begin{tabular}{cc}
		&True FOD \hskip0.3in \texttt{SH-ridge}	\hskip0.3in \texttt{SCSD8}\hskip0.5in \texttt{SCSD12}\hskip0.55in \texttt{SN-lasso}\\
		$90^{\circ}$& \includegraphics[width=0.8\textwidth]{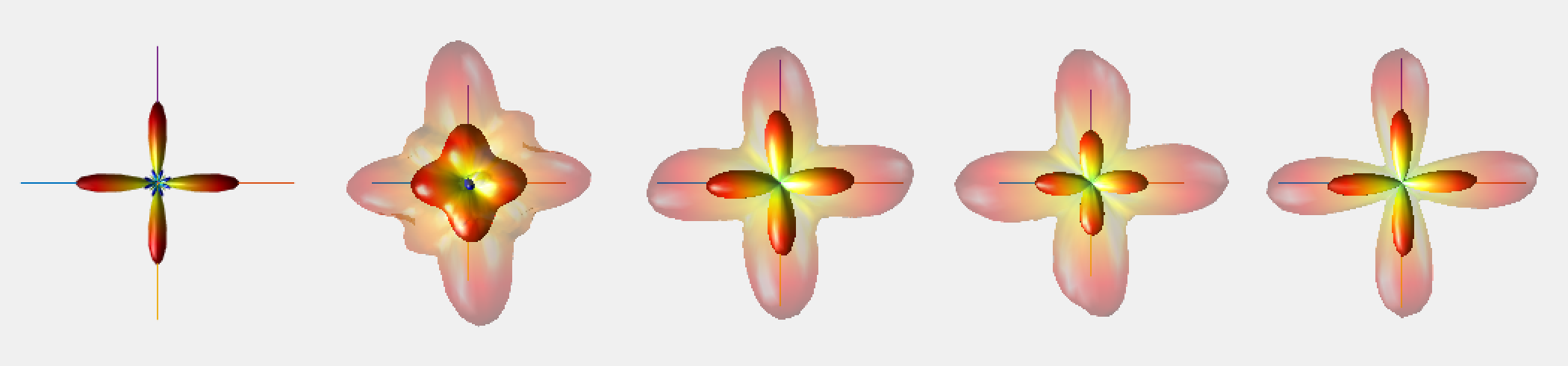}\\
		$75^{\circ}$&\includegraphics[width=0.8\textwidth]{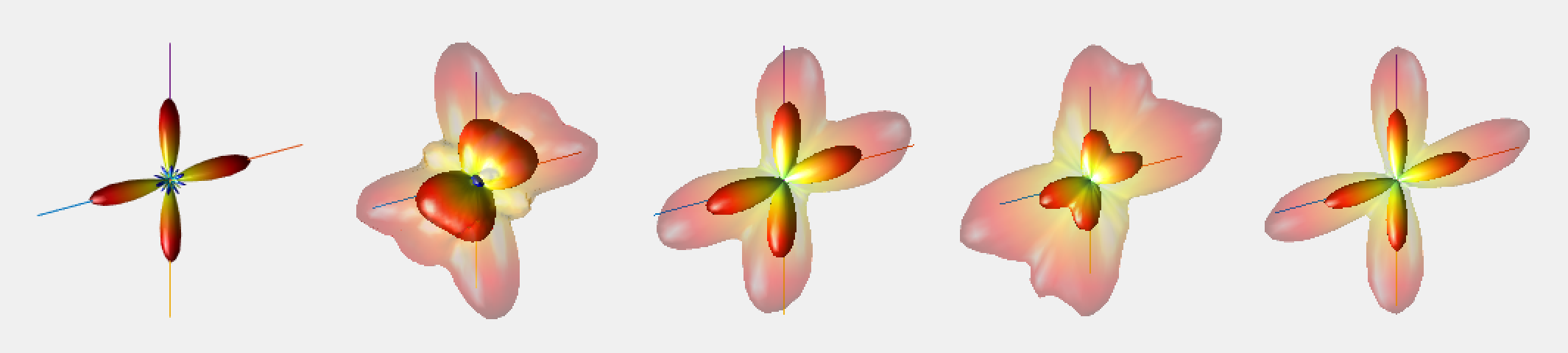}\\
		$60^{\circ}$&\includegraphics[width=0.8\textwidth]{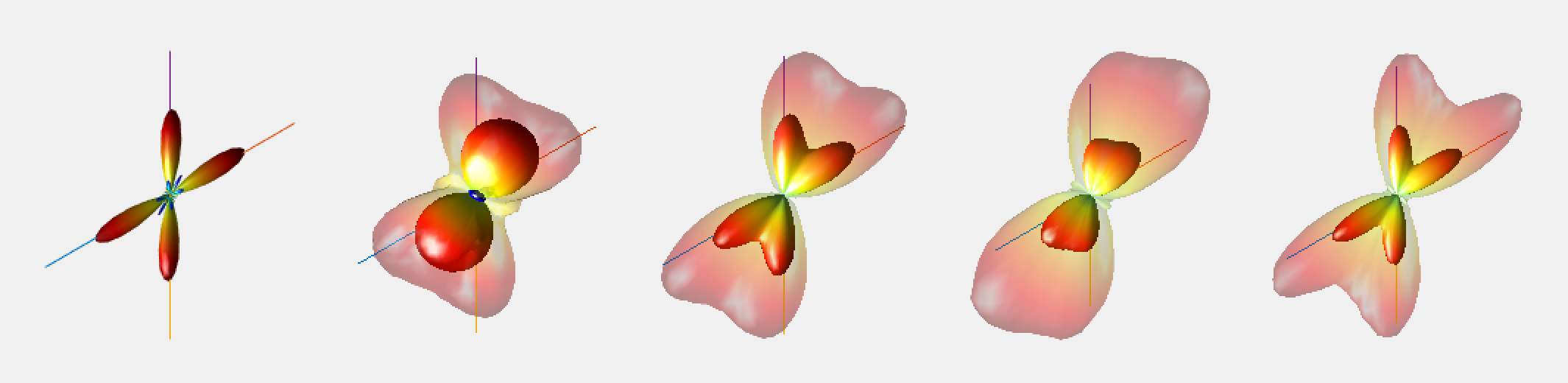}\\
	\end{tabular}
\end{figure}

\begin{center}
	\begin{figure}[H]
	\caption{\textbf{Two fiber crossing at $\mathbf{45^{\circ}}$ with $\mathbf{b=1000, 3000 s/mm^2,SNR=20, 50}$}. $n=41$ gradient directions and $ratio=10$. The lines indicate the true fiber directions; the opaque part  corresponds to mean estimated FOD across $100$ replicates; and the translucent part  corresponds to mean plus two standard deviations of the estimated FOD. 	\label{fig:b1_ratio10_lmax8_N41_45} }
	\begin{tabular}{cc}
		
		&True FOD \hskip0.3in \texttt{SH-ridge}	\hskip0.3in \texttt{SCSD12}\hskip0.5in \texttt{SCSD16}\hskip0.6in \texttt{SN-lasso}\\
			&$b=1000 s/mm^2,SNR=20$\\
		$45^{\circ}$&\includegraphics[width=0.8\textwidth]{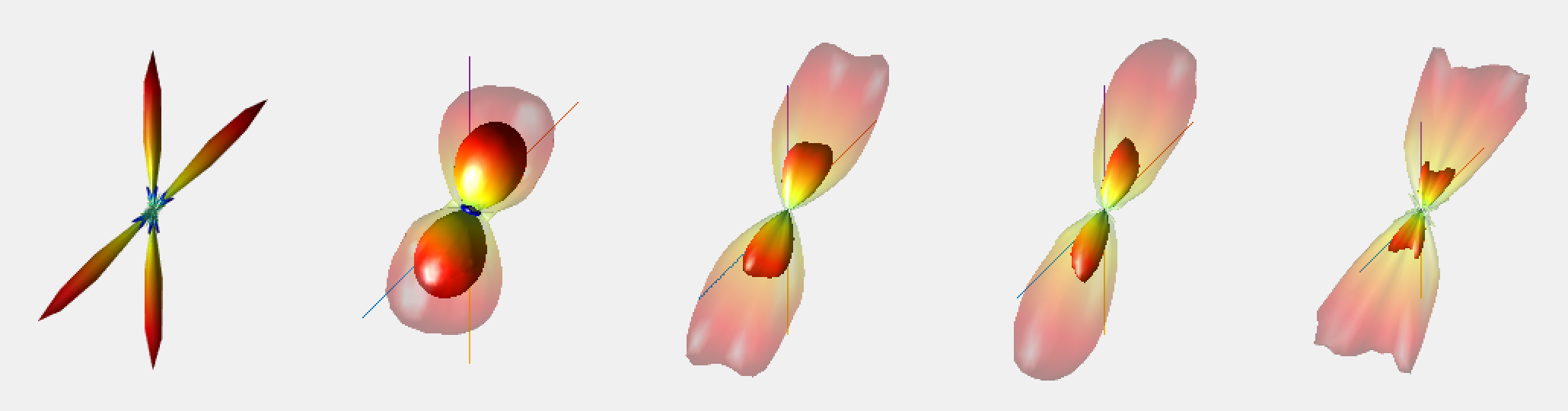}\\
	
		&$b=1000s/mm^2,SNR=50$\\
		$45^{\circ}$&\includegraphics[width=0.8\textwidth]{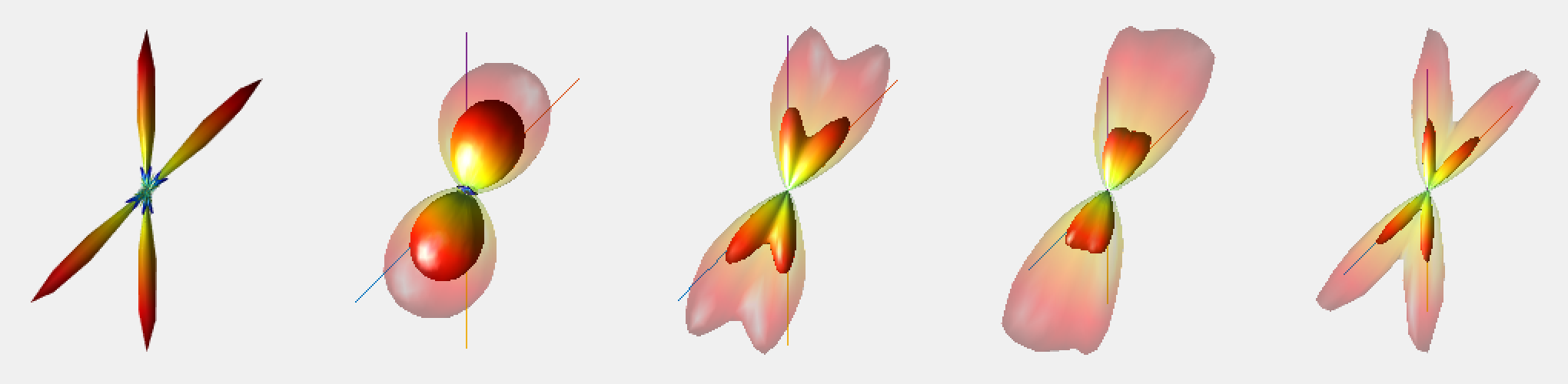}\\	
			&$b=3000s/mm^2, SNR=20$\\
			$45^{\circ}$& \includegraphics[width=0.8\textwidth]{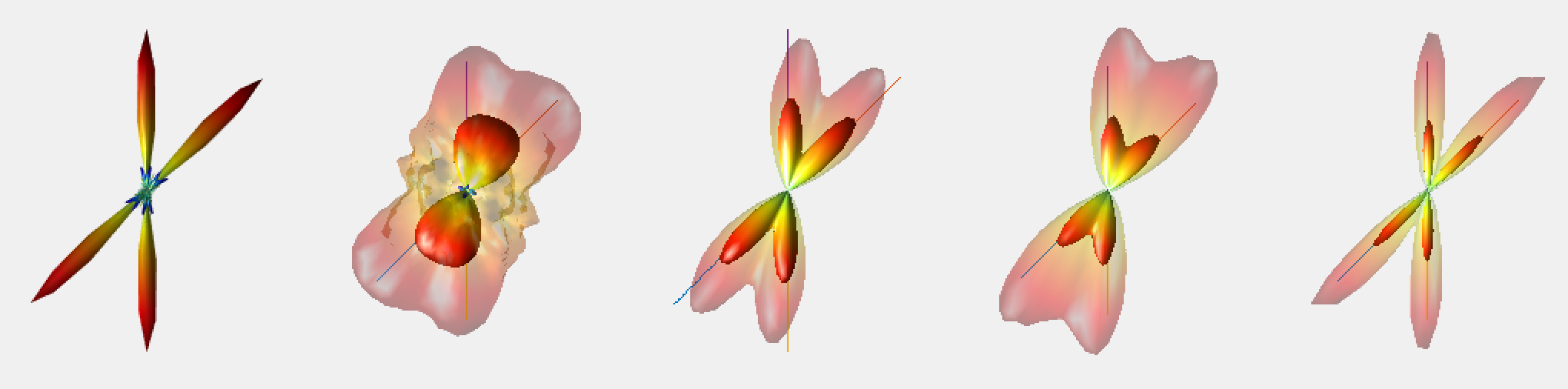}\\
		&$b=3000s/mm^2,SNR=50$\\
		$45^{\circ}$&\includegraphics[width=0.8\textwidth]{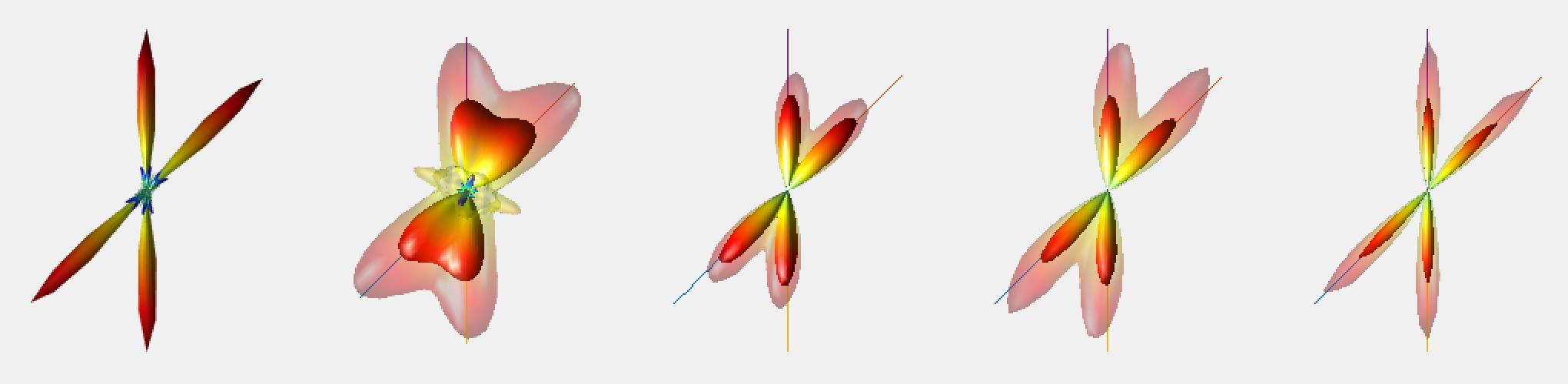}\\
			
	\end{tabular}
	\end{figure}
\end{center}

	As can be seen from Figure \ref{fig:b1_ratio10_lmax8_N41}, when the crossing angle is large-to-moderate ($90^{\circ}, 75^{\circ}, 60^{\circ}$), even at relatively low \textit{bvalue} and SNR ($b=1000s/mm^2,SNR=20$), \texttt{SN-lasso} leads to satisfactory reconstruction of FOD, particularly in terms of accurately identifying the peak directions. 
	\texttt{SCSD8}  and \texttt{SCSD12} also show various degrees of success in these cases, though their performances are not as good as that of \texttt{SN-lasso} (\texttt{SCSD16} estimates are very variable and thus omitted). Particularly,  \texttt{SN-lasso} estimates   have much more localized and sharper peaks.  
	While crossing at a  smaller angle ($45^{\circ}$; Figure \ref{fig:b1_ratio10_lmax8_N41_45}), none of these methods works well when \textit{bvalue} and SNR are both low ($b=1000s/mm^2,SNR=20$). However, when either parameter is increased ($b=3000s/mm^2$ and/or $SNR=50$), \texttt{SN-lasso} is able to accurately recover the  sharp features of FOD. \texttt{SCSD} estimators  are again much less localized (\texttt{SCSD8} estimates are overly smoothed and thus omitted). 
	
\begin{center}
	\begin{figure}[H]
	\caption{\textbf{Two fiber crossing at $\mathbf{30^{\circ}}$ with $\mathbf{b=3000, 5000 s/mm^2, SNR=20, 50}$}. $n=41$ gradient directions and $ratio=10$. The lines indicate the true fiber directions; the opaque part  corresponds to mean estimated FOD across $100$ replicates; and the translucent part  corresponds to mean plus two standard deviations of the estimated FOD. 	\label{fig:b1_ratio10_lmax8_N41_30} }
	\begin{tabular}{cc}
		
		&True FOD \hskip0.3in \texttt{SH-ridge}	\hskip0.3in \texttt{SCSD12}\hskip0.5in \texttt{SCSD16}\hskip0.6in \texttt{SN-lasso}\\
			&$b=3000s/mm^2, SNR=20$\\
			$30^{\circ}$&\includegraphics[width=0.8\textwidth]{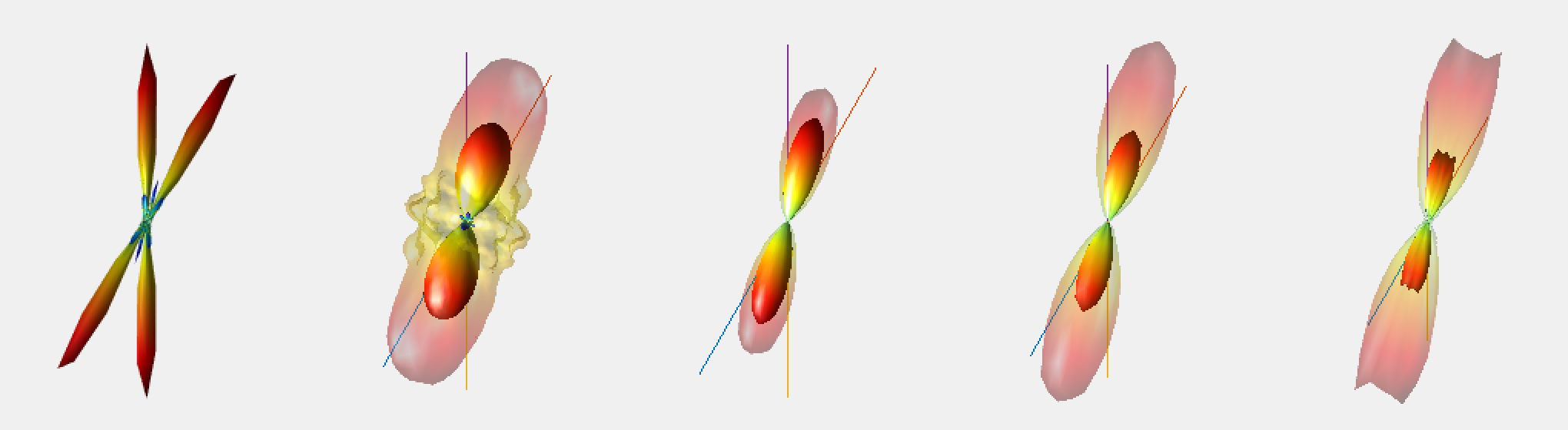}\\
		&$b=3000s/mm^2,SNR=50$\\
		$30^{\circ}$&\includegraphics[width=0.8\textwidth]{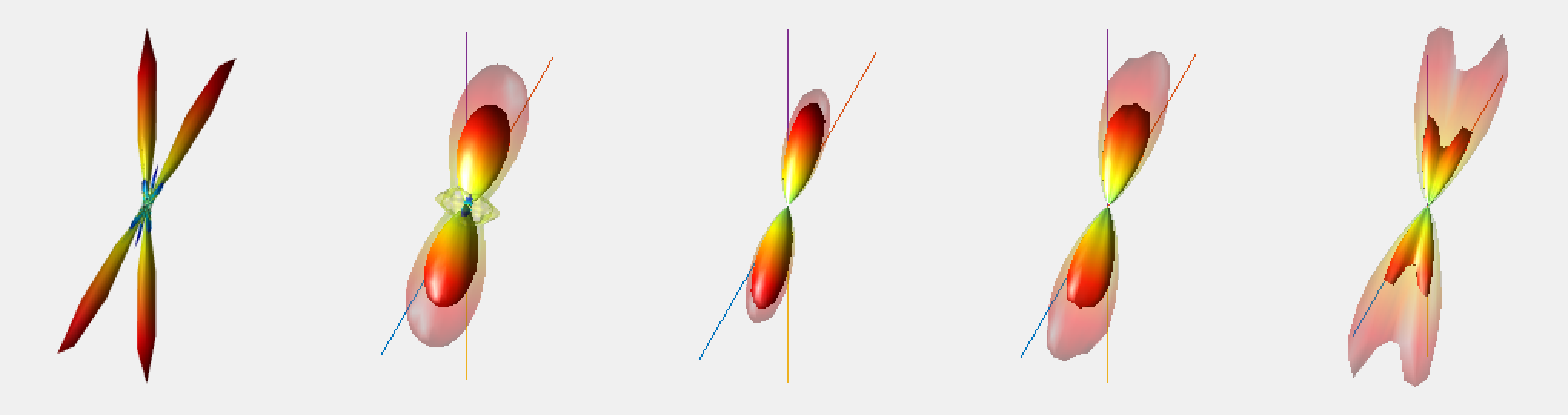}\\
		&$b=5000s/mm^2, SNR=20$\\
		$30^{\circ}$&\includegraphics[width=0.8\textwidth]{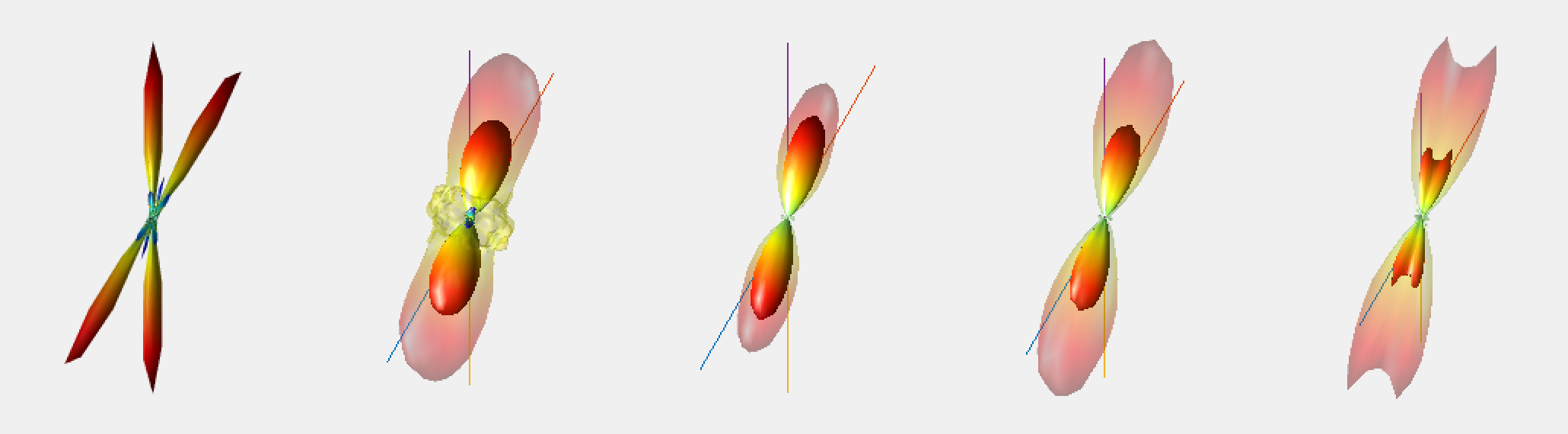}\\
		&$b=5000s/mm^2,SNR=50$\\
		$30^{\circ}$&\includegraphics[width=0.8\textwidth]{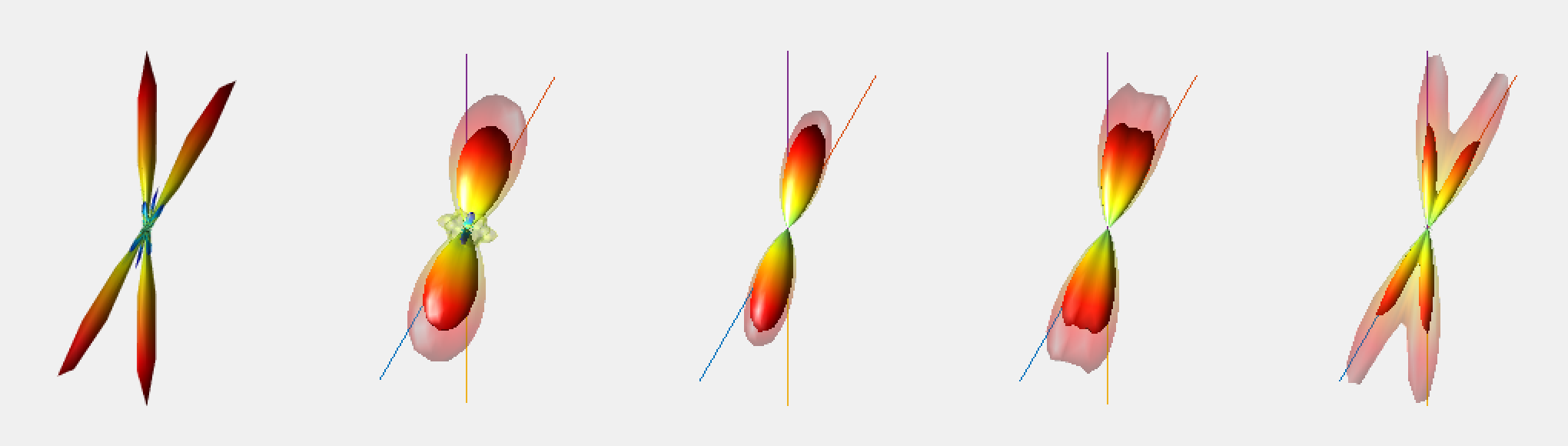}\\

	\end{tabular}
	\end{figure}
\end{center}

	For a small crossing angle ($30^{\circ}$; Figure \ref{fig:b1_ratio10_lmax8_N41_30}), \texttt{SN-lasso} works well when both \textit{bvalue} and SNR are high ($b=3000, 5000s/mm^2$ and $SNR=50$), while all other methods do a poor job in terms of  detecting crossing fibers irrespective of \textit{bvalue} and SNR. 
	
\begin{center}
	\begin{figure}[H]
		
		\caption{\textbf{Two fiber crossing at $\mathbf{45^{\circ}}$ with $\mathbf{b=1000, 3000 s/mm^2, SNR=20,50}$.}  $n=41$ gradient directions and $ratio=10$. 
			\label{fig:b1_ratio10_lmax8_N41_45_bar}}
		\begin{tabular}{c}
			(a)	Angular error in direction estimation as percentage of the separation angle\\
			\includegraphics[width=0.9\textwidth]{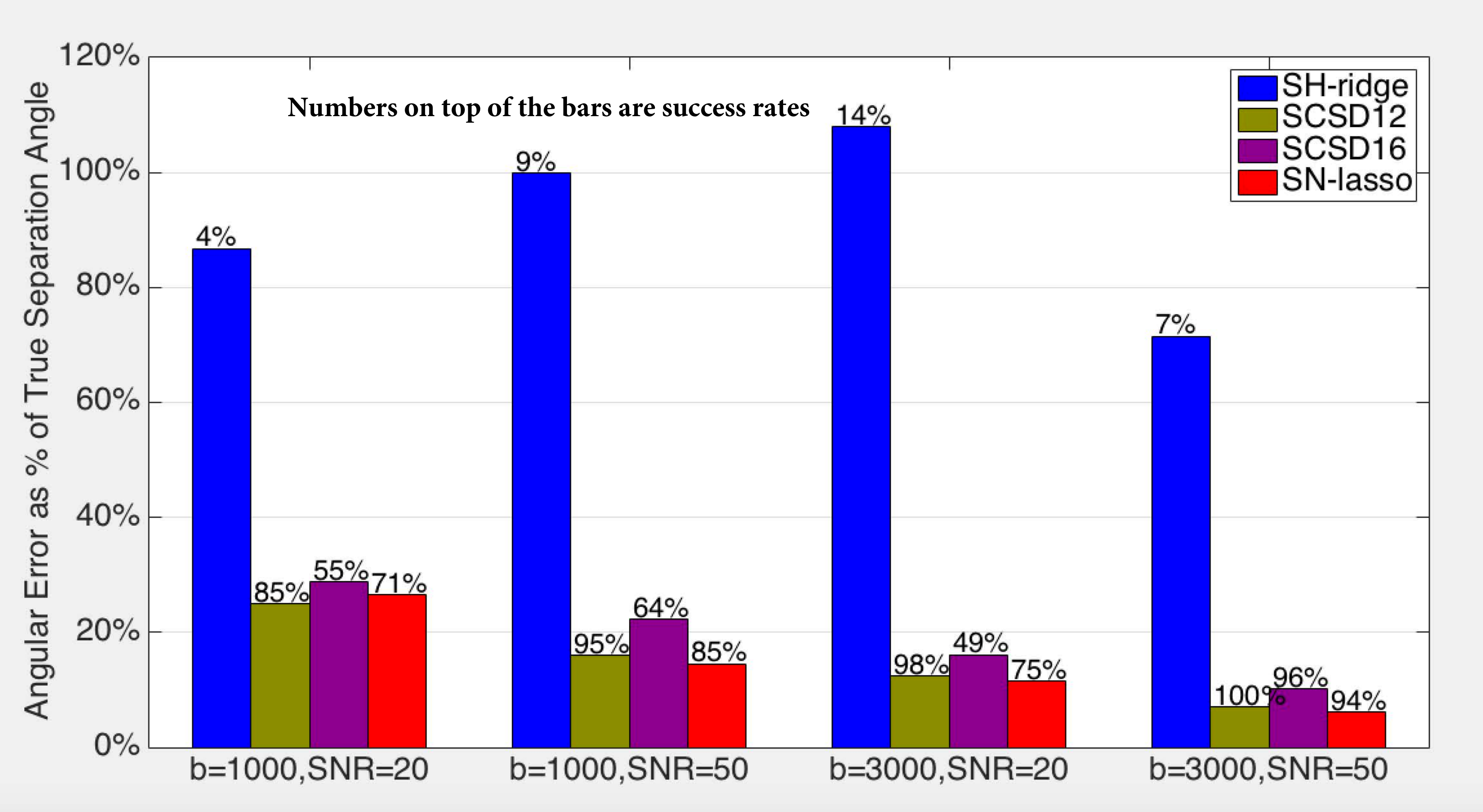}\\
			(b) Bias in separation angle estimation as percentage of the separation angle\\
			\includegraphics[width=0.9\textwidth]{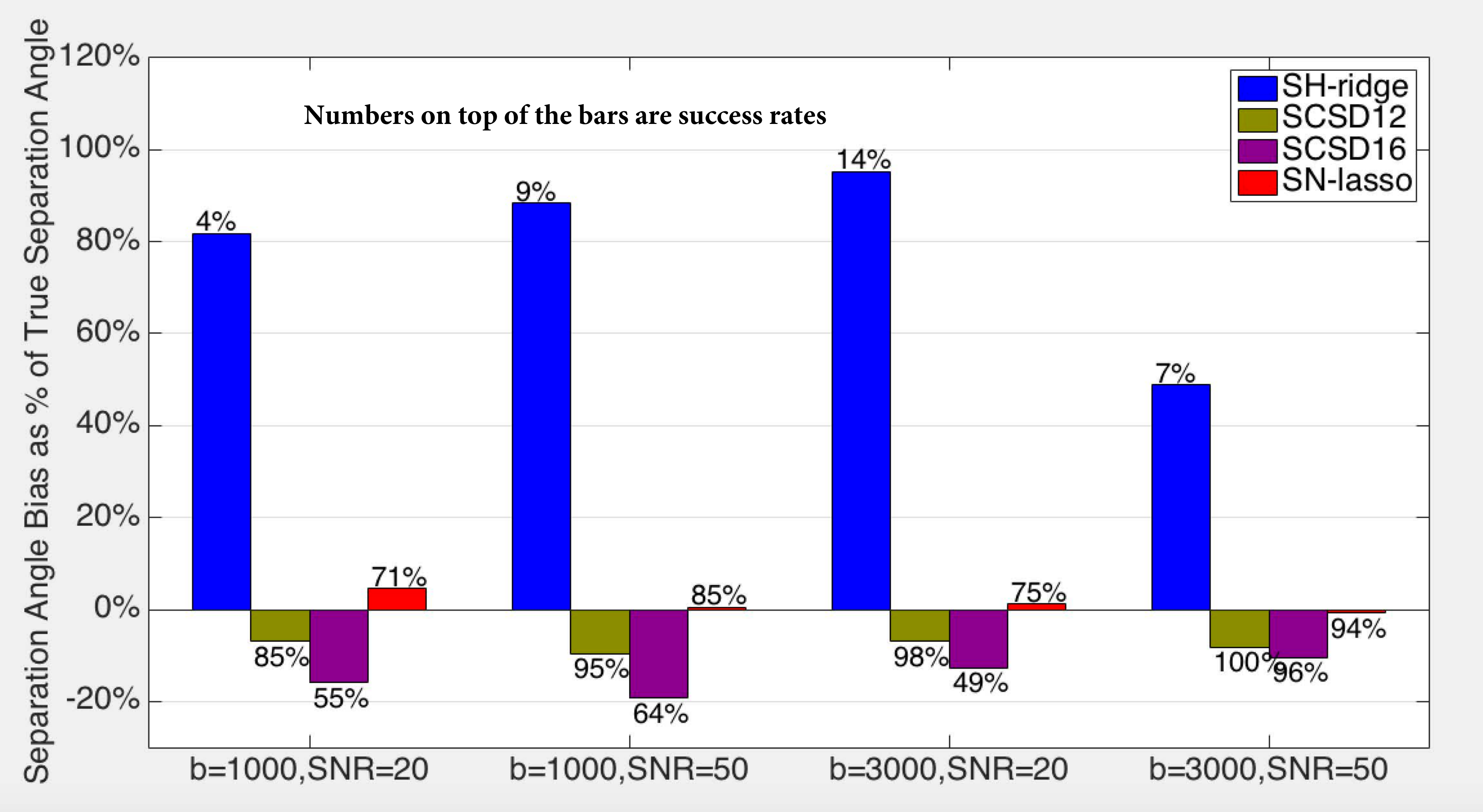}\\
		\end{tabular}
	\end{figure} 
\end{center}

\begin{center}
\begin{figure}[H]

	\caption{\textbf{Two fiber crossing at $\mathbf{30^{\circ}}$ with $\mathbf{b=3000, 5000 s/mm^2, SNR=20,50}$.}  $n=41$ gradient directions and $ratio=10$. 
	\label{fig:b1_ratio10_lmax8_N41_30_bar}}
		\begin{tabular}{c}
			(a)	Angular error in direction estimation as percentage of the separation angle\\
	\includegraphics[width=0.9\textwidth]{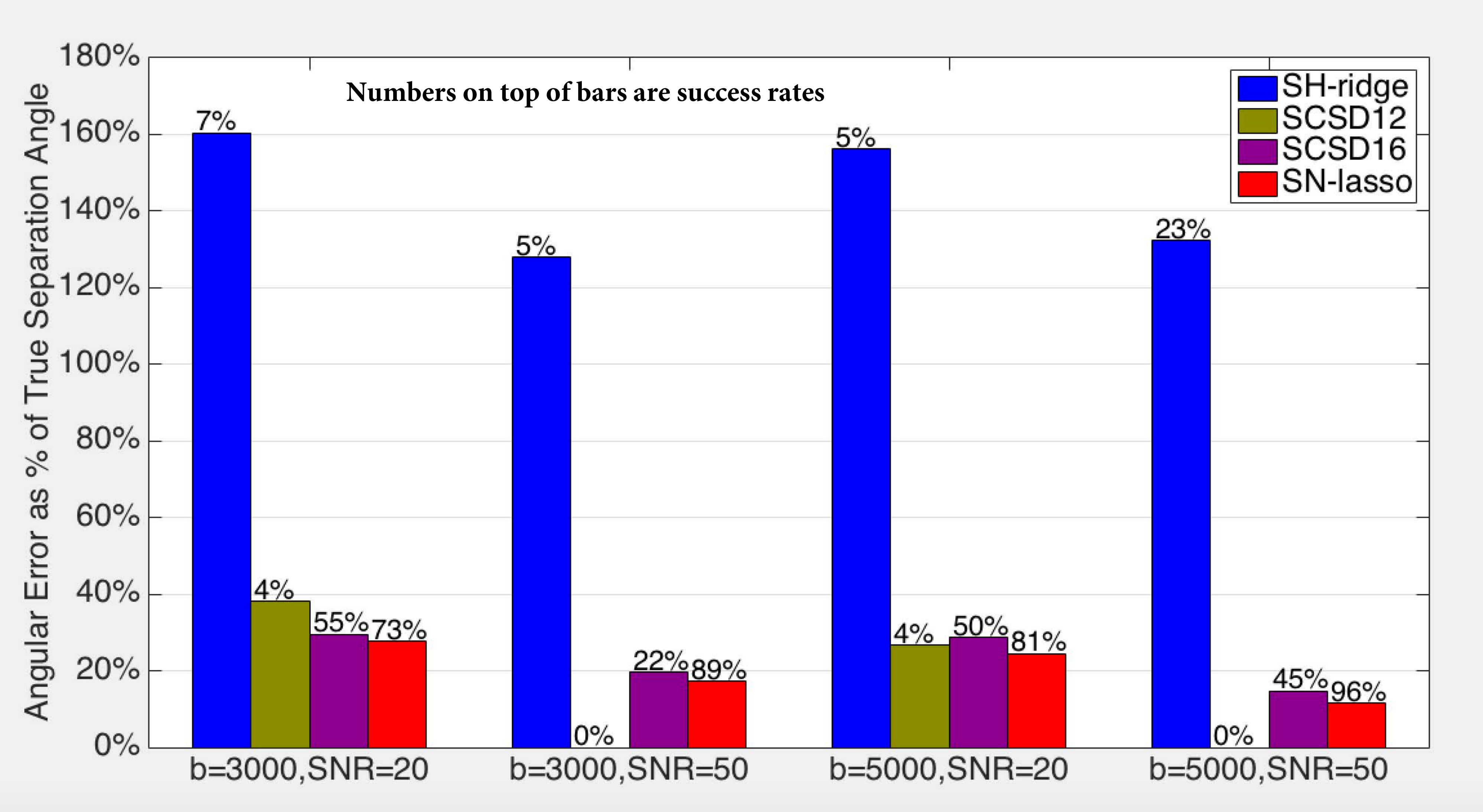}\\
	(b) Bias in separation angle estimation as percentage of the separation angle\\
		\includegraphics[width=0.9\textwidth]{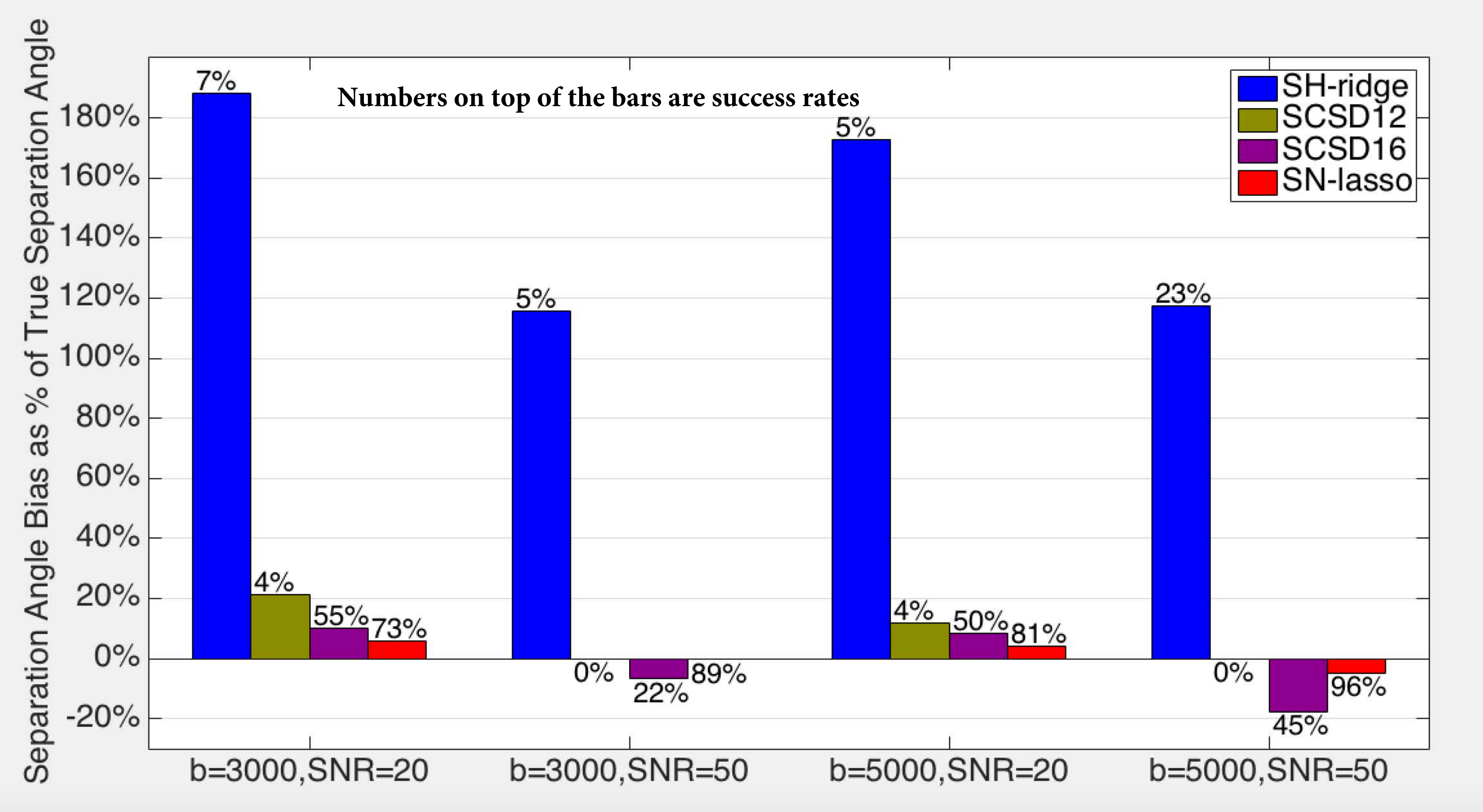}\\
	\end{tabular}
\end{figure} 
\end{center}

	In Figures \ref{fig:b1_ratio10_lmax8_N41_45_bar} and  \ref{fig:b1_ratio10_lmax8_N41_30_bar}, we depict success rate, angular error in direction estimation (averaged over the two directions)  and bias in separation angle estimation under $45^{\circ}$ and $30^{\circ}$ crossing, respectively. Under $45^{\circ}$ crossing, \texttt{SN-lasso} and \texttt{SCSD12} show similar performance in terms of success rate and angular error. However, \texttt{SN-lasso} shows much smaller bias in terms of separation angle estimation.  
		Under $30^{\circ}$ crossing,  \texttt{SN-lasso} has the best performance in all three aspects. More specifically, for all four combinations of \texttt{bvalue} and $SNR$, \texttt{SN-lasso} successfully identifies two peaks for at least 70\% of the replicates with an average angular error ranging from $2.7^{\circ} \sim 9 ^{\circ}$. For detailed numerical summaries, see Tables \ref{table:b1_ratio10_lmax8_N41}, \ref{table:Sep45_b13_ratio10_lmax16_N41}, \ref{table:Sep30_b35_ratio10_lmax16_N41} in the Supplementary Material.

These results  demonstrate that \texttt{SN-lasso} leads to satisfactory FOD estimation and direction identification  even when the crossing angle is as small as $30^{\circ}$ as long as the \textit{bvalue}  and SNR are sufficiently large. It also performs  the best among the competing methods with regard to these aspects. 
As large \textit{bvalue}  and SNR are being advocated with the advancement of MRI technologies  \citep{SetsompopEtAl2013, VanEssenEtAl2013}, \texttt{SN-lasso} holds great promise in resolving even subtle fiber crossing  patterns.  On the other hand, 	\texttt{SH-ridge} leads to overly smoothed estimates and consequent loss of directionality information, while \texttt{SCSD} estimators either tend to  have large variability (\texttt{SCSD12, SCSD16}) or overly smoothed estimates (\texttt{SCSD8}). Consistent with the observations in \cite{TournierEtal2007}, \texttt{SCSD8} works better for large-to-moderate crossing angles and \texttt{SCSD12, SCSD16} work better for small crossing angles.
However, without prior knowledge of the crossing angle, in practice it would be hard to choose which \texttt{SCSD} estimator to use.  In contrast, \texttt{SN-lasso} is able to  automatically adjust for different crossing angles  through the specification of the penalty parameter $\lambda$ which is determined in a data driven way.

\subsubsection*{Additional results: impact of various experimental parameters}

In addition to the 2-fiber setting, we also examine the isotropic ($K=0$),  1-fiber ($K=1$) and 3-fiber ($K=3$) cases. See Sections \ref{sec:0fiber}, \ref{sec:1fiber} and \ref{sec:3fiber} in Supplementary Material, respectively. A  crucial observation is that \texttt{SN-lasso} is able to automatically detect the isotropic diffusion by only selecting the constant function $\Phi_{00}$ and setting the coefficients of all the needlets to zero. 
In contrast, \texttt{SCSD} estimators  result in noisy estimates of FOD with a 
number of spurious peaks of similar heights (Figure \ref{fig:0fib_lmax8_N41}; Table \ref{table:0fib_lmax8_N41}).
\texttt{SN-lasso} and  \texttt{SCSD} estimators work well for the 1-fiber setting (i.e., no crossing), with nearly perfect success rates and small angular errors, whereas \texttt{SH-ridge} has 
significantly larger angular errors due to over-smoothing (Figure \ref{fig:1fib_b1_ratio10_lmax8} and Table \ref{table:1fib_b1_ratio10_lmax8}). 
 For the 3-fiber crossing,  \texttt{SN-lasso} again has visually the best FOD estimation  and  the highest overall success rate (Figures \ref{fig:3fib_b1_ratio10_lmax8_N81}, \ref{fig:3fib_b1_ratio10_lmax8_N321}, \ref{fig:3fib_b3_ratio10_lmax8_N81}; Tables  \ref{table:3fib_b1_ratio10_lmax8_N81}, \ref{table:3fib_b1_ratio10_lmax8_N321},\ref{table:3fib_b3_ratio10_lmax8_N81}).

In terms of the impact of  the number of gradient directions $n$, Figures \ref{fig:b1_ratio10_lmax8_N41} ($n=41$), \ref{fig:b1_ratio10_lmax8_N81} ($n=81$), \ref{fig:2fib_b1_ratio10_lmax8_N321} ($n=321$) and Tables \ref{table:b1_ratio10_lmax8_N41}, \ref{table:b1_ratio10_lmax8_N81}, \ref{table:2fib_b1_ratio10_lmax8_N321} indicate that, as $n$ increases, the variability of these estimators tends to decrease, the peak detection success rate tends to increase and the angular errors tend to decrease. In terms of the impact of the \textit{bvalue} $b$, the performance of all these estimators improves dramatically when $b$ increases from $1000s/mm^2$ to $3000s/mm^2$ (sections \ref{sec:b3000}  and \ref{sec:3fiber_3000}) due to the larger contrast 
provided by larger \textit{bvalue}. The relative performance of these estimators remains the same as previously discussed with \texttt{SN-lasso} leading to the best overall results.

\subsection{Real D-MRI Data Experiments}\label{sec:details_application}



\begin{figure}
	\centering
	\caption{
	\small{ROI I. (a) From left to right:  Fiber orientation colormap of the  $z$-slice $40$:  ROI I is indicated by the white box; Colormap, FA map and MD map of ROI I. (b): FOD estimates on ROI I:  FA map is drawn on the background such that voxels with small FA values are shown by dark background color; Moreover, the size of the FOD estimates is modulated by mean diffusivity (MD): The larger the MD is, the smaller is the size of the depicted FOD estimate.} \label{fig:realdata_regionb}}
	
	\begin{tabular}{cc}
		\multicolumn{2}{c}{ (a): Fiber orientation colormaps, FA map and MD map	}\\
		
		\multicolumn{2}{l}{Colormap of $z$-slice $40$ \hskip 0.05in Colormap of ROI I \hskip 0.25in FA map of ROI I \hskip 0.35in MD map of ROI I}\\
		
		\multicolumn{2}{c}{
			\includegraphics[scale=0.43]{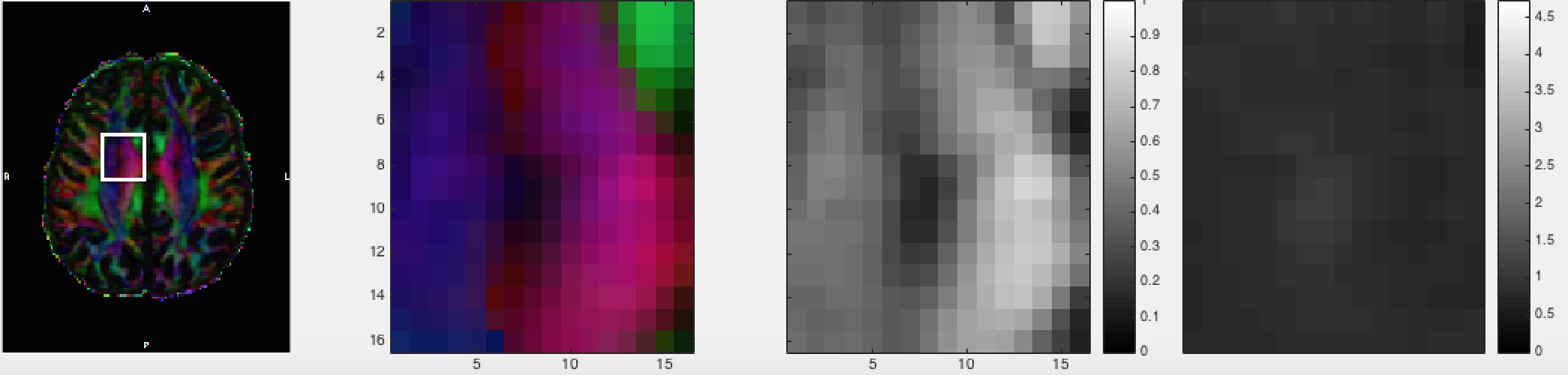}}\\
		\multicolumn{2}{c}{ (b): FOD estimates on ROI I	}\\			
		\texttt{SH-ridge}&\texttt{SCSD8}\\
		\includegraphics[scale=0.33]{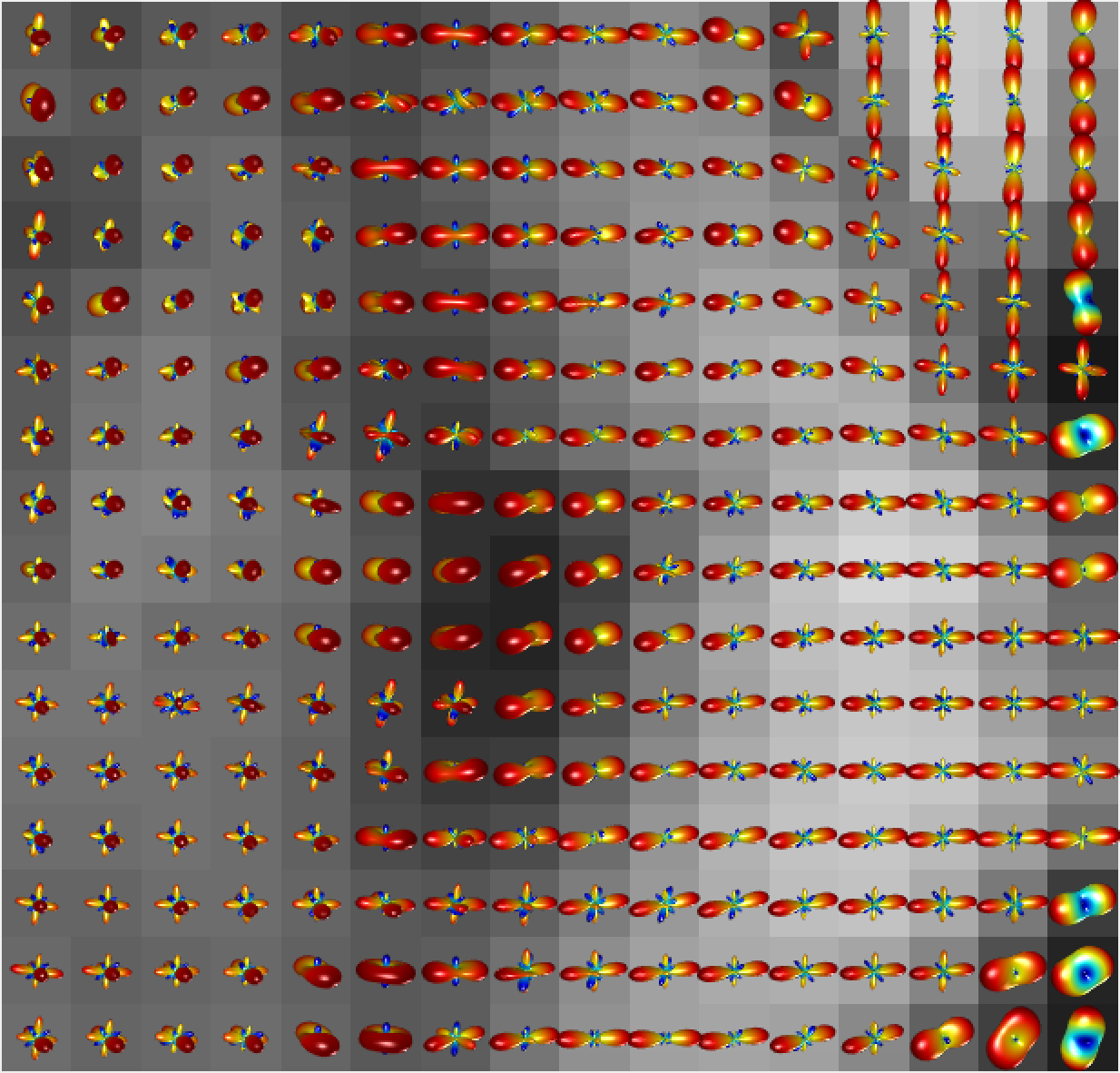}&	   	\includegraphics[scale=0.33]{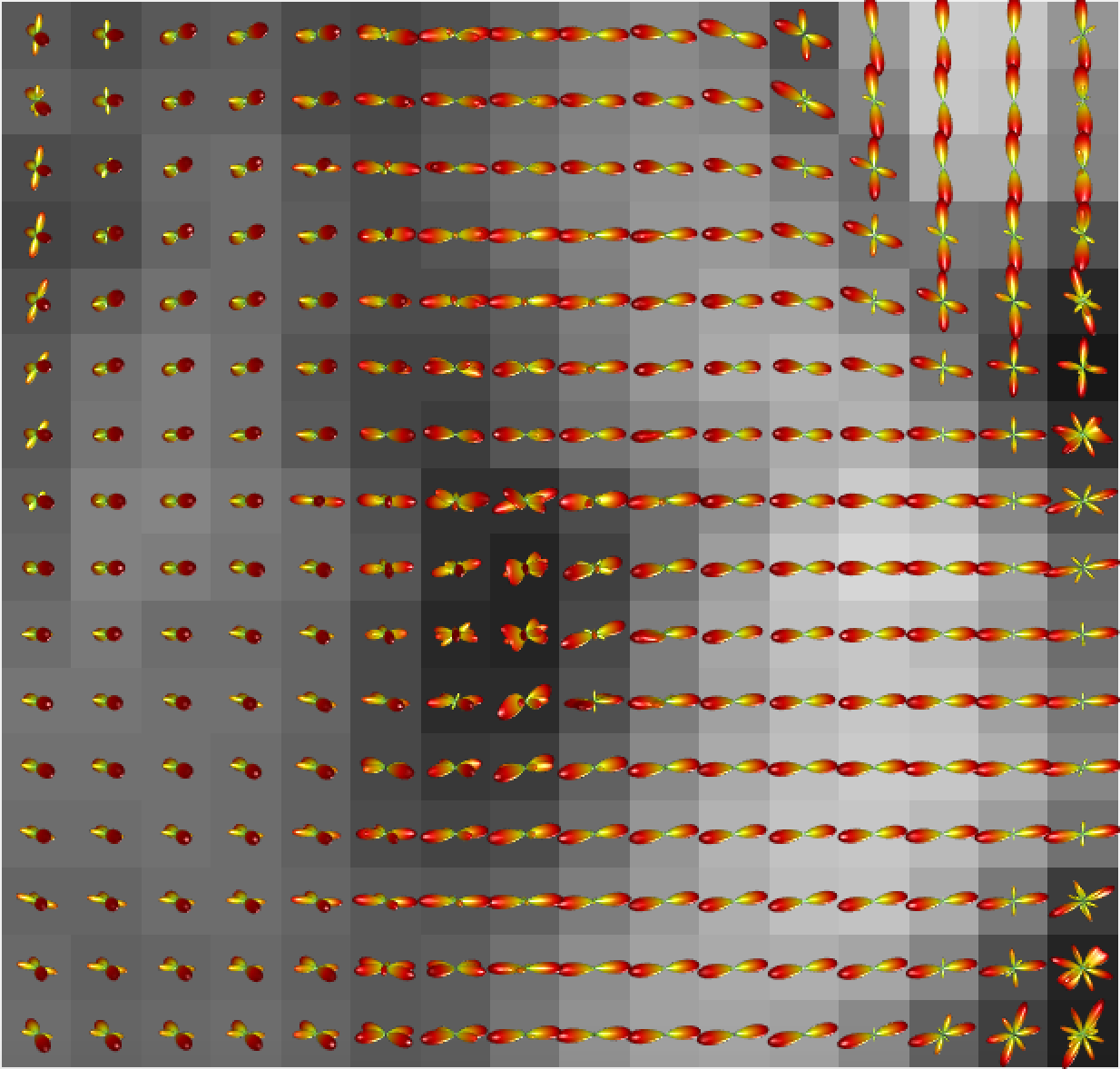}\\	
		\texttt{SCSD12}&\texttt{SN-lasso}\\
		\includegraphics[scale=0.33]{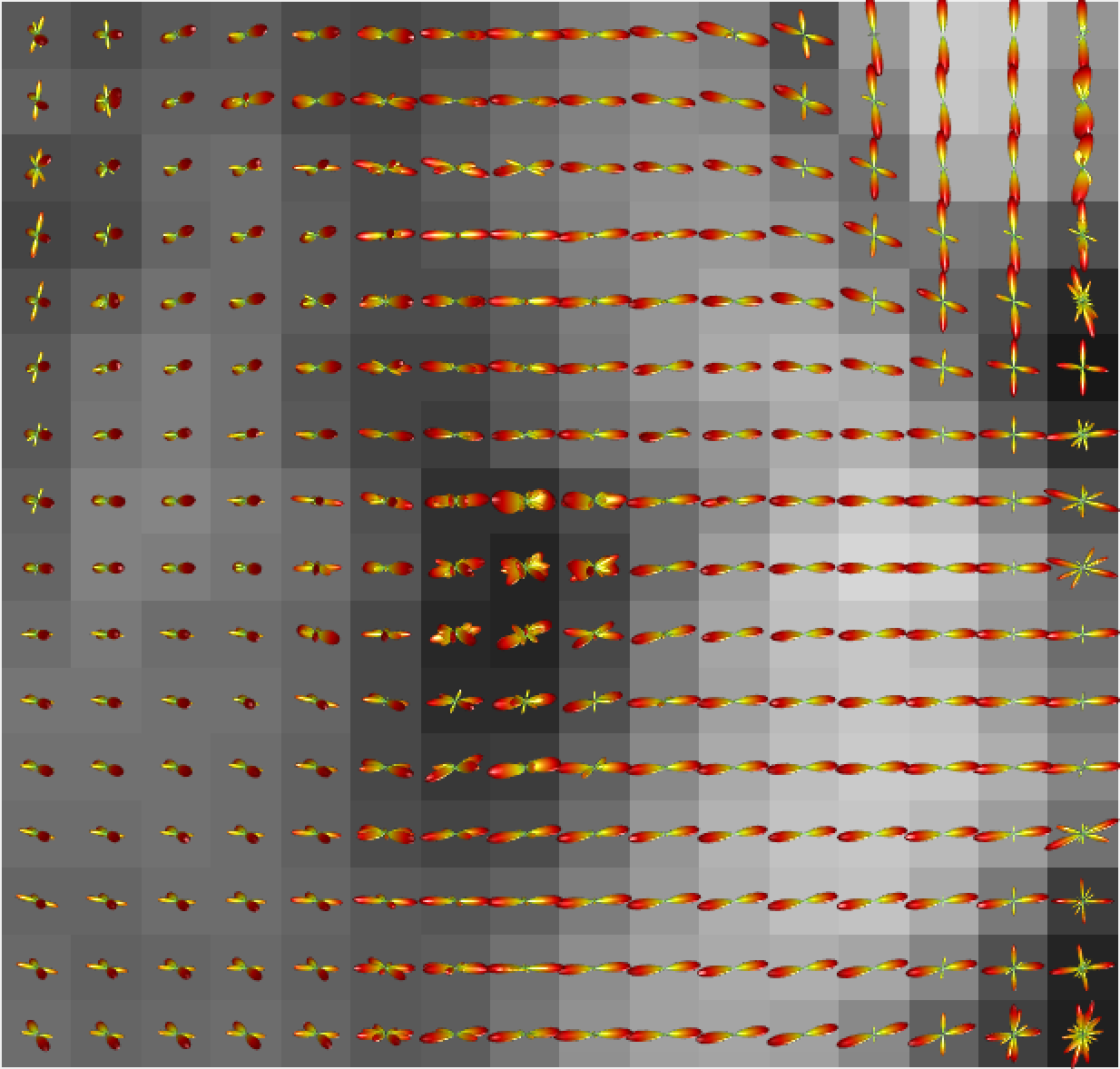}&
		\includegraphics[scale=0.33]{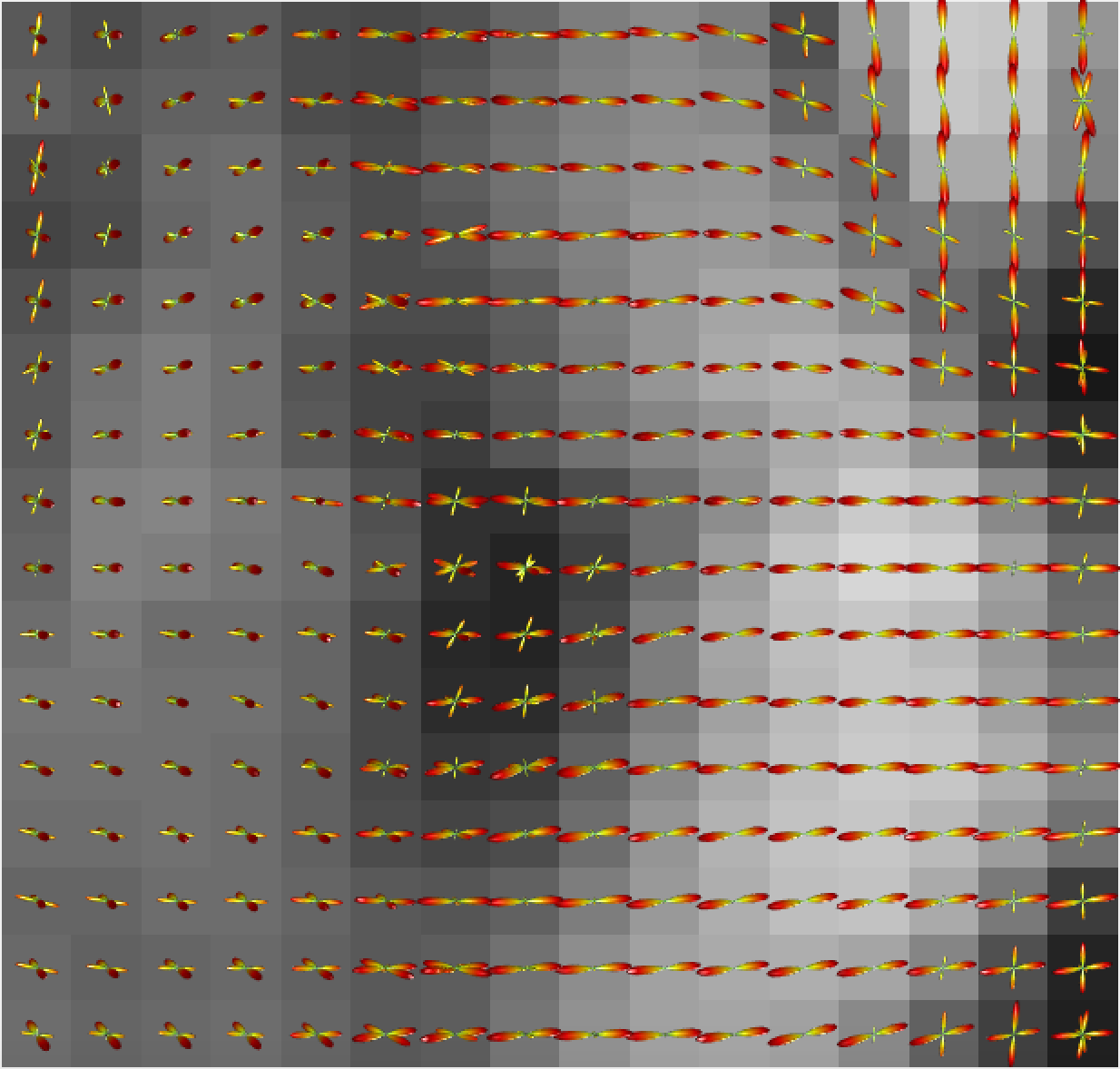}\\
	\end{tabular}
\end{figure}

We apply the FOD estimators to ROI I and the results are shown in  Figure \ref{fig:realdata_regionb}(b).
It can be seen that,  in the single fiber subregions, the \texttt{SN-lasso} estimates have sharp peaks   consistent with  the directions suggested by the fiber orientation colormaps. \texttt{SCSD8} and \texttt{SCSD12}  are also able to preserve directionality pattern in such regions.

\begin{figure}
	\centering
	\caption{
	\small{ROI I - Subregion 1: FOD estimates on a $5 \times 5$ subregion (columns 6--10  and rows 5--9). This subregion is indicated by the white boxes on the colormap, FA map and MD map of ROI I.} 
		\label{fig:realdata_regionb_sub1}}
	
	\begin{tabular}{cc}
		\multicolumn{2}{c}{ (a): Fiber orientation colormaps, FA map and MD map	}\\
			
			\multicolumn{2}{l}{Colormap of $z$-slice $40$ \hskip 0.05in Colormap of ROI I \hskip 0.25in FA map of ROI I \hskip 0.35in MD map of ROI I			}\\
			
			\multicolumn{2}{c}{
				\includegraphics[scale=0.43]{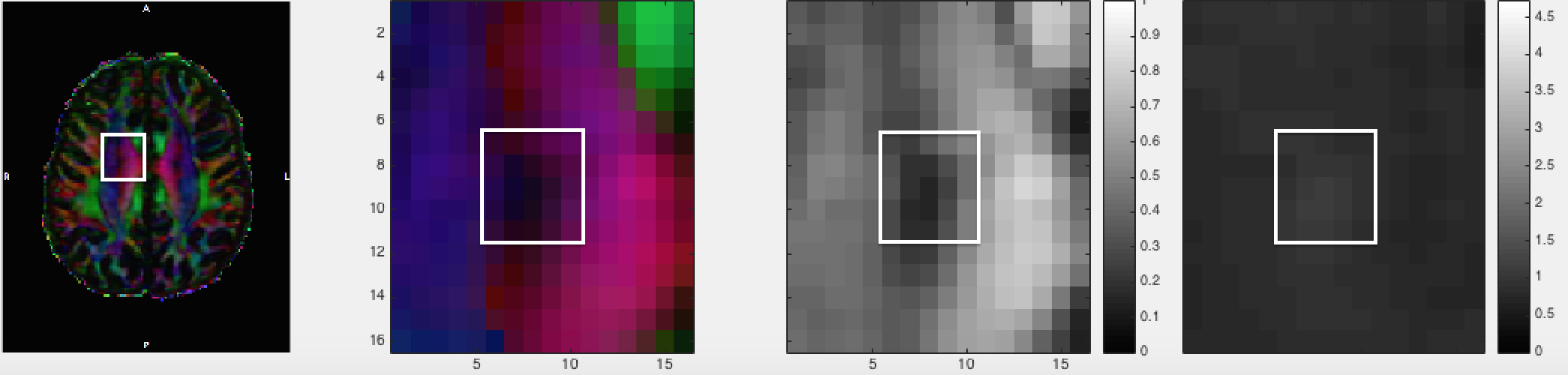}}\\
			\multicolumn{2}{c}{ (b): FOD estimates on the subregion indicated by the white box}\\			
		\texttt{SH-ridge}&\texttt{SCSD8}\\
		\includegraphics[scale=0.33]{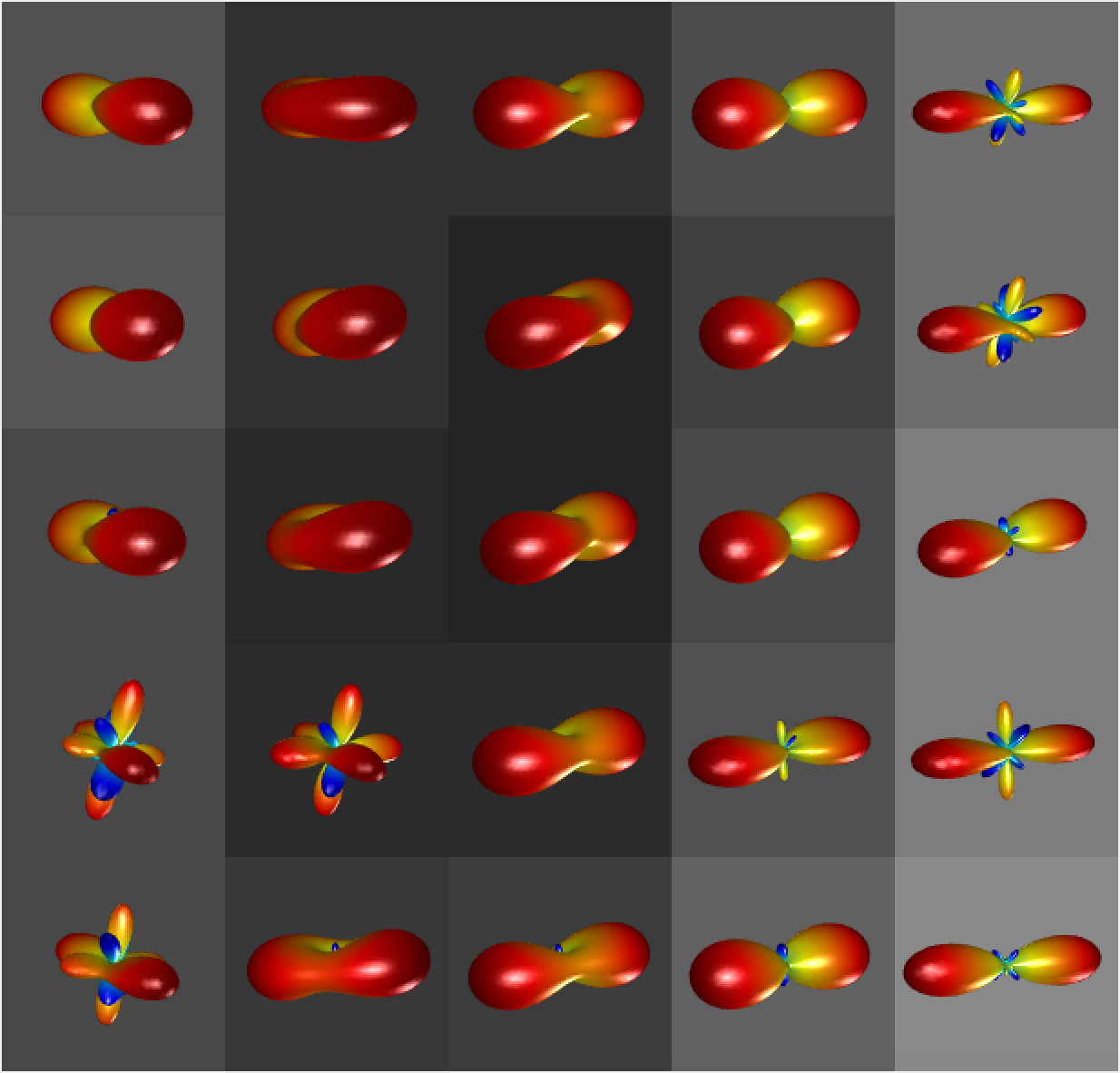}&	   	\includegraphics[scale=0.33]{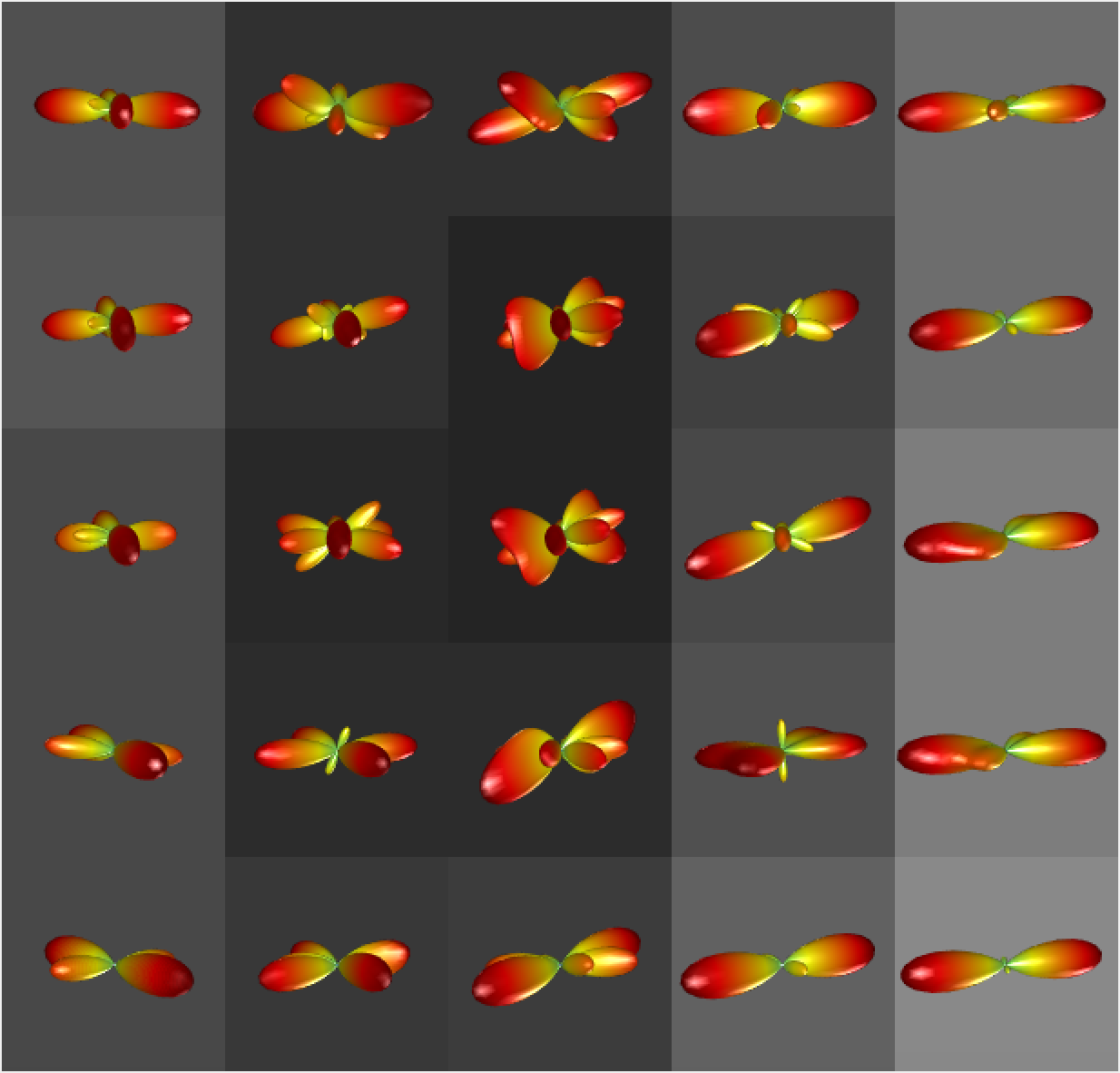}\\	
		\texttt{ SCSD12}&\texttt{SN-lasso}\\
		\includegraphics[scale=0.33]{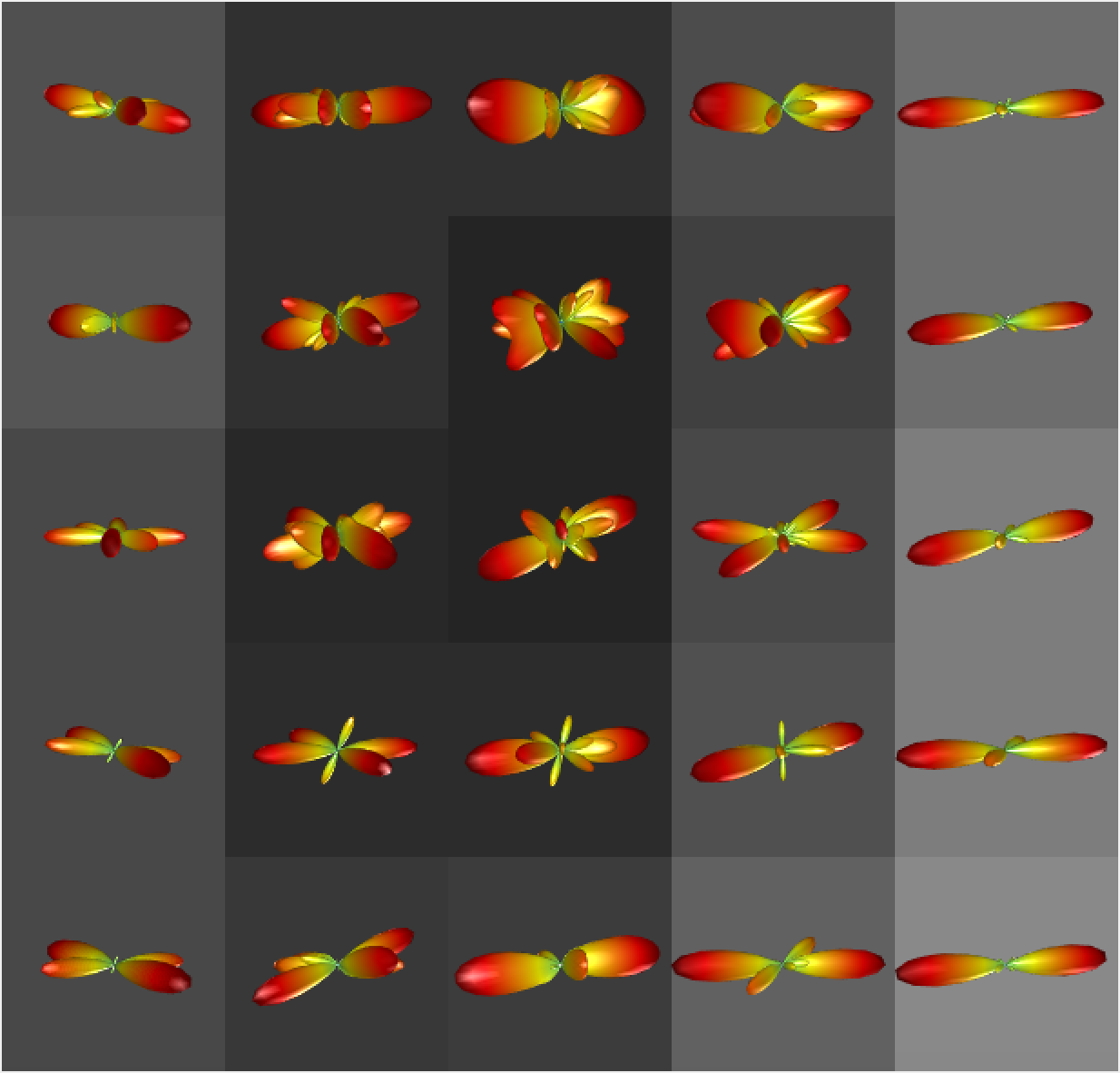}&
		\includegraphics[scale=0.33]{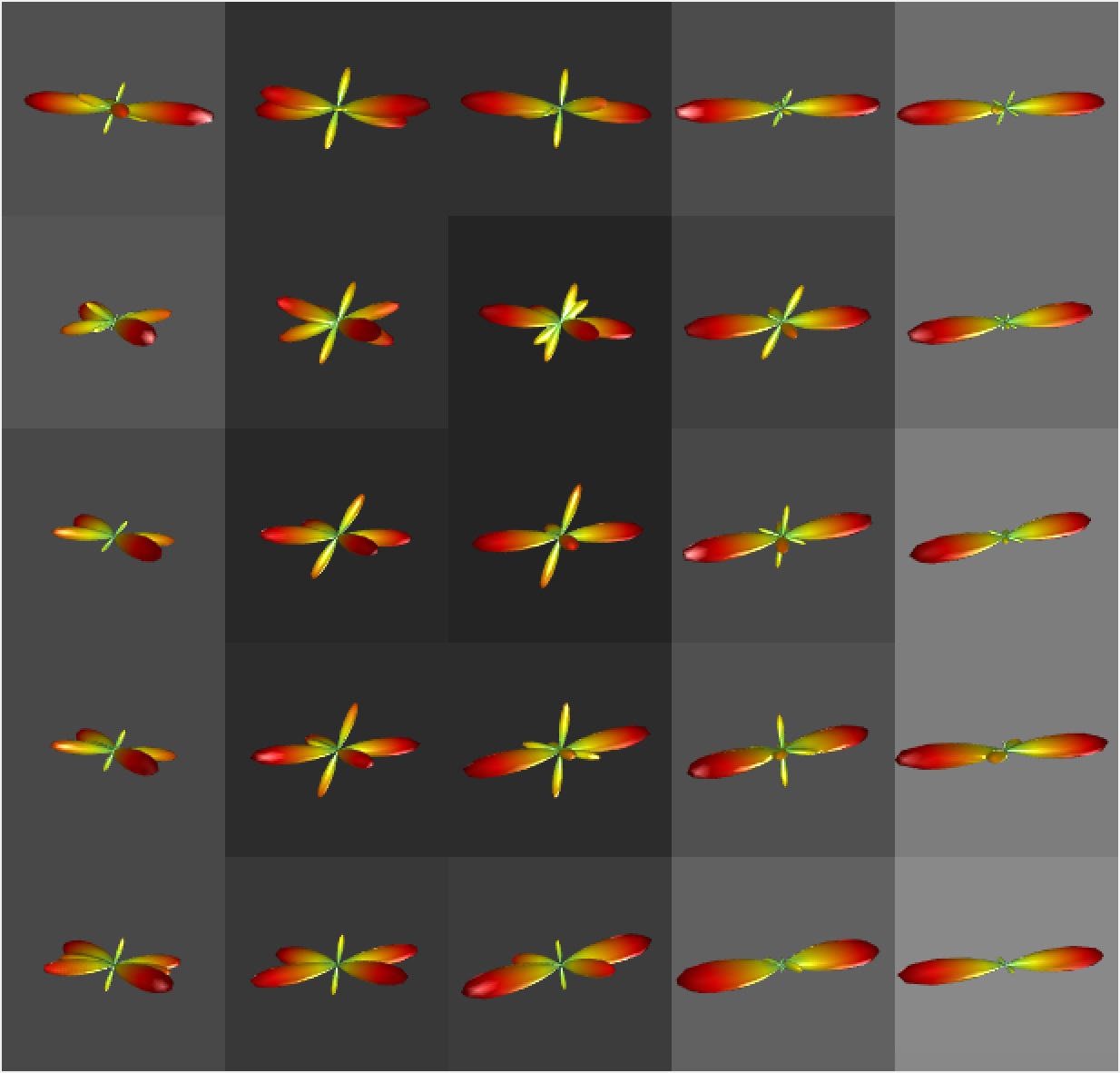}\\
		
	\end{tabular}
\end{figure}

\begin{figure}
	\centering
	\caption{
	\small{ROI I - Subregion 2: FOD estimates on a $5 \times 5$ subregion (columns 12-16  and rows 1 - 5). This subregion is  indicated by the white boxes on the colormap, FA map and MD map of ROI I.} 
		\label{fig:realdata_regionb_sub2}}
	
	\begin{tabular}{cc}
		\multicolumn{2}{c}{ (a): Fiber orientation colormaps, FA map and MD map	}\\
			
			\multicolumn{2}{l}{Colormap of $z$-slice $40$ \hskip 0.05in Colormap of ROI I \hskip 0.25in FA map of ROI I \hskip 0.35in MD map of ROI I			}\\
			
			\multicolumn{2}{c}{
				\includegraphics[scale=0.43]{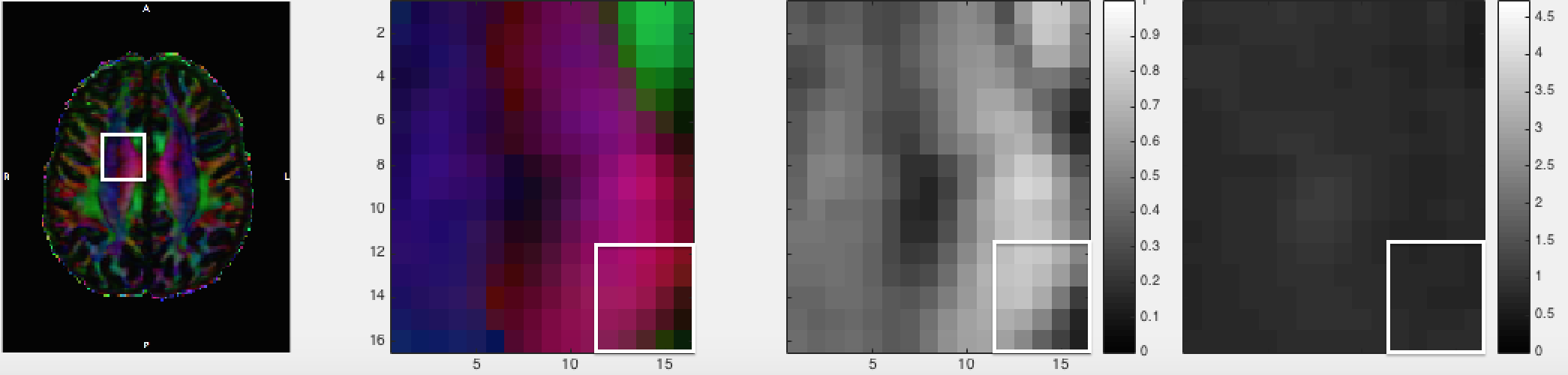}}\\
			\multicolumn{2}{c}{ (b): FOD estimates on the subregion indicated by the white box}\\			
		\texttt{SH-ridge}&\texttt{SCSD8}\\
		\includegraphics[scale=0.33]{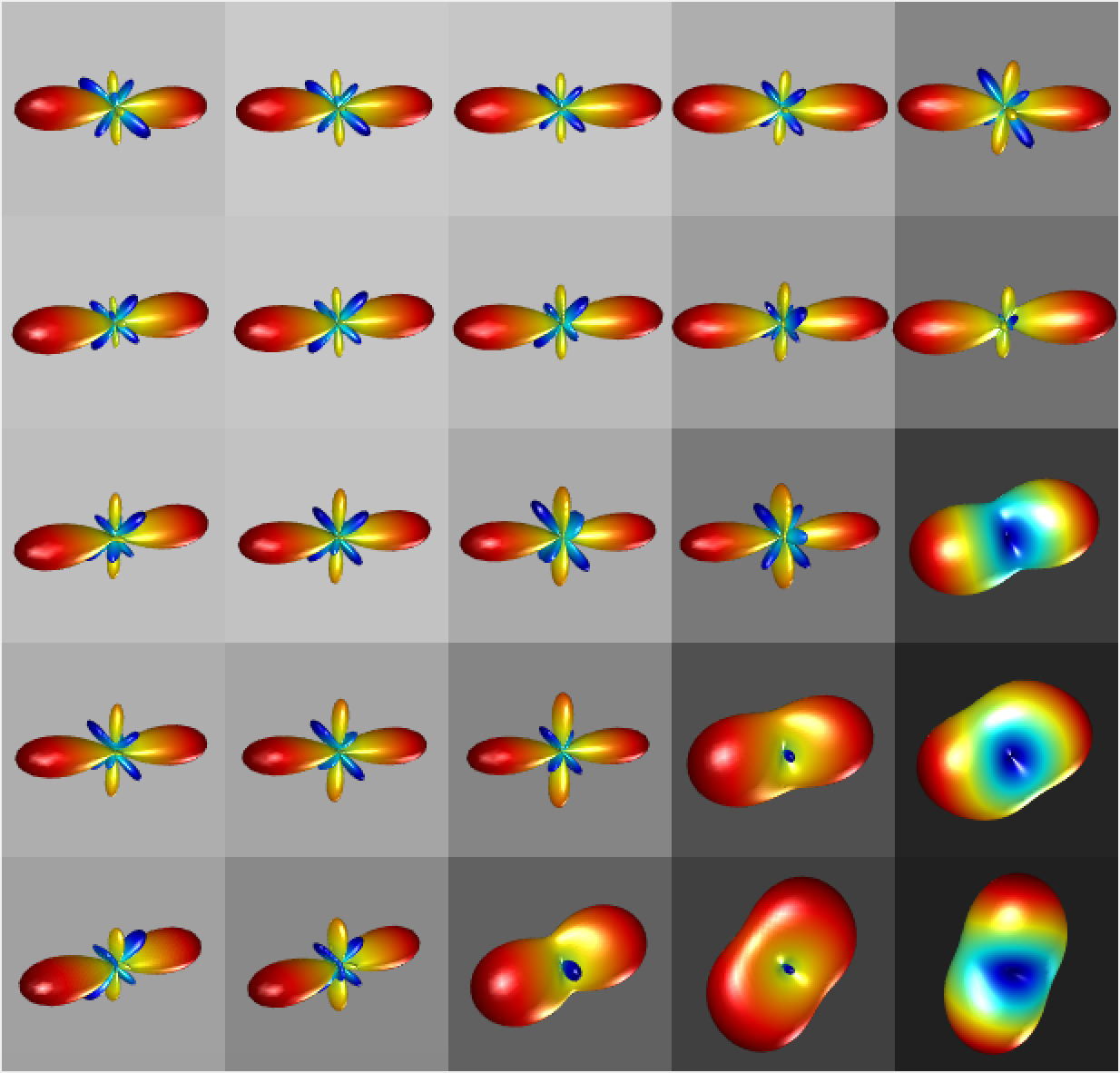}&	   	\includegraphics[scale=0.33]{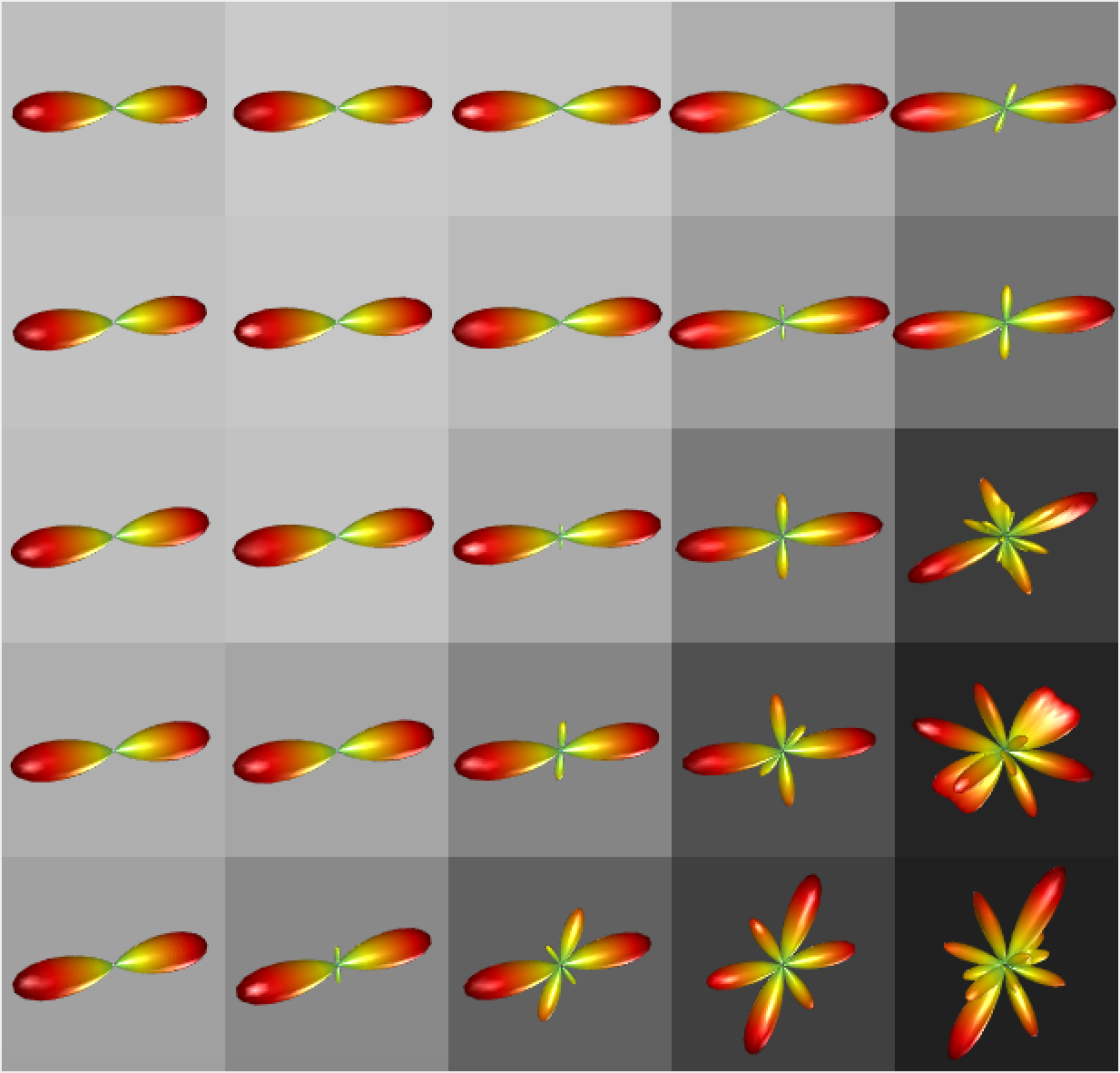}\\	
		\texttt{ SCSD12}&\texttt{SN-lasso}\\
		\includegraphics[scale=0.33]{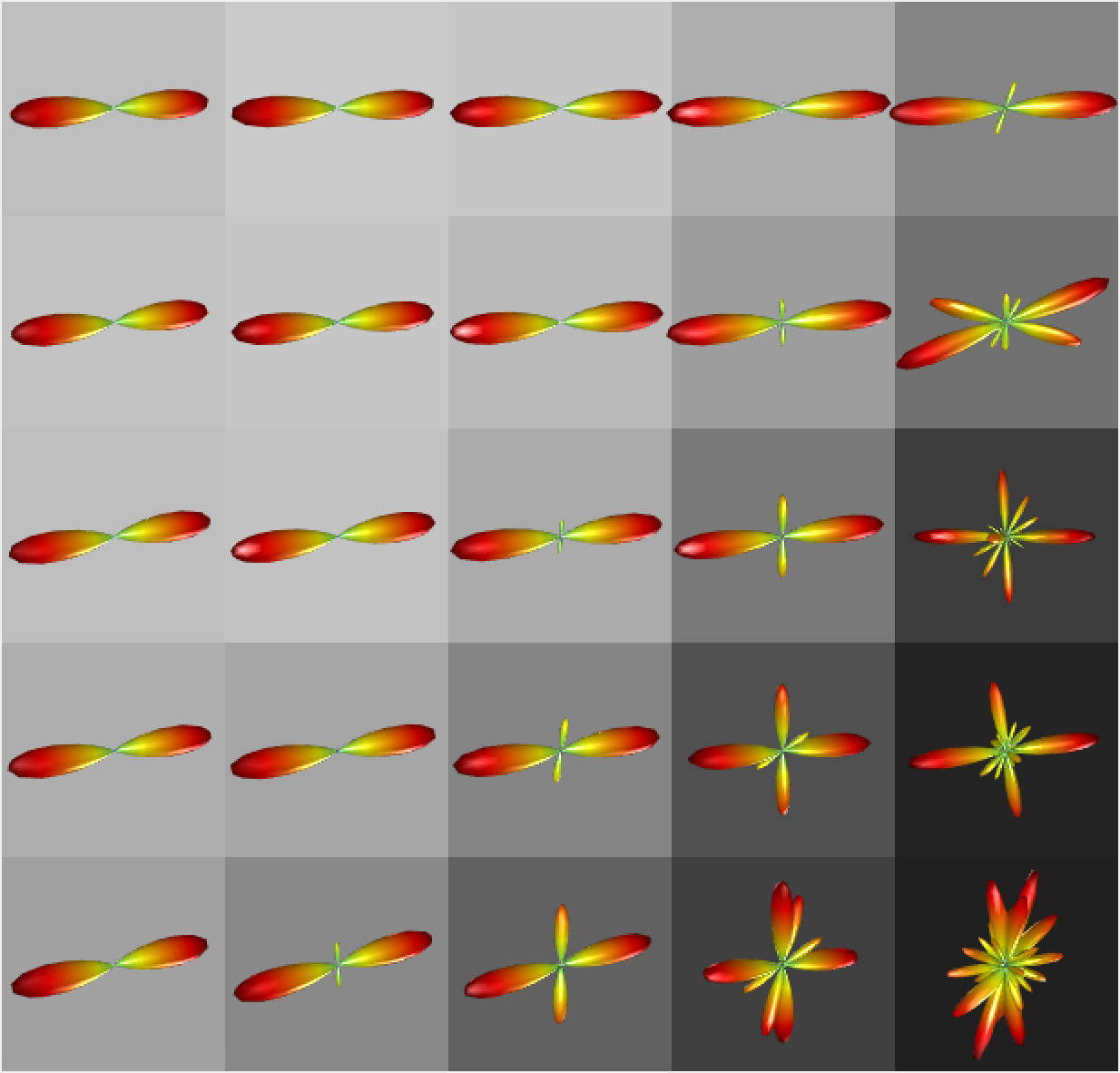}&
		\includegraphics[scale=0.33]{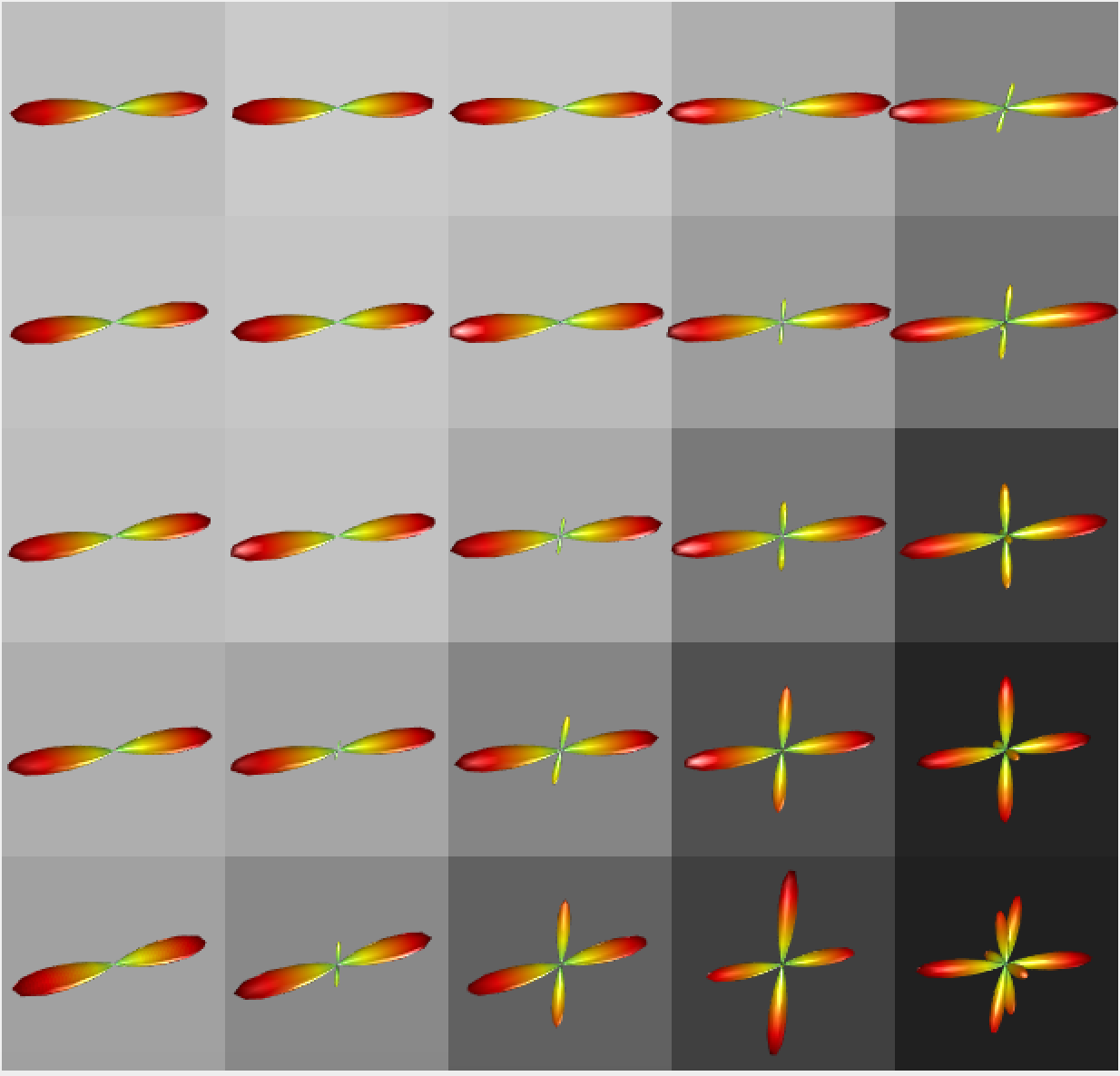}\\
		
	\end{tabular}
\end{figure}

However, in the crossing fiber regions,  SCSD estimators tend to be very noisy, while  \texttt{SH-ridge} leads to overly smoothed estimates causing loss of directionality information. In contrast, \texttt{SN-lasso} is able to correctly resolve crossing patterns as evidenced by the following observations.  
From Figure \ref{fig:realdata_regionb} (a), it can be seen that voxels in columns 6 to 10 (from left)  and rows 5 to 9 (from bottom) of ROI I have both small  FA and MD values  (dark on FA and MD maps). This  is likely due to crossing fibers rather than being CSF  since one defining feature of CSF is large MD values caused by faster water diffusion  \citep{alexander2007diffusion}. This is confirmed by the \texttt{SN-lasso} estimates  (Figure \ref{fig:realdata_regionb_sub1}(b), lower right panel) which show coherent fiber crossings in this subregion. The colormap suggests that these voxels represent fibers running in the left-to-right and into-to-out-of-the-page directions, in agreement with the SN-lasso estimates. 
On the other hand, the \texttt{SCSD} estimators result in FOD estimates that are incoherent among neighboring voxels and are with  spurious peaks (Figure \ref{fig:realdata_regionb_sub1}(b), upper right and lower left panels).  We zoom into another subregion of ROI I (Figure \ref{fig:realdata_regionb_sub2}), where the colormap indicates crossing between a horizontally orientated fiber (red-ish color) and a vertically orientated fiber (green-ish color) on the lower right corner. This is clearly captured by \texttt{SN-lasso}. The \texttt{SCSD} estimators again lead to noisy FOD estimates and \texttt{SH-ridge} estimates do not show clear directionality  in this crossing region. A third subregion is examined by Figure  \ref{fig:realdata_regionb_sub3} of the Supplementary Material which shows similar phenomena.

\begin{figure}
 	\centering
 	\caption{
 	\small{ROI II. (a) from left to right:  Fiber orientation colormap of the   $z$-slice $32$:  ROI II is indicated by the white box; Colormap, FA map and MD map of ROI II. (b): FOD estimates on ROI II:  FA map is drawn on the background such that voxels with small FA values are shown by dark background color; Moreover, the size of the FOD estimates is modulated by mean diffusivity (MD): The larger MD is, the smaller FOD estimate is depicted.}  \label{fig:realdata_regionc}}
 		
 		\begin{tabular}{cc}\\
 			\multicolumn{2}{c}{ (a): Fiber orientation colormaps, FA map and MD map	}\\
 			
 			\multicolumn{2}{l}{Colormap of $z$-slice $32$ \hskip 0.05in Colormap of ROI II \hskip 0.25in FA map of ROI II \hskip 0.35in MD map of ROI II			}\\
 			
 			\multicolumn{2}{c}{	\includegraphics[scale=0.43]{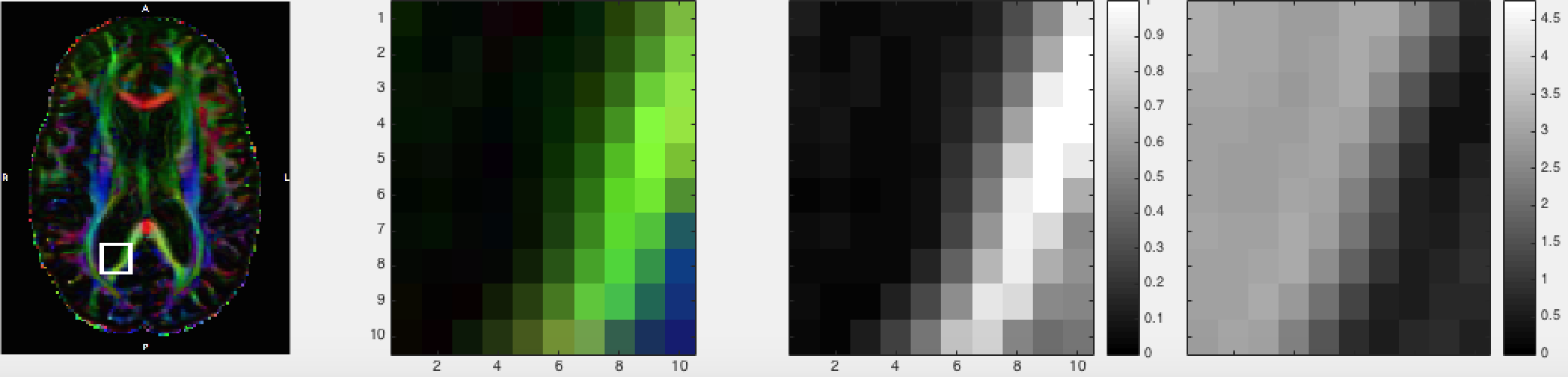}}\\
 			
 			\multicolumn{2}{c}{ (b): FOD estimates on ROI II	}\\			
 			\texttt{ SH-ridge}&\texttt{ SCSD8}\\
 			\includegraphics[scale=0.33]{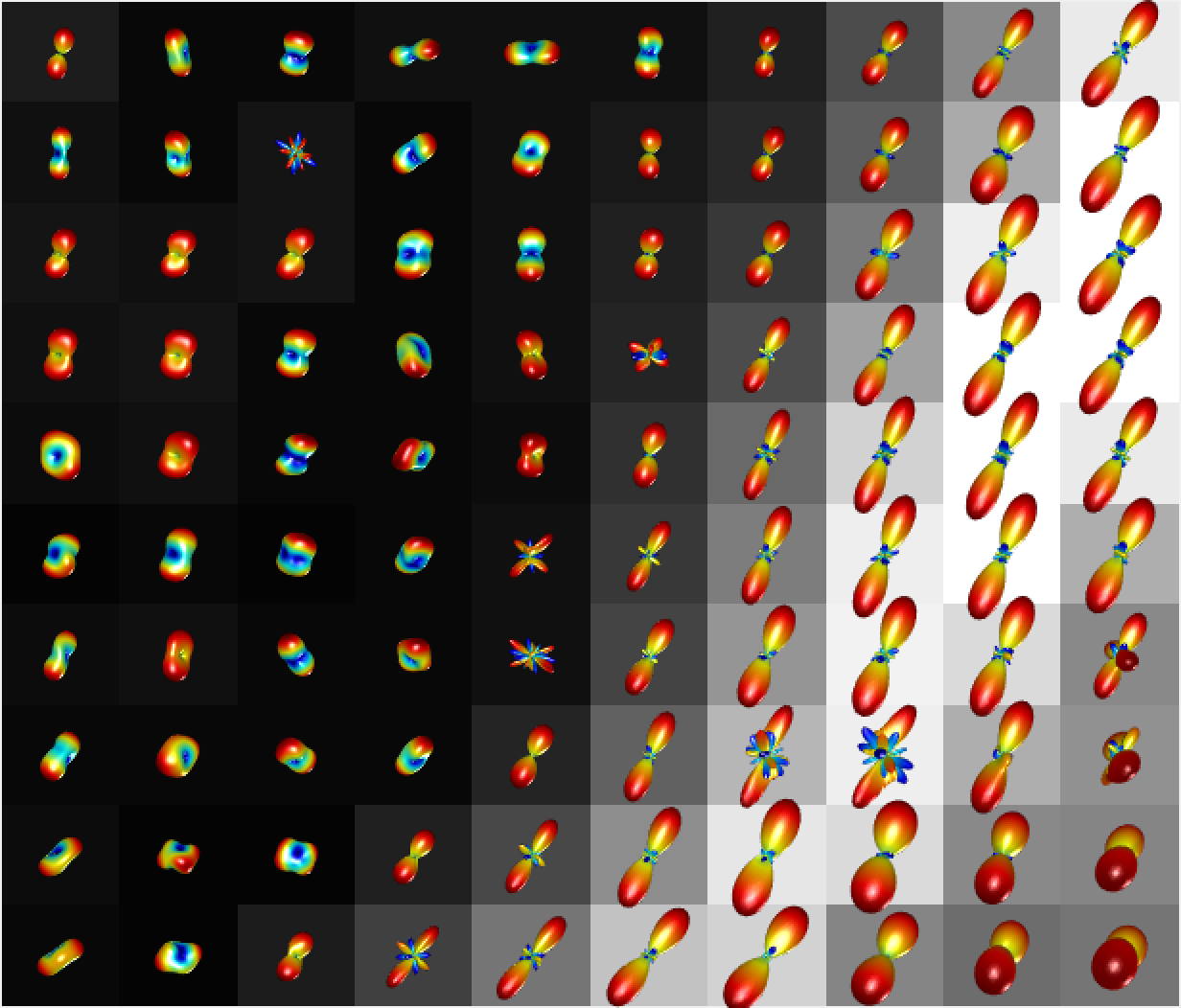}&	   	\includegraphics[scale=0.33]{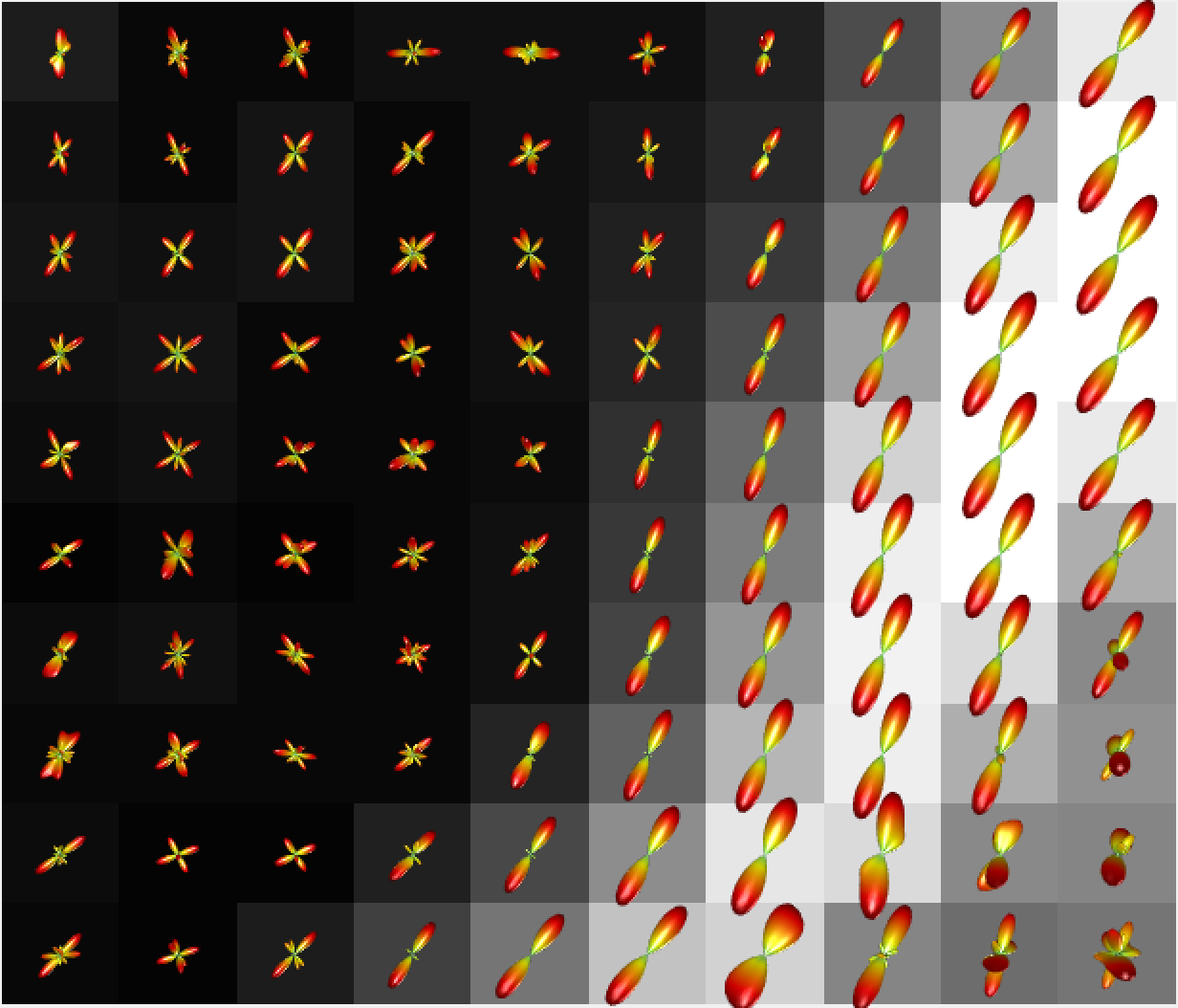}\\	
 			\texttt{SCSD12}&\texttt{SN-lasso}\\
 			\includegraphics[scale=0.33]{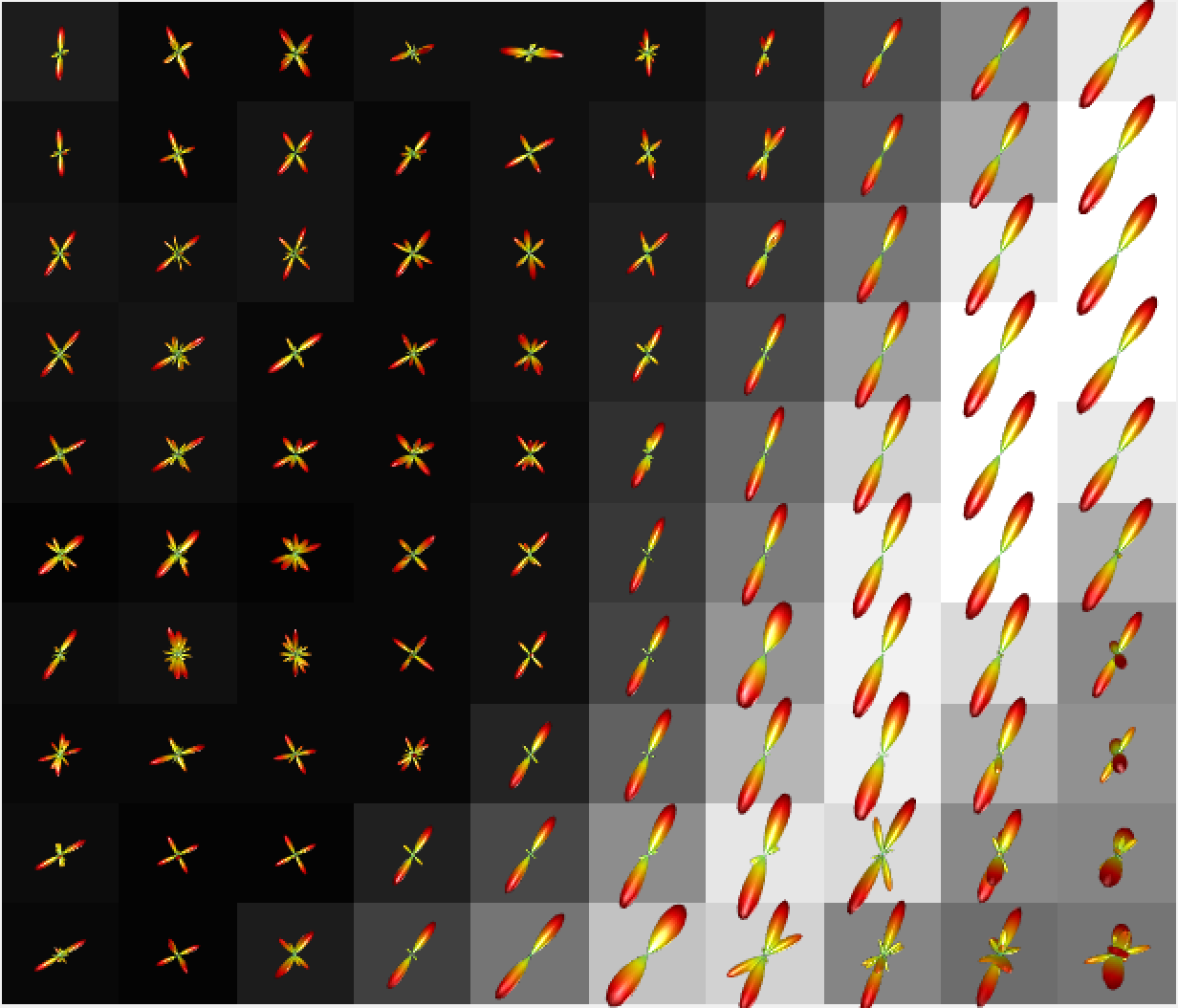}&
 			\includegraphics[scale=0.33]{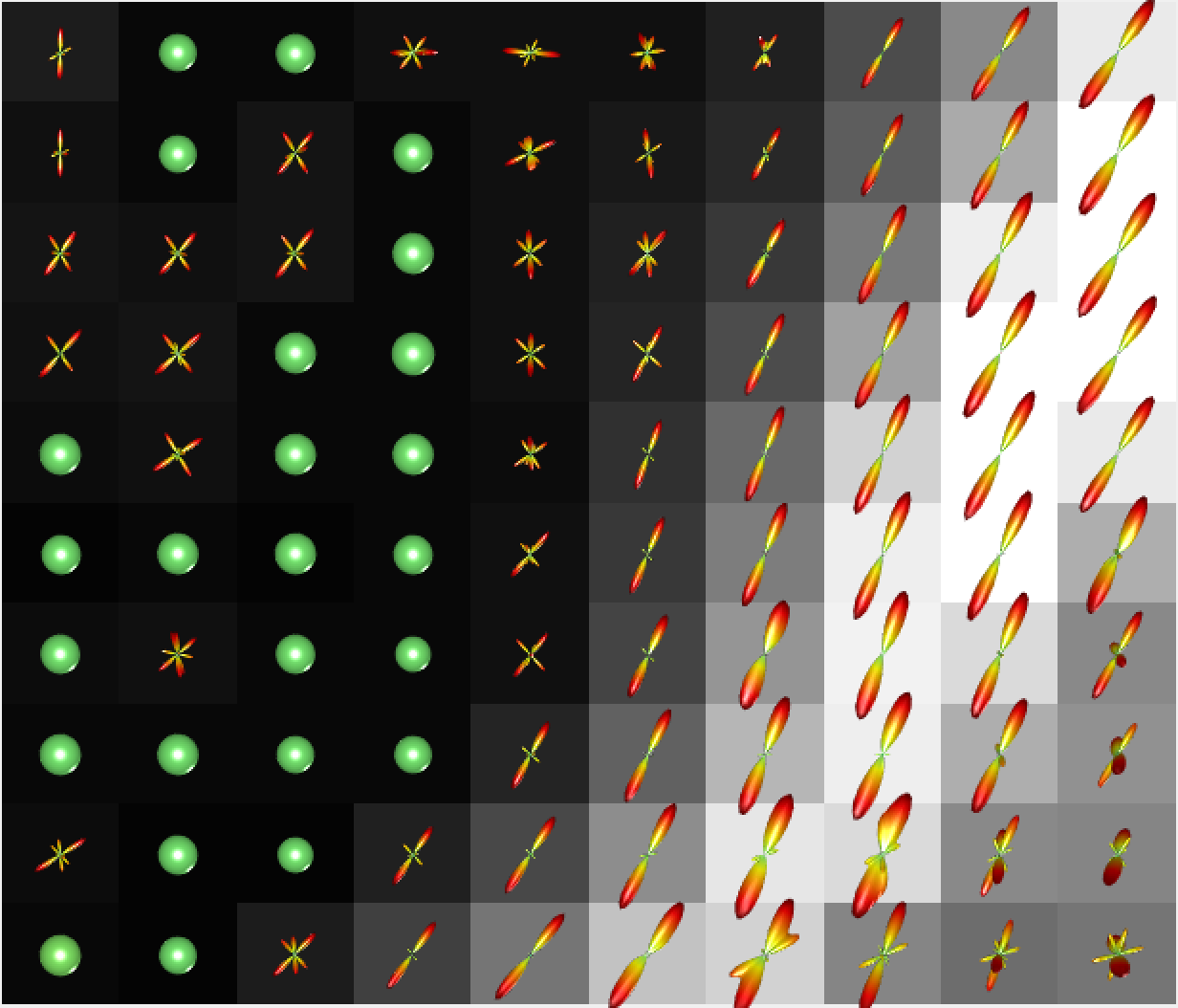}\\
 			
 		\end{tabular}
\end{figure}

The upper left part of ROI II is likely to be CSF (Figure \ref{fig:realdata_regionc}(a)), indicated by  large MD values and small FA values \citep{alexander2007diffusion}. This is corroborated by the \texttt{SN-lasso} estimates where FODs for many voxels in this subregion are estimated as being isotropic (represented by a green ball; Figure \ref{fig:realdata_regionc}(b), lower right panel). In contrast, the other three estimators fail to identify any isotropic voxels in this subregion (Figure \ref{fig:realdata_regionc}(b), the rest panels). The \texttt{SN-lasso} estimates also show clear single fiber in the lower right part of ROI II which is consistent with the bright greenish band on the colormap.

\section{Discussion}\label{sec:conclusion}

In this paper, we present a novel method for estimating FOD  that is accurate, has low variability and preserves sharp features, as well as being computationally efficient. The effectiveness of the proposed method derives from the utilization of  multiresolution spherical frame called needlets that admits stable and  parsimonious representation of FODs.   Statistical efficiency, in terms of high accuracy and low variability in the estimates, is gained through imposing a sparsity constraint on the needlet coefficients, together with a nonnegativity constraint on the estimated FOD.
The proposed method mitigates difficulties faced by  two well-known FOD estimators, namely \texttt{SH-ridge} \citep{TournierEtal2004}  and  SuperCSD \citep{TournierEtal2007}. Specifically, 
SH-ridge tends to over-smooth the peaks resulting in the loss of directionality information due to the global nature of the SH basis; and  SuperCSD tends to amplify spurious  peaks in estimated FODs due to high variability. On the other hand, the needlets frame used by the proposed method has excellent localization properties  (in both frequency and space),  and the $\ell_1$ penalized regression framework enables an adequate bias-variance trade off.


Experimental results on synthetic data suggest that the proposed method leads to better FOD estimation than two well known competing methods, particularly in terms of identification of major fiber directions.  These results are consistent with the well-established statistical notion that a nonlinear shrinkage strategy allied with sparse representation is more efficient for signal recovery than a linear shrinkage strategy
\citep{Johnstone2015,Tsybakov2009}.

The proposed method is able to successfully reconstruct an FOD when the fiber crossing angle is as small as $30^{\circ}$ provided that the \textit{bvalue} and SNR are sufficiently large (e.g., $b=3000, 5000 s/mm^2, SNR=50$).
Resolving small crossing angles at relatively low \textit{bvalues} has been reported  in literature to be a challenging setting. For example, in \cite{yeh2013sparse}, under $b=1500  s/mm^2, SNR=20, n=55$, the best performing method has around $20^{\circ}$ angular deviation at $45^{\circ}$ crossing.
On the other hand, with the fast advancement of D-MRI technologies, we are expecting larger \textit{bvalue} and SNR \citep{SetsompopEtAl2013,VanEssenEtAl2013}. Use of larger \textit{bvalue} for reconstruction of diffusion characteristics has also been previously advocated in the literature  \citep{JensenRH2016,JonesKT2013,RoineEtAl2015}. Therefore, the proposed method holds great promise in resolving even subtle fiber crossing patterns. 
Moreover, the proposed method has the distinct advantage of being able to identify isotropic diffusion with very high degrees of accuracy.

Another important insight gained from these experiments is that, for the representations considered in this paper,  employing larger \textit{bvalue} is more effective than using larger number of gradient directions for the purpose of  distinguishing among multiple fiber bundles with relatively small crossing angles.   The observations gained here may be used as a guideline for specifying the acquisition parameters in future D-MRI experiments. 

The proposed method is also examined  through a real data experiment that uses D-MRI data sets from the second phase of the ADNI project. It leads to realistic descriptions of crossing fibers with less noise than competing methods at a relatively low number of gradient directions, indicating practical value of our method for analyzing D-MRI data from many large scale studies. 

Our use of a multiresolution spherical wavelets bears some similarity with the dODF-sharpening
strategy of \cite{KezeleEtAl2010} where a spherical wavelets transform due to
\cite{StarckMAN2006} is employed.  However,  the method of \cite{KezeleEtAl2010} is not directly applicable here since we aim to estimate the FOD (or fiber ODF) rather than dODF. Moreover, the spherical wavelets constructed by \cite{StarckMAN2006} do not possess the spherical function representation characteristics 
and tight frame properties of needlets. Finally, unlike our method, 
where the needlets representation enables us to impose sparsity constraints, \cite{KezeleEtAl2010} use the spherical wavelets transform only as a band-pass filter.  Among alternative choices of spherical wavelets, the lifting scheme based wavelets constructed by  \cite{SchroderS1995} 
provide excellent spatial localization and ease of computation. However,   since these wavelets  are nonsmooth and consequently poorly localized in frequency, they are not optimal for FOD reconstruction since  FODs have sharp localized features, which require frequency localization of the basis functions for a parsimonious representation.

A key avenue for future research is the extension of this method to multiple $q$-shell data.  An emerging strength of such data is its ability to accommodate multiple cellular compartment models that separately quantify the contributions of free water, water bound within the myelin sheath, inter-axonal water, and other compartments to the D-MRI signal.  As such, multiple $q$-shell data may be best modeled to allow for different FOD representations for different compartments, as in the NODDI scheme \citep{ZhangEtAl2012}, rather than a single unified spherical function as here.  On the other hand, spherical needlets might provide accurate and robust representation of some of these signal components, possibly with specific coefficient penalties for specific components.  Future experimentation should determine settings in which spherical needlets are useful for modeling components of the multiple $q$-shell D-MRI signal.

While estimating the FOD at a voxel, our method only uses data from the same voxel.  Since there is  a natural connection between local fiber orientation and the shape of FODs, we may consider the incorporation of neighboring voxel data in the estimation of FODs to further improve statistical efficiency. We also want to conduct inferences such as constructing confidence balls for the estimated FODs, for which we shall build upon principles of resampling and recent developments in post-model-selection inference \citep{LockhartTTT2014,LeeSST2016,TibshiraniRTW2015}. 
Lastly, needlet representation can be extended to estimate other D-MRI related objects such as ODFs which are also spherical functions with sharp features. 
Thus we expect to see many applications and extensions of  the techniques proposed in this paper.

\section*{Acknowledgement}

This work is supported by the following grants: NIH 1R01EB021707, NIH P30DK072476, NSF DMS-1148643,  
NSF DMS-1407530, NSF IIS-1422218. \\

Data collection and sharing for this project was funded by the Alzheimer's Disease Neuroimaging Initiative
(ADNI) (National Institutes of Health Grant U01 AG024904) and DOD ADNI (Department of Defense award
number W81XWH-12-2-0012). ADNI is funded by the National Institute on Aging, the National Institute of
Biomedical Imaging and Bioengineering, and through generous contributions from the following: AbbVie,
Alzheimer's Association; Alzheimer's Drug Discovery Foundation; Araclon Biotech; BioClinica, Inc.; Biogen;
Bristol-Myers Squibb Company; CereSpir, Inc.; Cogstate; Eisai Inc.; Elan Pharmaceuticals, Inc.; Eli Lilly and
Company; EuroImmun; F. Hoffmann-La Roche Ltd and its affiliated company Genentech, Inc.; Fujirebio; GE
Healthcare; IXICO Ltd.; Janssen Alzheimer Immunotherapy Research \& Development, LLC.; Johnson \&
Johnson Pharmaceutical Research \& Development LLC.; Lumosity; Lundbeck; Merck \& Co., Inc.; Meso
Scale Diagnostics, LLC.; NeuroRx Research; Neurotrack Technologies; Novartis Pharmaceuticals
Corporation; Pfizer Inc.; Piramal Imaging; Servier; Takeda Pharmaceutical Company; and Transition
Therapeutics. The Canadian Institutes of Health Research is providing funds to support ADNI clinical sites
in Canada. Private sector contributions are facilitated by the Foundation for the National Institutes of Health
(\url{www.fnih.org}). The grantee organization is the Northern California Institute for Research and Education,
and the study is coordinated by the Alzheimer's Therapeutic Research Institute at the University of Southern
California. ADNI data are disseminated by the Laboratory for Neuro Imaging at the University of Southern
California.





\clearpage
\newpage

\setcounter{page}{1}
\setcounter{section}{0}
\renewcommand{\thesection}{S.\arabic{section}}
\setcounter{subsection}{0}
\renewcommand{\thesubsection}{S.\arabic{section}.\arabic{subsection}}
\setcounter{equation}{0}
\renewcommand{\theequation}{S.\arabic{equation}}
\setcounter{figure}{0}
\renewcommand{\thefigure}{S.\arabic{figure}}
\setcounter{table}{0}
\renewcommand{\thetable}{S.\arabic{table}}
\setcounter{proposition}{0}
\renewcommand{\theproposition}{S.\arabic{proposition}}
\setcounter{lemma}{0}
\renewcommand{\thelemma}{S.\arabic{lemma}}
\setcounter{corollary}{0}
\renewcommand{\thecorollary}{S.\arabic{corollary}}

\begin{center}
\Large{\bf Supplementary Material for ``Estimating fiber orientation distribution from diffusion MRI with spherical needlets''}
\end{center}
\section{Symmetrized needlets and the design matrix \label{sec:design}} 
Let  $\{\Phi_{lm};l=0,1,2,\ldots;m=-l,-l+1,\ldots,l-1,l\}$ denote the complex-valued SH functions where $l$ denotes the harmonic order and $m$ denotes the phase order. It forms an orthonormal basis for the square integrable complex functions on the unite sphere $\mathbb{S}^2$. The real symmetrized SH functions are defined as follows \citep{DescoteauxAFD2007}. For $l=0,2,4,\ldots$ and $m=-l,-l+1,\ldots,l-1,l$
	\begin{equation}
		\widetilde{\Phi}_{lm}=
		\begin{cases}
			\sqrt{2}\cdot \text{Re}\big(\Phi_{lm}\big), & \text{if}\ -l\leq m<0 \\
			\Phi_{l0}, & \text{if}\ m=0 \\
			\sqrt{2}\cdot \text{Img}\big(\Phi_{lm}\big), & \text{if}\ 0<m\leq l
		\end{cases},
	\end{equation}
	where $\text{Re}\big(\Phi_{lm}\big)$ and  $\text{Img}\big(\Phi_{lm}\big)$ are the real and imaginary parts of $\Phi_{lm}$, respectively. The real symmetrized SH functions form an orthonormal basis for the square integrable real symmetric functions on the unit sphere.

Let $\{\zeta_{jk}\}$ denote the \textit{quadrature} points according to the HEALPix construction of spherical needlets and $\{\widetilde{\psi}_{jk}\}$ denote the needlet functions.
	The symmetrized needlet functions can be easily constructed utilizing the following two facts: (i) For each level $j$, the \textit{quadrature} points appear in pairs that are symmetric to the origin, i.e. for each $k$, there is a $k'$ such that $\zeta_{jk'}=-\zeta_{jk}$. (ii) The needlet functions corresponding to the pair of symmetric \textit{quadrature} points satisfy $\widetilde{\psi}_{jk}(\mathbf{x}) = \widetilde{\psi}_{jk'}(-\mathbf{x})$ for  $\mathbf{x}\in\mathbb{S}^2$. The symmetrized needlets is defined as follows: 
	$$\psi_{jk}(\mathbf{x}) = \frac{1}{2}\big(\widetilde{\psi}_{jk}(\mathbf{x}) + \widetilde{\psi}_{jk'}(\mathbf{x})\big), ~~\zeta_{jk}\in \mathbb{S}^2_+.$$
	
\begin{figure}[H]
			\begin{center}
					\caption{Left panel: Real symmetric spherical harmonics; Right Panel: Real symmetric spherical needlets. \label{fig:SH_SN}}
					\begin{tabular}{cc}
				\includegraphics[width=4in]{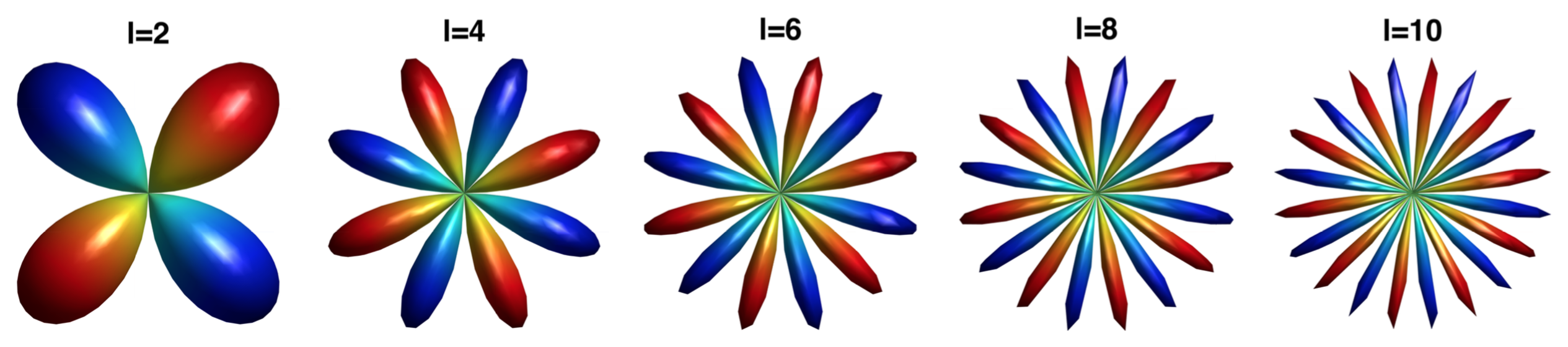}&~~
					\includegraphics[width=2in]{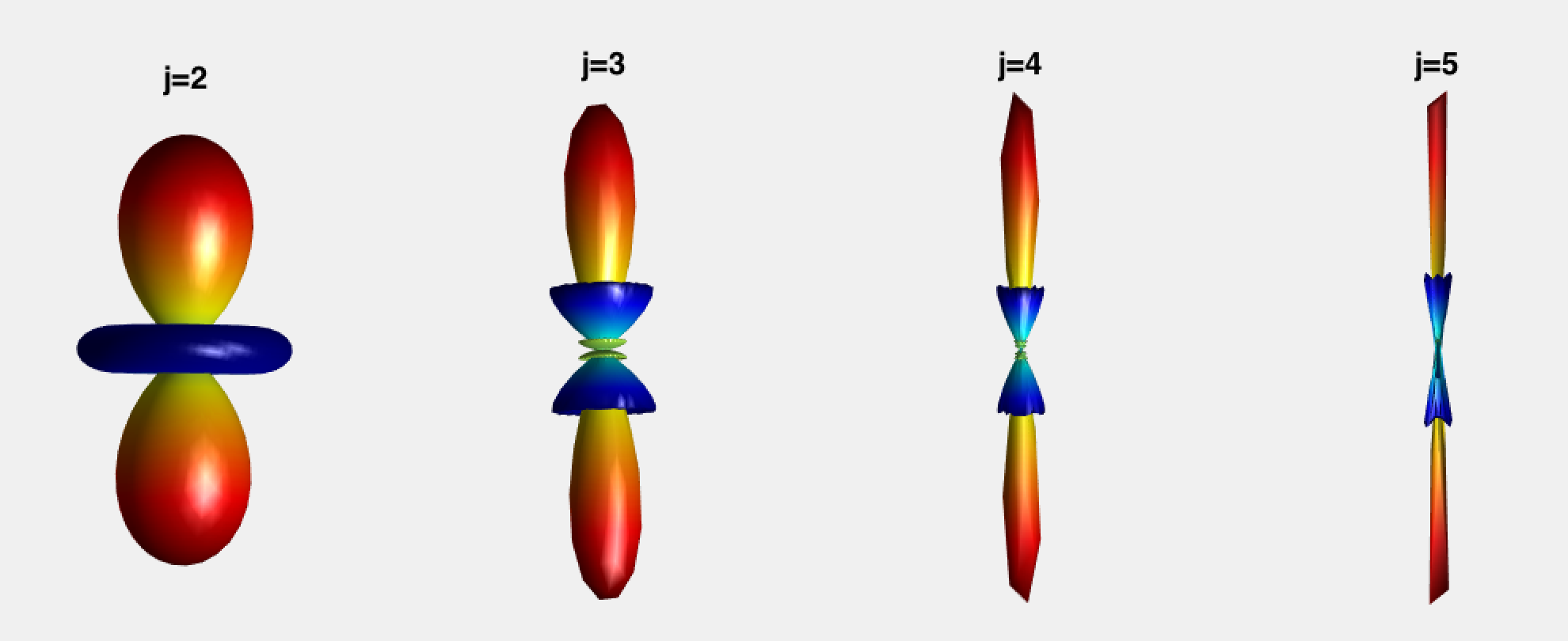}
					\end{tabular}			
			\end{center}
		\end{figure}

	Real symmetric SH functions and symmetrized needelts corresponding to various frequency levels are depiected in Figure   \ref{fig:SH_SN}.
	
	Recall that $\Phi_{00}$, the constant function on unit sphere,  is included to make the needlets frame tight. Hence the total number of basis functions in the (symmetried) needlets frame up to the frequency level $j_{max}$ is $1+\frac{1}{2}\sum_{j=1}^{j_{max}} N_{j,pix}$, where $N_{j,pix}$ is the number of spherical needlets at frequency level $j$ and $N_{j,pix} = 12\times(2^{j-1})^2$ according to the HEALPix construction of spherical needlets.

Let $\mathbf{\Phi}_l$ denote the vector consisting of real symmetric SH functions upto level $l$ and  $\mathbf{\Psi}_j$ denote the vector consisting of real symmetric spherical needlets (SN)  functions upto level $j$. Let $L_l=|\mathbf{\Phi}_{l}|$ and $N_j=|\mathbf{\Psi}_j|$.
Let $\mathbb{SH}_{l}$ and $\mathbb{SN}_{j}$  denote the linear space generated by $\mathbf{\Phi}_l$ and $\mathbf{\Psi}_j$, respectively. 

For $l_{\max} \geq 1$, let $j_{max} = \lceil\log_2(l_{\max})+1\rceil$. For example, when $l_{\max}=8$ ($L=45$ SH basis functions), the corresponding $j_{\max}=4$ ($N=511$ needlets).
 Then by needelt construction (equation (\ref{eq:needlet_construct}) of the main text): 
$$
\mathbb{SH}_{l_{\max}} \leq \mathbb{SN}_{j_{\max}} \leq \mathbb{SH}_{l_{\max+1}}, ~~~ L:=L_{l_{\max}} < N:= N_{j_{\max}} < \tilde{L}:=L_{l_{\max}+1}. 
$$
where ``$\leq$" stands for the ``subspace" symbol.   

If a spherical function $F \in \mathbb{SH}_{l_{\max}}$, then it can be linearly represented by all three collections $\mathbf{\Phi}_{l_{\max}}, \mathbf{\Psi}_{j_{\max}}, \mathbf{\Phi}_{l_{\max+1}}$. Denote their respective coefficients as $\mathbf{f}, \boldsymbol{\beta}, \mathbf{\tilde{f}}$. Note that the needlet coefficients $\boldsymbol{\beta}$ are defined through inner products according to equation (\ref{eq:needlet_coef}) of the main text. (As needlets are not linearly independent,  linear representation in needlets is not unique and thus we need to explicitly define the needlet coefficients.)  

Since $\mathbb{SN}_{j_{\max}} \leq \mathbb{SH}_{l_{\max+1}}$, $\mathbf{\Psi}_{j_{\max}}$ can be linearly represented by $\mathbf{\Phi}_{l_{\max}+1}$:
$$
\mathbf{\Psi}_{j_{\max}}=\mathbf{\tilde{C}}_{N \times \tilde{L}}\mathbf{\Phi}_{l_{\max}+1}. 
$$
Thus, $
\boldsymbol{\beta}_{N \times 1} =\mathbf{\tilde{C}}_{N \times \tilde{L}} \mathbf{\tilde{f}}_{\tilde{L} \times 1}. 
$

Moreover, due to orthogonality of the SH functions, $\mathbf{\tilde{f}}^T=(\mathbf{f}^T, \mathbf{0}^T)$.  Let $\mathbf{C}^\ast$ denote the matrix consisting of the first $L$ columns of  $\mathbf{\tilde{C}}$, then

$$
\boldsymbol{\beta}_{N \times 1} =\mathbf{C}^{\ast}_{N \times L} \mathbf{f}_{L \times 1}. 
$$ 

Note that $rank(\mathbf{C}^{\ast})=L$, and thus 

$$
\mathbf{f}_{L \times 1}= \mathbf{C}_{L \times N}\boldsymbol{\beta}_{N \times 1}, 
$$
with 
$$
\mathbf{C}=(\mathbf{C}^{\ast, T} \mathbf{C}^{\ast})^{-1} \mathbf{C}^{\ast, T}
$$
being the transition matrix from needlet coefficients to SH coefficients.  Finally, the first $L$ SH coefficients of the first $N$ needlets, $\mathbf{C}^{\ast}$, can be computed analytically according to the needlets construction formula equation (\ref{eq:needlet_construct}) of the main text.

\section{Algorithms details} 

\subsection{The ADMM algorithm}\label{subsec:ADMM}
We describe the ADMM algorithm \citep{BoydEtAl2011}  for non-negativity constrained $\ell_1$ penalization problem in this section.
	For $$x\in\mathbb{R}^n,\;A\in\mathbb{R}^{m\times n},\;b\in\mathbb{R}^m,\;C\in\mathbb{R}^{l\times n}\;\textrm{and}\;d\in\mathbb{R}^l,$$ the constrained lasso problem can be formulated as: 
	$$
	{\rm argmin}_{x: Cx\leq d} \frac{1}{2}||Ax-b||_2^2+\lambda||x||_1.
	$$
	The corresponding ADMM form of this problem is to minimize: 
	$$
	\frac{1}{2}||Ax-b||_2^2+\lambda||z||_1+I_+(w)
	$$
	subject to
	$$ x-z=0,\quad Cx+w-d=0,$$
	where $z\in\mathbb{R}^n$, $w\in\mathbb{R}^l$, $I_+(w)=0$ if $w\geq0$(elementwise) and $I_+(w)=\infty$ if otherwise. 
	
	The augmented Lagrangian has the form:
	$$L_\rho(x,z,w,u,t)=\frac{1}{2}||Ax-b||_2^2+\lambda||z||_1+I_+(w)+\frac{\rho}{2}||x-z+u||_2^2+\frac{\rho}{2}||Cx+w-d+t||_2^2,$$
	where $u\in\mathbb{R}^n$ and $t\in\mathbb{R}^l$.

	The ADMM algorithm performs iterative updates as follows:
	\begin{itemize}
		\item
		\textbf{x-minimization}
		\begin{itemize}
			\item
			$x^{k+1}=\arg\min_x\frac{1}{2}||Ax-b||_2^2+\frac{\rho}{2}||x-z^k+u^k||_2^2+\frac{\rho}{2}||Cx+w^k-d+t^k||_2^2$.
			\begin{itemize}
				\item
				$x^{k+1}=\left(A^TA+\rho I_n+\rho C^TC\right)^{-1}\left(A^Tb+\rho\left(z^k-u^k\right)+\rho C^T\left(d-w^k-t^k\right)\right)$, when $m+l\geq n$.
				\item
				$x^{k+1}=\rho^{-1}\left[I_n-B^T\left(BB^T+\rho I_{m+l}\right)^{-1}B\right]\left(A^Tb+\rho\left(z^k-u^k\right)+\rho C^T\left(d-w^k-t^k\right)\right)$, where $B=\left(\begin{array}{c}A\\ \sqrt{\rho}C\end{array}\right)\in\mathbb{R}^{(m+l)\times n}$, when $m+l<n$.
			\end{itemize}
		\end{itemize}
		\item
		\textbf{z/w-minimization}
		\begin{itemize}
			\item
			$z^{k+1}=\arg\min_z\frac{\rho}{2}||x^{k+1}+u^k-z||_2^2+\lambda||z||_1=S_{\lambda/\rho}(x^{k+1}+u^k)$,\\
			where $S_{\lambda/\rho}(x^{k+1}+u^k)_i=(|x^{k+1}_i+u^k_i|-\frac{\lambda}{\rho})_+\textrm{sign}(x^{k+1}_i+u^k_i)$.
			\item
			$w^{k+1}=\arg\min_w\frac{\rho}{2}||w+Cx^{k+1}-d+t^k||_2^2+I_+(w)=\max\left\{0,\;-Cx^{k+1}+d-t^k\right\}$,\\
			where the maximum is taken elementwisely.
		\end{itemize}
		\item
		\textbf{dual-update}
		\begin{itemize}
			\item
			$u^{k+1}=u^k+x^{k+1}-z^{k+1}$.
			\item
			$t^{k+1}=t^k+Cx^{k+1}+w^{k+1}-d$.
		\end{itemize}
	\end{itemize}

	The stopping criterion of the algorithm is as follows :
	\begin{itemize}
		\item
		$||r^k||_2=\left(||x^k-z^k||_2^2+||Cx^k+w^k-d||_2^2\right)^{\frac{1}{2}}$.
		\item
		$||s^k||_2=\rho||z^k-z^{k-1}-C^T\left(w^k-w^{k-1}\right)||_2$.
		\item
		$\epsilon^{\textrm{pri}}=\sqrt{n+l}\epsilon^{\textrm{abs}}+\epsilon^{\textrm{rel}}\max\left\{\left(||x^k||_2^2+||Cx^k||_2^2\right)^{\frac{1}{2}}, \left(||z^k||_2^2+||w^k||_2^2\right)^{\frac{1}{2}}, ||d||_2\right\}$.
		\item
		$\epsilon^{\textrm{dual}}=\sqrt{n}\epsilon^{\textrm{abs}}+\epsilon^{\textrm{rel}}\rho||u^k+C^Tt^k||_2$.
		\item
		Terminate when $||r^k||_2\leq\epsilon^{\textrm{pri}}$ and $||s^k||_2\leq\epsilon^{\textrm{dual}}$.
	\end{itemize}

	We set $\epsilon^{\textrm{abs}}=10^{-4}$ and $\epsilon^{\textrm{rel}}=10^{-2}$ to strike a good balance between convergence time and accuracy. 
We also set $\rho=\lambda$ following the suggestions in  \citep{BoydEtAl2011}.

\subsection{Choice of penalty parameter}\label{subsec:penalty}

We consider a criterion that is based on the residual sum of
squares (RSS). Note that, $RSS$ itself is not able to achieve bias-variance trade-off: It will
always lead to the largest model (i.e., selecting the smallest $\lambda$). Instead, we choose the
largest $\lambda$ where the RSS starts to flatten out. Specifically, we fit model (\ref{eq:wavelet_penalized_regression})
with a sequence of decreasing $\lambda$s on a grid $\Lambda=\{\lambda_1>\cdots>\lambda_K\}$ and calculate $RSS_k$ and the  change of $RSS_k$ on the logarithm scale:

$$
\delta_k:=\left | \frac{\log RSS_{k}-\log RSS_{k-1}}{\log \lambda_{k} -\log \lambda_{k-1}} \right|, ~~ k=2,\cdots, K.
$$

Once the averaged relative change across $T$ consecutive steps is smaller than a pre-specified threshold $\epsilon$ we will stop the algorithm and return
$$
\lambda_{opt}:=\lambda_{k^*}, ~~ where~~~ k^*=\min\{T \leq k \leq K: 
\frac{1}{T} \sum_{l=0}^{T-1} \delta_{k-l}
<\epsilon\}
$$
as the chosen $\lambda$ (minimum of an empty set is defined as $K$).
In the numerical studies, we set $\Lambda$  as $P=500$ equally spaced  grid points on the
log-scale on $10^{-5} \sim 10^{-2}$. We set  $T=5\% \times 500=25$ and the threshold $\epsilon=2\times 10^{-4}$. The results are not very sensitive to the choice of $P,T, \epsilon$. For example, with $P=50$ grid points, $T=2$ and $\epsilon$ ranges from $2\times 10^{-4}$ to $5\times 10^{-3}$, we obtain qualitatively similar results.

\subsection{Peak detection algorithm \label{subsec:peak}}

Here we describe a peak detection algorithm based on grid search.
Given an estimated FOD evaluated on a fine equal-angular grid, the algorithm has three steps: 
\begin{itemize}
	\item Step I:  For each grid point, we compare the (estimated) FOD value on this point with those of its $k$ nearest neighboring grid points according to the arc-length distance. If its FOD value is no smaller than that of any of its neighbors, this grid point would be identified as a local maximal. 
	\item Step II: We identify the highest peak (global maximal) and eliminate any local peak whose height is less than a pre-specified percentage $\alpha$ of that of the highest peak. This step aims to eliminate false local peaks due to noise.
	\item Step III: We cluster the local maximals that are within a (arc-length) distance of a pre-specified threshhold $A$ together and use their mean location as a detected peak location. This step is to assure that each plateau (if any) in the (estimated) FOD would be counted as one peak.
\end{itemize}
In practice, we choose $k$ such that the neighborhood of each grid point spans about $25$ degrees, because based on our numerical studies, it is found that  it is hard to distinguish fiber directions with separation angles less than $30$ degrees. For the value of $\alpha$, it is obvious that large $\alpha$-values result in less detected peaks and smaller $\alpha$-values result in more detected peaks. We found that the peak detection algorithm 
leads to pretty stable results for $\alpha$ between $0.1$ and $0.4$. We choose $\alpha=0.25$ in our numerical experiments. $A$ is set such that local maximals within $5$ degrees of each other are clustered together and counted as one peak.

\section{Experimental details}\label{subsec:simul_additional}

\subsection{SuperCSD algorithm \label{subsec:scsd}}

The SuperCSD algorithm is as follows \citep{TournierEtal2007} :  Consider an $l^s_{\max}$ order SH presentation. Let $L_s$ be the corresponding number of SH basis functions.
 Let $\mathbf{P}_{n_s \times L_s}$ and $\mathbf{\Phi}_{n \times L_s}^s$ be the evaluation matrices of these $L_s$ SH basis functions, on a dense evaluation grid with $n_s$ grid points (e.g., $n_s=2562$ equiangular grid points), and on the $n$ gradient directions, respectively. Let $\mathbf{R}^s$ be the diagonal matrix corresponding to the $l^s_{\max}$ order SH representation coefficients of the response function.
\begin{enumerate}
	\item \textbf{Initial step}: Get an initial estimator $\widehat{\mathbf{f}}_0$ by \texttt{SH-ridge}. 
	\item \textbf{Filter step}: In the $l^s_{\max}$ order SH representation of $\widehat{\mathbf{f}}_0$, set the spherical harmonics  coefficients  over order $l=4$ to zero to reduce high frequency noise.
	 
	\item \textbf{$(k+1)$th updating step}: Define
	$$\hat{\mathbf{F}}^k= \mathbf{P} \hat{\mathbf{f}}^k$$ as the estimated FOD  on the dense evaluation grid from the $k$th step.  Let
	
	\begin{equation}\label{eq:super_csd}
	\widehat{\mathbf{f}}^{k+1} = \arg\min_{\mathbf{f}} \parallel
	\mathbf{y} - \mathbf{\Phi}^s {\mathbf{R}^s}  \mathbf{f}\parallel_2^2 +
	\lambda\parallel \mathbf{P}^k \mathbf{f}\parallel_2^2, 
	\end{equation}
	
	where $\mathbf{P}^k$ is an $n_s\times L_s$ matrix,   
	\begin{equation}\label{eq:super_csd_L}
	\mathbf{P}^k_{i,(l,m)} := \begin{cases}
	\mathbf{P}_{i,(l,m)} & \text{if } \hat{F}^k_{i} \leq \tau  \\
	0      & \text{if } \hat{F}^k_{i} > \tau
	\end{cases}, ~~~i=1,\cdots, n_s, ~~l=0,2,\cdots, l^s_{\max},~~ m=-l,\cdots,0,\cdots, l,
	\end{equation}
	  that penalizes small (including negative) values of the estimated FOD. 
	\item Repeat step 3 until $\mathbf{P}^k$ stabilizes.
\end{enumerate}
 The recommended values  for $\tau, \lambda$ by  \citep{TournierEtal2007} are $0.1, 1$, respectively. As for $l^s_{\max}$, the results from \citep{TournierEtal2007} suggest relatively small level, e.g.,  $l^s_{\max}=8$ for large separation angles, and relatively large level,  e.g.,  $l^s_{\max}=12$ for small separation angles.  
\subsection{Synthesis data: additional results}
(For figures in this and the subsequent sections, please see \url{http://anson.ucdavis.edu/~jie/FOD_estimation_needlet_supp.pdf}.)

In the tables, we report the performance of each estimator using the following numerical metrics: (i) The success rate of the peak detection algorithm applied to the $100$ estimated FODs, where `success' means that the algorithm identifies the correct number of fiber bundles. These are shown under ``Correct'' in the tables. We also report the proportion of replicates where the number of fiber bundles are either over-estimated or under-estimated. These are described under ``Over'' and ``Under'', respectively, in the tables;  (ii) Across the successful replicates identified in (i), the mean angular error(s) between the identified peak(s) and the true fiber  direction(s), as well as the mean estimated separation angle(s) between pairs of fiber bundles. These are reported under ``Error1", ``Error2", ``Mean Sep." etc. in the tables; (iii) The mean and standard deviation (across $100$ replicates) of the Hellinger distance  between the estimated FOD and $F^*$ -- the true FOD $F$ projected on to an SH basis with $l_{\max}=8$.  These are shown in the columns ``Mean H-dist.'' and ``SD H-dist.'' in the tables.

\subsubsection{$0$-fiber simulation: $n=41$ \label{sec:0fiber}}

\begin{figure}[H]
	\centering
	\caption{0-fiber: $n=41$ gradient directions. From left to right: SH $l_{\max}=8$ representation of true FOD,  \texttt{SH-ridge},  \texttt{SCSD8}, \texttt{SCSD12}, \texttt{SN-lasso}.  From top to bottom: $b=1000s/mm^2, 3000s/mm^2, 5000s/mm^2$ .  The opaque part in the plots corresponds to mean estimated FOD across $100$ replicates, and the translucent part in the plots corresponds to mean plus two standard deviations of the estimated FOD.
		\label{fig:0fib_lmax8_N41}    }
\end{figure}

\begin{table}[H]                                                                      
	\centering                    
	\caption{0-fiber: $n=41$ gradient directions \label{table:0fib_lmax8_N41}      }    

\begin{tabular}{c|ccc|cc}\hline\hline                                                                                             	\multicolumn{6}{c}{$b=1000s/mm^2$}\\      
	Estimator & Correct & Under & Over & Mean H-dist. & SD H-dist. \\                            
	\hline                                                       
	SH-ridge & 0.00 & 0.00 & 1.00 & 0.00 & 0.0000 \\       
	\hline                                                 
	SCSD8 & 0.00 & 0.00 & 1.00 & 0.49 & 0.0477 \\          
	\hline                                                 
	SCSD12 & 0.00 & 0.00 & 1.00 & 0.67 & 0.0681 \\         
	\hline                                                 
	SN-lasso & 1.00 & 0.00 & 0.00 & 0.00 & 0.0000 \\        
	\hline \hline                                 
	
		\multicolumn{6}{c}{$b=3000s/mm^2$}\\      
			Estimator & Correct & Under & Over & Mean H-dist. & SD H-dist. \\                            
			\hline                                                       
			SH-ridge & 0.00 & 0.00 & 1.00 & 0.00 & 0.0000 \\       
			\hline                                                 
			SCSD8 & 0.00 & 0.00 & 1.00 & 0.03 & 0.0003 \\          
			\hline                                                 
			SCSD12 & 0.97 & 0.00 & 0.03 & 0.05 & 0.1114 \\         
			\hline                                                 
			SN-lasso & 1.00 & 0.00 & 0.00 & 0.00 & 0.0000 \\         
			\hline \hline                       
			\multicolumn{6}{c}{$b=5000s/mm^2$}\\         
			Estimator & Correct & Under & Over & Mean H-dist. & SD H-dist. \\                            
			\hline                                                       
			SH-ridge & 0.01 & 0.00 & 0.99 & 0.00 & 0.0000 \\       
			\hline                                                 
			SCSD8 & 0.00 & 0.00 & 1.00 & 0.03 & 0.0003 \\          
			\hline                                                 
			SCSD12 & 1.00 & 0.00 & 0.00 & 0.03 & 0.0006 \\         
			\hline                                                 
			SN-lasso & 1.00 & 0.00 & 0.00 & 0.00 & 0.0000 \\       
			\hline \hline                                        
	\end{tabular}
\end{table}

\subsubsection{$1$-fiber simulation: $b=1000s/mm^2$ and $ratio=10$ \label{sec:1fiber}}


\begin{figure}[H]
	\centering
	\caption{1-fiber: $b=1000s/mm^2$ and $ratio=10$. From left to right: SH $l_{\max}=8$ representation of true FOD,  \texttt{SH-ridge},  \texttt{SCSD8}, \texttt{SCSD12}, \texttt{SN-lasso}; Top panel: $n=41$, bottom panel: $n=81$.
		The lines indicate the true fiber direction,  the opaque part in the plots corresponds to mean estimated FOD across $100$ replicates, and the  translucent  part in the plots corresponds to mean plus two standard deviations of the estimated FOD.
		\label{fig:1fib_b1_ratio10_lmax8}    }
\end{figure}

\begin{table}[H]                                                                      
	\centering                    
	\caption{1-fiber: $b=1000s/mm^2$ and $ratio=10$.\label{table:1fib_b1_ratio10_lmax8}      }    
\begin{tabular}{c|ccc|c|cc}\hline\hline                               
	\multicolumn{7}{c}{$n=41$}\\                                                                      
	Estimator & Correct & Under & Over & Error & Mean H-dist. & SD H-dist. \\                            
	\hline                                                           
	SH-ridge & 0.91 & 0.00 & 0.09 & 3.58 & 0.62 & 0.0098 \\                
		\hline                                                           
		SCSD8 & 1.00 & 0.00 & 0.00 & 2.37 & 0.60 & 0.0025 \\            
		\hline                                                           
		SCSD12 & 0.97 & 0.00 & 0.03 & 2.30 & 0.52 & 0.0337 \\           
		\hline                                                           
		SN-lasso & 1.00 & 0.00 & 0.00 & 2.58 & 0.50 & 0.0136 \\                
		\hline\hline                                                                                                                                            
 
		\multicolumn{7}{c}{$n=81$}\\                                                                      
		Estimator & Correct & Under & Over & Error & Mean H-dist. & SD H-dist. \\                            
		\hline                                                                                     
		SH-ridge & 0.92 & 0.00 & 0.08 & 2.33 & 0.62 & 0.0098 \\                
			\hline                                                           
			SCSD8 & 1.00 & 0.00 & 0.00 & 1.38 & 0.60 & 0.0023 \\            
			\hline                                                           
			SCSD12 & 1.00 & 0.00 & 0.00 & 1.43 & 0.51 & 0.0049 \\           
			\hline                                                           
			SN-lasso & 1.00 & 0.00 & 0.00 & 1.69 & 0.49 & 0.0089 \\                
		\hline \hline                                                           
	\end{tabular}                                                                                
\end{table}


\subsubsection{$2$-fiber simulation: $n=41$ and $ratio=10$}

 \begin{center}
 	\begin{figure}[H]
 		
 		\caption{\textbf{Two fiber crossing at $\mathbf{90^{\circ}, 75^{\circ}, 60^{\circ}}$ with $\mathbf{b=1000 s/mm^2,SNR=20}$}.  $n=41$ gradient directions and $ratio=10$.
 			\label{fig:b1_ratio10_lmax8_N41_bar}}
 	\end{figure} 
 \end{center}
 
 \begin{table}[H]                                                                      
 	\centering                    
 	\caption{\textbf{Two fiber crossing at $\mathbf{90^{\circ}, 75^{\circ}, 60^{\circ}}$ with $\mathbf{b=1000 s/mm^2,SNR=20}$}.  $n=41$ gradient directions and $ratio=10$. \label{table:b1_ratio10_lmax8_N41}      }                                              
 	
 	\begin{tabular}{c|ccc|ccc|cc}\hline\hline                               
 		\multicolumn{9}{c}{Separation angle:  $sep=90^{\circ}$}\\                                                                      
 		Estimator & Correct & Under & Over & Error-1  & Error-2 & Mean Sep. & Mean H-dist. & SD H-dist. \\
 		\hline                                                                                 
 		SH-ridge & 0.18 & 0.74 & 0.08 & 11.68 & 12.97 & 76.15 & 0.62 & 0.0098 \\                     
 		\hline                                                                                 
 		SCSD8 & 0.64 & 0.00 & 0.36 & 8.16 & 8.99 & 87.01 & 0.60 & 0.0542 \\                   
 		\hline                                                                                 
 		SCSD12 & 0.44 & 0.00 & 0.56 & 11.38 & 12.02 & 87.10 & 0.63 & 0.1185 \\                
 		\hline                                                                                 
 		SN-lasso & 0.85 & 0.00 & 0.15 & 8.96 & 8.74 & 85.66 & 0.54 & 0.0879 \\                       
 		\hline\hline
 		
 		\multicolumn{9}{c}{Separation angle:  $sep=75^{\circ}$}\\                                                                      
 		Estimator & Correct & Under & Over & Error-1  & Error-2 & Mean Sep. & Mean H-dist. & SD H-dist. \\
 		\hline
 		SH-ridge & 0.08 & 0.84 & 0.08 & 14.35 & 20.35 & 84.06 & 0.62 & 0.0091 \\                     
 		\hline                                                                                 
 		SCSD8 & 0.82 & 0.00 & 0.18 & 9.50 & 9.14 & 71.27 & 0.56 & 0.0638 \\                   
 		\hline                                                                                 
 		SCSD12 & 0.19 & 0.00 & 0.81 & 8.68 & 11.78 & 75.70 & 0.63 & 0.0942 \\                 
 		\hline                                                                                 
 		SN-lasso & 0.85 & 0.00 & 0.15 & 8.06 & 9.40 & 74.67 & 0.51 & 0.0935 \\                       
 		\hline	\hline
 		
 		\multicolumn{9}{c}{Separation angle:  $sep=60^{\circ}$}\\                                                                      
 		Estimator & Correct & Under & Over & Error-1  & Error-2 & Mean Sep. & Mean H-dist. & SD H-dist. \\
 		\hline
 		SH-ridge & 0.04 & 0.90 & 0.06 & 22.83 & 43.95 & 82.24 & 0.61 & 0.0070 \\                     
 		\hline                                                                                 
 		SCSD8 & 0.94 & 0.05 & 0.01 & 9.38 & 9.01 & 55.81 & 0.55 & 0.0741 \\                   
 		\hline                                                                                 
 		SCSD12 & 0.27 & 0.00 & 0.73 & 9.97 & 13.67 & 50.20 & 0.61 & 0.0863 \\                 
 		\hline                                                                                 
 		SN-lasso & 0.88 & 0.02 & 0.10 & 10.56 & 10.27 & 59.63 & 0.51 & 0.1097 \\                     
 		\hline\hline
 		
 		\end{tabular}                                                                          
 		
 		\end{table}               
 		
 		\begin{table}[H]                                                                      
 			\centering            
 				\caption{\textbf{Two fiber crossing at $\mathbf{45^{\circ}}$ with $\mathbf{b=1000, 3000 s/mm^2, SNR=20,50}$.}  $n=41$ gradient directions and $ratio=10$.  \label{table:Sep45_b13_ratio10_lmax16_N41}                }                                                                                                              
 			\begin{tabular}{c|ccc|ccc|cc}                                                   \hline\hline                                                                              	\multicolumn{9}{c} {$b=1000s/mm^2,SNR=20$}\\
 				& Correct & Under & Over & Error-1 & Error-2 & Mean Sep. & Mean H-dist. & SD H-dist \\
 				\hline                                                                                 
 				SH-ridge & 0.04 & 0.92 & 0.04 & 7.28 & 70.84 & 81.80 & 0.66 & 0.0073 \\                      
 				\hline                                                                                 
 				SCSD12 & 0.85 & 0.13 & 0.02 & 10.97 & 11.55 & 41.87 & 0.70 & 0.0460 \\                
 				\hline                                                                                 
 				SCSD16 & 0.55 & 0.06 & 0.39 & 12.10 & 13.78 & 37.83 & 0.73 & 0.0566 \\                
 				\hline                                                                                 
 				SN-lasso & 0.71 & 0.10 & 0.19 & 9.76 & 14.14 & 47.05 & 0.74 & 0.1073 \\                      
 				\hline\hline                                                                                	\multicolumn{9}{c} {$b=1000s/mm^2,SNR=50$}\\
 						& Correct & Under & Over & Error-1 & Error-2 & Mean Sep. & Mean H-dist. & SD H-dist \\ 
 					\hline                                                                                 
 					SH-ridge & 0.09 & 0.88 & 0.03 & 21.86 & 68.07 & 84.79 & 0.66 & 0.0037 \\                     
 					\hline                                                                                 
 					SCSD12 & 0.95 & 0.04 & 0.01 & 6.88 & 7.56 & 40.65 & 0.66 & 0.0288 \\                  
 					\hline                                                                                 
 					SCSD16 & 0.64 & 0.01 & 0.35 & 9.97 & 10.01 & 36.30 & 0.71 & 0.0565 \\                 
 					\hline                                                                                 
 					SN-lasso & 0.85 & 0.02 & 0.13 & 5.58 & 7.51 & 45.23 & 0.64 & 0.0729 \\                       
 					\hline\hline
 				\multicolumn{9}{c}{$b=3000s/mm^2,SNR=20$}\\             
 						& Correct & Under & Over & Error-1 & Error-2 & Mean Sep. & Mean H-dist. & SD H-dist  \\
 					\hline                                                                                 
 					SH-ridge & 0.14 & 0.69 & 0.17 & 12.25 & 84.96 & 87.88 & 0.66 & 0.0222 \\                     
 					\hline                                                                                 
 					SCSD12 & 0.98 & 0.01 & 0.01 & 5.01 & 6.13 & 41.88 & 0.65 & 0.0215 \\                  
 					\hline                                                                                 
 					SCSD16 & 0.49 & 0.00 & 0.51 & 5.86 & 8.62 & 39.28 & 0.68 & 0.0481 \\                  
 					\hline                                                                                 
 					SN-lasso & 0.75 & 0.00 & 0.25 & 4.18 & 6.18 & 45.53 & 0.63 & 0.0777 \\                       
 					\hline\hline                                                                           	\multicolumn{9}{c}{$b=3000s/mm^2,SNR=50$}\\        
 					          		& Correct & Under & Over & Error-1 & Error-2 & Mean Sep. & Mean H-dist. & SD H-dist \\
 					          	\hline                                                                                 
 					          	SH-ridge & 0.07 & 0.41 & 0.52 & 32.90 & 31.45 & 66.98 & 0.66 & 0.0111 \\                     
 					          	\hline                                                                                 
 					          	SCSD12 & 1.00 & 0.00 & 0.00 & 2.36 & 3.96 & 41.28 & 0.64 & 0.0059 \\                  
 					          	\hline                                                                                 
 					          	SCSD16 & 0.96 & 0.00 & 0.04 & 4.06 & 5.06 & 40.27 & 0.64 & 0.0323 \\                  
 					          	\hline                                                                                 
 					          	SN-lasso & 0.94 & 0.00 & 0.06 & 1.65 & 3.88 & 44.64 & 0.56 & 0.0358 \\                       
 					          	\hline\hline                                                            
 			\end{tabular}

 		\end{table}

 		\begin{table}[H]                                                                      
 			\centering             
 				\caption{\textbf{Two fiber crossing at $\mathbf{30^{\circ}}$ with $\mathbf{b=3000, 5000 s/mm^2, SNR=20, 50}$}. $n=41$ gradient directions and $ratio=10$. \label{table:Sep30_b35_ratio10_lmax16_N41}          }                                                                                                             
 			\begin{tabular}{c|ccc|ccc|cc}                                                   
 				\hline\hline	\multicolumn{9}{c}{$b=3000s/mm^2,SNR=20$}\\                                                                                     
 					& Correct & Under & Over & Error-1 & Error-2 & Mean Sep. & Mean H-dist. & SD H-dist \\\hline                                                                                
 			SH-ridge & 0.07 & 0.73 & 0.20 & 22.18 & 74.10 & 86.47 & 0.66 & 0.0262 \\                     
 			\hline                                                                                 
 			SCSD12 & 0.04 & 0.96 & 0.00 & 8.63 & 14.23 & 36.38 & 0.66 & 0.0161 \\                 
 			\hline                                                                                 
 			SCSD16 & 0.55 & 0.43 & 0.02 & 7.51 & 10.26 & 33.00 & 0.68 & 0.0397 \\                 
 			\hline                                                                                 
 			SN-lasso & 0.73 & 0.20 & 0.07 & 7.55 & 9.05 & 31.75 & 0.69 & 0.0810 \\                       
 				\hline\hline
 				\multicolumn{9}{c}{$b=3000s/mm^2,SNR=50$}\\                                                                       	& Correct & Under & Over & Error-1 & Error-2 & Mean Sep. & Mean H-dist. & SD H-dist \\
 				\hline                                                                                 
 				SH-ridge & 0.05 & 0.61 & 0.34 & 9.21 & 67.56 & 64.64 & 0.65 & 0.0161 \\                      
 				\hline                                                                                 
 				SCSD12 & 0.00 & 1.00 & 0.00 & - & - & - & 0.66 & 0.0082 \\                            
 				\hline                                                                                 
 				SCSD16 & 0.22 & 0.78 & 0.00 & 5.96 & 5.86 & 27.99 & 0.65 & 0.0208 \\                  
 				\hline                                                                                 
 				SN-lasso & 0.89 & 0.10 & 0.01 & 4.47 & 5.95 & 29.90 & 0.60 & 0.0598 \\            \\                       
 				\hline\hline
 				         	\multicolumn{9}{c}{$b=5000s/mm^2,SNR=20$}\\  
 				   	& Correct & Under & Over & Error-1 & Error-2 & Mean Sep. & Mean H-dist. & SD H-dist \\
 				   	\hline                                                                                 
 				   	SH-ridge & 0.05 & 0.54 & 0.41 & 9.12 & 84.56 & 81.85 & 0.66 & 0.0241 \\                      
 				   	\hline                                                                                 
 				   	SCSD12 & 0.04 & 0.96 & 0.00 & 6.66 & 9.37 & 33.49 & 0.65 & 0.0138 \\                  
 				   	\hline                                                                                 
 				   	SCSD16 & 0.50 & 0.47 & 0.03 & 7.27 & 10.01 & 32.50 & 0.66 & 0.0370 \\                 
 				   	\hline                                                                                 
 				   	SN-lasso & 0.81 & 0.13 & 0.06 & 6.16 & 8.53 & 31.28 & 0.66 & 0.0771 \\                       
 				   	\hline\hline   
 					\multicolumn{9}{c}{$b=5000s/mm^2,SNR=50$}\\     	
 					& Correct & Under & Over & Error-1 & Error-2 & Mean Sep. & Mean H-dist. & SD H-dist \\
 					\hline                                                                                 
 					SH-ridge & 0.23 & 0.44 & 0.33 & 23.05 & 56.33 & 65.26 & 0.66 & 0.0113 \\                     
 					\hline                                                                                 
 					SCSD12 & 0.00 & 1.00 & 0.00 & - & - & - & 0.65 & 0.0070 \\                            
 					\hline                                                                                 
 					SCSD16 & 0.45 & 0.55 & 0.00 & 4.23 & 4.59 & 24.68 & 0.64 & 0.0132 \\                  
 					\hline                                                                                 
 					SN-lasso & 0.96 & 0.04 & 0.00 & 2.71 & 4.23 & 28.59 & 0.57 & 0.0458  \\                       
 					\hline\hline                                                                                 	                                                                         	                                                                           
 			\end{tabular}

 		\end{table}

\subsubsection{$2$-fiber simulation: $n=81$, $b=1000s/mm^2$, and $ratio=10$}	
 		\begin{figure}[H]
 			\centering
 			\caption{
 				\small{Two fiber crossing: $n=81$ gradient directions, $b=1000s/mm^2$ and $ratio=10$. From left to right: SH $l_{\max}=8$ representation of true FOD,  \texttt{SH-ridge},  \texttt{SCSD8}, \texttt{SCSD12}, \texttt{SN-lasso}; From top to bottom: Separation angle $sep=90^{\circ}$, $sep=75^{\circ}$, $sep=60^{\circ}$, $sep=45^{\circ}$. The lines indicate the true fiber directions, the opaque part in the plots corresponds to mean estimated FOD across $100$ replicates, and the translucent part in the plots corresponds to mean plus two standard deviations of the estimated FOD.}
 				\label{fig:b1_ratio10_lmax8_N81}    }
 		\end{figure}

 		\begin{table}[H]                                                                      
 			\centering                    
 			\caption{Two fiber crossing: $n=81$ gradient directions, $b=1000s/mm^2$ and $ratio=10$.\label{table:b1_ratio10_lmax8_N81}      }                                              
 			
 			\begin{tabular}{c|ccc|ccc|cc}\hline\hline                               
 				\multicolumn{9}{c}{Separation angle:  $sep=90^{\circ}$}\\                                                                      
 				Estimator & Correct & Under & Over & Error-1  & Error-2 & Mean Sep. & Mean H-dist. & SD H-dist. \\
 				\hline                                                                                 
 				SH-ridge & 0.43 & 0.53 & 0.04 & 9.36 & 9.97 & 78.30 & 0.62 & 0.0095 \\                       
 				\hline                                                                                 
 				SCSD8 & 0.66 & 0.00 & 0.34 & 6.03 & 6.11 & 86.95 & 0.58 & 0.0523 \\                   
 				\hline                                                                                 
 				SCSD12 & 0.49 & 0.00 & 0.51 & 7.59 & 8.54 & 88.06 & 0.60 & 0.1365 \\                  
 				\hline                                                                                 
 				SN-lasso & 0.88 & 0.00 & 0.12 & 5.90 & 6.62 & 86.28 & 0.49 & 0.0618 \\                       
 				\hline\hline
 				
 				\multicolumn{9}{c}{Separation angle:  $sep=75^{\circ}$}\\                                                                      
 				Estimator & Correct & Under & Over & Error-1  & Error-2 & Mean Sep. & Mean H-dist. & SD H-dist. \\
 				\hline
 				SH-ridge & 0.09 & 0.86 & 0.05 & 8.01 & 9.67 & 79.72 & 0.62 & 0.0075 \\                       
 				\hline                                                                                 
 				SCSD8 & 0.87 & 0.00 & 0.13 & 6.82 & 7.43 & 71.23 & 0.54 & 0.0428 \\                   
 				\hline                                                                                 
 				SCSD12 & 0.21 & 0.00 & 0.79 & 6.01 & 6.38 & 72.16 & 0.58 & 0.0871 \\                  
 				\hline                                                                                 
 				SN-lasso & 0.94 & 0.00 & 0.06 & 6.11 & 6.75 & 73.47 & 0.47 & 0.0621 \\                       
 				\hline\hline
 				
 				\multicolumn{9}{c}{Separation angle:  $sep=60^{\circ}$}\\                                                                      
 				Estimator & Correct & Under & Over & Error-1  & Error-2 & Mean Sep. & Mean H-dist. & SD H-dist. \\
 				\hline
 				SH-ridge & 0.03 & 0.94 & 0.03 & 7.37 & 7.71 & 65.43 & 0.61 & 0.0046 \\                       
 				\hline                                                                                 
 				SCSD8 & 1.00 & 0.00 & 0.00 & 7.54 & 6.80 & 55.56 & 0.52 & 0.0404 \\                   
 				\hline                                                                                 
 				SCSD12 & 0.30 & 0.00 & 0.70 & 8.59 & 8.93 & 49.50 & 0.58 & 0.0819 \\                  
 				\hline                                                                                 
 				SN-lasso & 0.92 & 0.02 & 0.06 & 7.72 & 8.23 & 58.58 & 0.47 & 0.0787 \\                       
 				\hline\hline
 				
 				\multicolumn{9}{c}{Separation angle:  $sep=45^{\circ}$}\\                                                                      
 				Estimator & Correct & Under & Over & Error-1  & Error-2 & Mean Sep. & Mean H-dist. & SD H-dist. \\
 				\hline		
 				SH-ridge & 0.02 & 0.97 & 0.01 & 49.91 & 38.99 & 87.46 & 0.60 & 0.0054 \\                     
 				\hline                                                                                 
 				SCSD8 & 0.03 & 0.97 & 0.00 & 11.99 & 17.06 & 43.19 & 0.56 & 0.0254 \\                 
 				\hline                                                                                 
 				SCSD12 & 0.90 & 0.06 & 0.04 & 9.75 & 9.82 & 41.14 & 0.57 & 0.0680 \\                  
 				\hline                                                                                 
 				SN-lasso & 0.73 & 0.26 & 0.01 & 7.64 & 8.72 & 43.95 & 0.53 & 0.0712 \\                       
 				\hline\hline                                                                                          
 				\end{tabular}                                                                          
 				
 				\end{table}

\subsubsection{$2$-fiber simulation: $n=321$, $b=1000s/mm^2$, and $ratio=10$}
	
	 \begin{figure}[H]
	 	\centering
	 	\caption{Two fiber crossing: $n=321$ gradient directions, $b=1000s/mm^2$ and $ratio=10$. From left to right: SH $l_{\max}=8$ representation of true FOD,  \texttt{SH-ridge},  \texttt{SCSD8}, \texttt{SCSD12}, \texttt{SN-lasso}; From top to bottom: separation angle $sep=90^{\circ}$, $sep=75^{\circ}$, $sep=60^{\circ}$, $sep=45^{\circ}$. The lines indicate the true fiber directions, the opaque part in the plots corresponds to mean estimated FOD across $100$ replicates, and the  translucent  part in the plots corresponds to mean plus two standard deviations of the estimated FOD.
	 		\label{fig:2fib_b1_ratio10_lmax8_N321}    }
	 \end{figure}

	\begin{table}[H]                                                                      
		\centering                    
		\caption{Two fiber crossing: $n=321$ gradient directions, $b=1000s/mm^2$ and $ratio=10$.\label{table:2fib_b1_ratio10_lmax8_N321}      }                                              
		
		\begin{tabular}{c|ccc|ccc|cc}\hline\hline                               
			\multicolumn{9}{c}{Separation angle:  $sep=90^{\circ}$}\\                                                                      
			Estimator & Correct & Under & Over & Error-1  & Error-2 & Mean Sep. & Mean H-dist. & SD H-dist. \\
			\hline 
		SH-ridge & 0.96 & 0.00 & 0.04 & 3.60 & 5.49 & 86.48 & 0.61 & 0.0119 \\                       
				\hline                                                                                 
				SCSD8 & 1.00 & 0.00 & 0.00 & 2.67 & 3.73 & 88.22 & 0.54 & 0.0068 \\                   
				\hline                                                                                 
				SCSD12 & 0.83 & 0.00 & 0.17 & 2.94 & 3.93 & 88.04 & 0.47 & 0.0775 \\                  
				\hline                                                                                 
				SN-lasso & 1.00 & 0.00 & 0.00 & 2.47 & 4.03 & 87.57 & 0.45 & 0.0349 \\                                                           
				\hline\hline
				
				\multicolumn{9}{c}{Separation angle:  $sep=75^{\circ}$}\\                                                                      
		 		Estimator & Correct & Under & Over & Error-1  & Error-2 & Mean Sep. & Mean H-dist. & SD H-dist. \\
		 		\hline 
				SH-ridge & 0.65 & 0.32 & 0.03 & 5.11 & 5.95 & 76.59 & 0.60 & 0.0131 \\                       
				\hline                                                                                 
				SCSD8 & 1.00 & 0.00 & 0.00 & 3.23 & 4.30 & 71.95 & 0.51 & 0.0135 \\                   
				\hline                                                                                 
				SCSD12 & 0.58 & 0.00 & 0.42 & 2.81 & 3.84 & 73.36 & 0.48 & 0.0790 \\                  
				\hline                                                                                 
				SN-lasso & 1.00 & 0.00 & 0.00 & 2.83 & 4.12 & 73.79 & 0.42 & 0.0363 \\                       
				\hline\hline
				
				\multicolumn{9}{c}{Separation angle:  $sep=60^{\circ}$}\\                                                                      
		 		Estimator & Correct & Under & Over & Error-1  & Error-2 & Mean Sep. & Mean H-dist. & SD H-dist. \\
		 		\hline 
				SH-ridge & 0.17 & 0.80 & 0.03 & 4.57 & 7.42 & 58.97 & 0.60 & 0.0076 \\                       
				\hline                                                                                 
				SCSD8 & 1.00 & 0.00 & 0.00 & 4.34 & 5.21 & 54.42 & 0.51 & 0.0201 \\                   
				\hline                                                                                 
				SCSD12 & 0.58 & 0.00 & 0.42 & 5.35 & 5.54 & 55.47 & 0.49 & 0.0895 \\                  
				\hline                                                                                 
				SN-lasso & 1.00 & 0.00 & 0.00 & 4.59 & 5.80 & 58.16 & 0.42 & 0.0518 \\                       
				\hline\hline
				
				\multicolumn{9}{c}{Separation angle:  $sep=45^{\circ}$}\\                                                                      
		 		Estimator & Correct & Under & Over & Error-1  & Error-2 & Mean Sep. & Mean H-dist. & SD H-dist. \\
		 		\hline
				SH-ridge & 0.01 & 0.99 & 0.00 & 7.91 & 79.54 & 84.68 & 0.60 & 0.0067 \\                      
				\hline                                                                                 
				SCSD8 & 0.00 & 1.00 & 0.00 & - & - & - & 0.55 & 0.0105 \\                             
				\hline                                                                                 
				SCSD12 & 0.96 & 0.03 & 0.01 & 6.90 & 7.94 & 41.65 & 0.52 & 0.0589 \\                  
				\hline                                                                                 
				SN-lasso & 0.73 & 0.27 & 0.00 & 5.82 & 7.41 & 41.32 & 0.49 & 0.0424 \\                       
			\hline\hline
		\end{tabular}                                                                                                               
	\end{table}  
	
	 \subsubsection{$2$-fiber simulation: $b=3000s/mm^2$ and $ratio=10$ \label{sec:b3000}}

	 \begin{figure}[H]
	 	\centering
	 	\caption{Two fiber crossing: $n=41$ gradient directions, $b=3000s/mm^2$ and $ratio=10$. From left to right: SH $l_{\max}=8$ representation of true FOD,  \texttt{SH-ridge},  \texttt{SCSD8}, \texttt{SCSD12}, \texttt{SN-lasso}; From top to bottom: separation angle $sep=90^{\circ}$, $sep=75^{\circ}$, $sep=60^{\circ}$, $sep=45^{\circ}$. The lines indicate the true fiber directions, the opaque part in the plots corresponds to mean estimated FOD across $100$ replicates, and the  translucent  part in the plots corresponds to mean plus two standard deviations of the estimated FOD.
	 		\label{fig:2fib_b3_ratio10_lmax8_N41}    }
	 \end{figure}

	\begin{table}[H]                                                                      
		\centering                    
		\caption{Two fiber crossing: $n=41$ gradient directions, $b=3000s/mm^2$ and $ratio=10$.\label{table:2fib_b3_ratio10_lmax8_N41}      }                                              
		
		\begin{tabular}{c|ccc|ccc|cc}\hline\hline                               
			\multicolumn{9}{c}{Separation angle:  $sep=90^{\circ}$}\\                                                                      
			Estimator & Correct & Under & Over & Error-1  & Error-2 & Mean Sep. & Mean H-dist. & SD H-dist. \\
			\hline 
			SH-ridge & 0.47 & 0.00 & 0.53 & 3.56 & 4.89 & 87.32 & 0.60 & 0.0242 \\                       
				\hline                                                                                 
				SCSD8 & 1.00 & 0.00 & 0.00 & 2.25 & 4.21 & 87.96 & 0.53 & 0.0074 \\                   
				\hline                                                                                 
				SCSD12 & 0.98 & 0.00 & 0.02 & 2.43 & 4.27 & 87.56 & 0.45 & 0.0224 \\                  
				\hline                                                                                 
				SN-lasso & 0.98 & 0.00 & 0.02 & 2.41 & 4.57 & 87.17 & 0.43 & 0.0271 \\
				\hline\hline
				
				\multicolumn{9}{c}{Separation angle:  $sep=75^{\circ}$}\\                                                                      
		 		Estimator & Correct & Under & Over & Error-1  & Error-2 & Mean Sep. & Mean H-dist. & SD H-dist. \\
		 		\hline 
				SH-ridge & 0.47 & 0.01 & 0.52 & 6.32 & 4.89 & 81.46 & 0.59 & 0.0245 \\                       
				\hline                                                                                 
				SCSD8 & 1.00 & 0.00 & 0.00 & 3.02 & 4.13 & 73.12 & 0.51 & 0.0065 \\                   
				\hline                                                                                 
				SCSD12 & 0.93 & 0.00 & 0.07 & 2.83 & 3.96 & 74.13 & 0.43 & 0.0436 \\                  
				\hline                                                                                 
				SN-lasso & 0.97 & 0.00 & 0.03 & 3.08 & 4.26 & 74.88 & 0.41 & 0.0260 \\   
				\hline\hline
				
				\multicolumn{9}{c}{Separation angle:  $sep=60^{\circ}$}\\                                                                      
		 		Estimator & Correct & Under & Over & Error-1  & Error-2 & Mean Sep. & Mean H-dist. & SD H-dist. \\
		 		\hline
				SH-ridge & 0.42 & 0.19 & 0.39 & 6.70 & 11.69 & 65.17 & 0.59 & 0.0274 \\                      
				\hline                                                                                 
				SCSD8 & 1.00 & 0.00 & 0.00 & 3.44 & 4.38 & 56.66 & 0.49 & 0.0146 \\                   
				\hline                                                                                 
				SCSD12 & 0.87 & 0.00 & 0.13 & 3.33 & 4.25 & 58.57 & 0.43 & 0.0634 \\                  
				\hline                                                                                 
				SN-lasso & 0.96 & 0.00 & 0.04 & 3.36 & 4.10 & 58.97 & 0.39 & 0.0395 \\ 
				\hline\hline
				
				\multicolumn{9}{c}{Separation angle:  $sep=45^{\circ}$}\\                                                                      
		 		Estimator & Correct & Under & Over & Error-1  & Error-2 & Mean Sep. & Mean H-dist. & SD H-dist. \\
		 		\hline
				SH-ridge & 0.14 & 0.69 & 0.17 & 12.25 & 84.96 & 87.88 & 0.59 & 0.0303 \\                     
				\hline                                                                                 
				SCSD8 & 0.18 & 0.82 & 0.00 & 6.60 & 7.87 & 37.81 & 0.54 & 0.0113 \\                   
				\hline                                                                                 
				SCSD12 & 0.98 & 0.01 & 0.01 & 5.01 & 6.13 & 41.88 & 0.48 & 0.0523 \\                  
				\hline                                                                                 
				SN-lasso & 0.97 & 0.02 & 0.01 & 4.92 & 6.56 & 43.13 & 0.46 & 0.0505 \\
			\hline\hline
		\end{tabular}
	\end{table}

	 \begin{figure}[H]
	 	\centering
	 	\caption{Two fiber crossing: $n=81$ gradient directions, $b=3000s/mm^2$ and $ratio=10$. From left to right: SH $l_{\max}=8$ representation of true FOD,  SH-ridge,  SCSD8, SCSD12, SN-lasso; From top to bottom: separation angle $sep=90^{\circ}$, $sep=75^{\circ}$, $sep=60^{\circ}$, $sep=45^{\circ}$. The lines indicate the true fiber directions, the opaque part in the plots corresponds to mean estimated FOD across $100$ replicates, and the translucent part in the plots corresponds to mean plus two standard deviations of the estimated FOD.
	 		\label{fig:2fib_b3_ratio10_lmax8_N81}    }
	 \end{figure}
	 
	 \begin{table}[H]                                                                      
	 	\centering                    
	 	\caption{Two fiber crossing: $n=81$ gradient directions, $b=3000s/mm^2$ and $ratio=10$.\label{table:2fib_b3_ratio10_lmax8_N81}      }                                              
	 	
	 	\begin{tabular}{c|ccc|ccc|cc}\hline\hline                               
	 		\multicolumn{9}{c}{Separation angle:  $sep=90^{\circ}$}\\                                                                      
	 		Estimator & Correct & Under & Over & Error-1  & Error-2 & Mean Sep. & Mean H-dist. & SD H-dist. \\
	 		\hline 
	 			SH-ridge & 0.85 & 0.00 & 0.15 & 1.92 & 3.69 & 88.18 & 0.59 & 0.0087 \\                       
	 				\hline                                                                                 
	 				SCSD8 & 1.00 & 0.00 & 0.00 & 1.11 & 3.16 & 88.09 & 0.52 & 0.0042 \\                   
	 				\hline                                                                                 
	 				SCSD12 & 1.00 & 0.00 & 0.00 & 1.07 & 3.30 & 87.93 & 0.43 & 0.0074 \\                  
	 				\hline                                                                                 
	 				SN-lasso & 1.00 & 0.00 & 0.00 & 1.50 & 3.66 & 87.36 & 0.42 & 0.0147 \\  
	 				\hline\hline
	 				
	 				\multicolumn{9}{c}{Separation angle:  $sep=75^{\circ}$}\\                                                                      
	 		 		Estimator & Correct & Under & Over & Error-1  & Error-2 & Mean Sep. & Mean H-dist. & SD H-dist. \\
	 		 		\hline 	
	 		 		SH-ridge & 0.85 & 0.00 & 0.15 & 4.90 & 4.32 & 82.02 & 0.57 & 0.0113 \\                       
	 				\hline                                                                                 
	 				SCSD8 & 1.00 & 0.00 & 0.00 & 2.14 & 3.50 & 72.92 & 0.50 & 0.0039 \\                   
	 				\hline                                                                                 
	 				SCSD12 & 1.00 & 0.00 & 0.00 & 2.08 & 3.38 & 73.85 & 0.41 & 0.0084 \\                  
	 				\hline                                                                                 
	 				SN-lasso & 1.00 & 0.00 & 0.00 & 2.16 & 3.49 & 75.26 & 0.40 & 0.0135 \\  
	 				\hline\hline
	 				
	 				\multicolumn{9}{c}{Separation angle:  $sep=60^{\circ}$}\\                                                                      
	 		 		Estimator & Correct & Under & Over & Error-1  & Error-2 & Mean Sep. & Mean H-dist. & SD H-dist. \\
	 		 		\hline 
	 				SH-ridge & 0.40 & 0.03 & 0.57 & 3.79 & 5.65 & 62.99 & 0.58 & 0.0116 \\                       
	 				\hline                                                                                 
	 				SCSD8 & 1.00 & 0.00 & 0.00 & 2.54 & 3.47 & 56.60 & 0.48 & 0.0098 \\                   
	 				\hline                                                                                 
	 				SCSD12 & 0.98 & 0.00 & 0.02 & 2.32 & 3.37 & 59.33 & 0.39 & 0.0200 \\                  
	 				\hline                                                                                 
	 				SN-lasso & 0.99 & 0.00 & 0.01 & 2.54 & 3.40 & 59.65 & 0.38 & 0.0194 \\
	 				\hline\hline
	 				
	 				\multicolumn{9}{c}{Separation angle:  $sep=45^{\circ}$}\\                                                                      
	 		 		Estimator & Correct & Under & Over & Error-1  & Error-2 & Mean Sep. & Mean H-dist. & SD H-dist. \\
	 		 		\hline
	 		 		SH-ridge & 0.10 & 0.89 & 0.01 & 22.04 & 77.25 & 88.64 & 0.58 & 0.0112 \\                     
	 				\hline                                                                                 
	 				SCSD8 & 0.20 & 0.80 & 0.00 & 6.44 & 6.55 & 34.58 & 0.54 & 0.0073 \\                   
	 				\hline                                                                                 
	 				SCSD12 & 1.00 & 0.00 & 0.00 & 3.61 & 4.78 & 42.15 & 0.45 & 0.0285 \\                  
	 				\hline                                                                                 
	 				SN-lasso & 1.00 & 0.00 & 0.00 & 3.77 & 4.64 & 42.42 & 0.44 & 0.0297 \\
	 		\hline\hline
	 	\end{tabular}
	 \end{table}

\subsubsection{$3$-fiber simulation:  $b=1000s/mm^2$ and $ratio=10$ \label{sec:3fiber}}


\begin{figure}[H]
	\centering
	\caption{Three fiber crossing: $n=81$ gradient directions, $b=1000s/mm^2$ and $ratio=10$. From left to right: SH $l_{\max}=8$ representation of true FOD,  \texttt{SH-ridge},  \texttt{SCSD8}, \texttt{SCSD12}, \texttt{SN-lasso}; From top to bottom: separation angle $sep=90^{\circ}$, $sep=75^{\circ}$, $sep=60^{\circ}$. The lines indicate the true fiber directions, the opaque part in the plots corresponds to mean estimated FOD across $100$ replicates, and the  translucent  part in the plots corresponds to mean plus two standard deviations of the estimated FOD.
		\label{fig:3fib_b1_ratio10_lmax8_N81}    }
\end{figure}

\begin{table}[H]                                                                      
	\centering                    
	\caption{Three fiber crossing: $n=81$ gradient directions, $\textit{bvalue}=1000s/mm^2$ and $ratio=10$.\label{table:3fib_b1_ratio10_lmax8_N81}      }                                              
	{\scriptsize
	\begin{tabular}{c|ccc|ccc|ccc|cc}\hline\hline                               
		\multicolumn{12}{c}{Separation angle:  $sep=90^{\circ}$}\\                                                                      
		Estimator & Correct & Under & Over & Error1  & Error2 & Error3 & Mean & Mean & Mean & Mean  & SD  \\
		& &  &  &   &  &  & Sep1 & Sep2 & Sep3 & H-dist. & H-dist. \\
		\hline                                                                                                   
		                       
				SH-ridge & 0.08 & 0.92 & 0.00 & 13.30 & 12.84 & 9.41 & 82.46 & 84.94 & 85.81 & 0.63 & 0.0153 \\                                   
				\hline                                                                                                                      
				SCSD8 & 0.11 & 0.00 & 0.89 & 12.38 & 13.30 & 10.47 & 84.54 & 86.08 & 87.85 & 0.68 & 0.0440 \\                              
				\hline                                                                                                                      
				SCSD12 & 0.07 & 0.00 & 0.93 & 11.50 & 9.70 & 9.41 & 86.85 & 85.83 & 86.92 & 0.69 & 0.0268 \\                               
				\hline                                                                                                                      
				SN-lasso & 0.46 & 0.28 & 0.26 & 12.90 & 12.42 & 11.09 & 84.92 & 84.95 & 84.96 & 0.67 & 0.0461 \\                                                                   
				\hline\hline
				
				\multicolumn{12}{c}{Separation angle:  $sep=75^{\circ}$}\\                                                                      
		 		Estimator & Correct & Under & Over & Error-1  & Error-2 & Error-3 & Sep-1. & Sep-2. & Sep-3. & H-dist. & SD H-dist. \\
		 		\hline
				SH-ridge & 0.04 & 0.96 & 0.00 & 11.43 & 14.34 & 7.39 & 84.18 & 80.80 & 82.31 & 0.60 & 0.0128 \\                                   
				\hline                                                                                                                      
				SCSD8 & 0.09 & 0.00 & 0.91 & 16.92 & 16.26 & 19.20 & 67.33 & 68.33 & 74.68 & 0.65 & 0.0601 \\                              
				\hline                                                                                                                      
				SCSD12 & 0.04 & 0.00 & 0.96 & 14.96 & 11.93 & 12.02 & 74.44 & 80.87 & 70.89 & 0.67 & 0.0439 \\                             
				\hline                                                                                                                      
				SN-lasso & 0.36 & 0.54 & 0.10 & 12.57 & 13.85 & 10.60 & 74.38 & 73.18 & 72.48 & 0.63 & 0.0516 \\                                  
				\hline\hline
				
				\multicolumn{12}{c}{Separation angle:  $sep=60^{\circ}$}\\                                                                      
		 		Estimator & Correct & Under & Over & Error-1  & Error-2 & Error-3 & Sep-1. & Sep-2. & Sep-3. & H-dist. & SD H-dist. \\
		 		\hline
				SH-ridge & 0.00 & 1.00 & 0.00 & - & - & - & - & - & - & 0.58 & 0.0039 \\                                                          
				\hline                                                                                                                      
				SCSD8 & 0.44 & 0.54 & 0.02 & 13.98 & 17.72 & 18.71 & 53.36 & 55.45 & 55.84 & 0.59 & 0.0507 \\                              
				\hline                                                                                                                      
				SCSD12 & 0.03 & 0.00 & 0.97 & 20.13 & 27.51 & 15.47 & 39.02 & 58.89 & 57.18 & 0.65 & 0.0589 \\                             
				\hline                                                                                                                      
				SN-lasso & 0.74 & 0.19 & 0.07 & 15.95 & 15.62 & 20.77 & 58.19 & 61.85 & 61.16 & 0.61 & 0.0746 \\                                   
		\hline\hline
	\end{tabular}
}
\end{table}


\begin{figure}[H]
	\centering
	\caption{Three fiber crossing: $n=321$ gradient directions, $b=1000s/mm^2$ and $ratio=10$. From left to right: SH $l_{\max}=8$ representation of true FOD,  \texttt{SH-ridge},  \texttt{SCSD8}, \texttt{SCSD12}, \texttt{SN-lasso}; From top to bottom: separation angle $sep=90^{\circ}$, $sep=75^{\circ}$, $sep=60^{\circ}$. The lines indicate the true fiber directions, the opaque part in the plots corresponds to mean estimated FOD across $100$ replicates, and the  translucent  part in the plots corresponds to mean plus two standard deviations of the estimated FOD.
		\label{fig:3fib_b1_ratio10_lmax8_N321}    }
\end{figure}

\begin{table}[H]                                                                      
	\centering                    
	\caption{Three fiber crossing: $n=321$ gradient directions, $b=1000s/mm^2$ and $ratio=10$.\label{table:3fib_b1_ratio10_lmax8_N321}      }                                              
	{\scriptsize
		\begin{tabular}{c|ccc|ccc|ccc|cc}\hline\hline                               
			\multicolumn{12}{c}{Separation angle:  $sep=90^{\circ}$}\\                                                                      
			Estimator & Correct & Under & Over & Error1  & Error2 & Error3 & Mean & Mean & Mean & Mean  & SD  \\
			& &  &  &   &  &  & Sep1 & Sep2 & Sep3 & H-dist. & H-dist. \\
			\hline                                                                                                   
		SH-ridge & 0.66 & 0.34 & 0.00 & 5.86 & 6.83 & 5.70 & 86.67 & 86.34 & 86.75 & 0.59 & 0.0166 \\                                     
			\hline                                                                                                                      
			SCSD8 & 0.72 & 0.00 & 0.28 & 5.46 & 7.36 & 6.32 & 87.39 & 87.17 & 87.69 & 0.53 & 0.0570 \\                                 
			\hline                                                                                                                      
			SCSD12 & 0.50 & 0.00 & 0.50 & 4.42 & 6.52 & 6.32 & 86.96 & 87.29 & 87.01 & 0.52 & 0.0847 \\                                
			\hline                                                                                                                      
			SN-lasso & 0.93 & 0.01 & 0.06 & 6.64 & 8.43 & 6.66 & 85.48 & 86.14 & 85.63 & 0.48 & 0.0562 \\                                                                      
			\hline\hline
			
			\multicolumn{12}{c}{Separation angle:  $sep=75^{\circ}$}\\                                                                      
	 		Estimator & Correct & Under & Over & Error-1  & Error-2 & Error-3 & Sep-1. & Sep-2. & Sep-3. & H-dist. & SD H-dist. \\
	 		\hline
			SH-ridge & 0.35 & 0.65 & 0.00 & 6.25 & 6.65 & 7.29 & 77.47 & 77.57 & 77.10 & 0.58 & 0.0155 \\                                     
			\hline                                                                                                                      
			SCSD8 & 0.51 & 0.00 & 0.49 & 7.99 & 8.68 & 7.60 & 70.95 & 71.37 & 71.95 & 0.54 & 0.0794 \\                                 
			\hline                                                                                                                      
			SCSD12 & 0.20 & 0.00 & 0.80 & 8.60 & 8.14 & 6.86 & 75.54 & 72.95 & 72.02 & 0.54 & 0.0771 \\                                
			\hline                                                                                                                      
			SN-lasso & 0.92 & 0.02 & 0.06 & 7.47 & 8.11 & 6.74 & 72.77 & 74.00 & 74.10 & 0.44 & 0.0579 \\                                     
			\hline\hline
			
			\multicolumn{12}{c}{Separation angle:  $sep=60^{\circ}$}\\                                                                      
	 		Estimator & Correct & Under & Over & Error-1  & Error-2 & Error-3 & Sep-1. & Sep-2. & Sep-3. & H-dist. & SD H-dist. \\
	 		\hline
			SH-ridge & 0.00 & 1.00 & 0.00 & - & - & - & - & - & - & 0.59 & 0.0020 \\                                                          
			\hline                                                                                                                      
			SCSD8 & 0.82 & 0.17 & 0.01 & 9.17 & 10.03 & 8.28 & 53.12 & 54.65 & 54.03 & 0.51 & 0.0431 \\                                
			\hline                                                                                                                      
			SCSD12 & 0.08 & 0.00 & 0.92 & 8.70 & 10.96 & 8.07 & 57.26 & 55.73 & 55.56 & 0.58 & 0.0716 \\                               
			\hline                                                                                                                      
			SN-lasso & 0.66 & 0.33 & 0.01 & 9.23 & 10.79 & 10.11 & 58.89 & 60.06 & 60.06 & 0.52 & 0.0740 \\                                   
		\hline\hline
		\end{tabular}
	}
\end{table}

\subsubsection{$3$-fiber simulation:  $b=3000s/mm^2$ and $ratio=10$
	\label{sec:3fiber_3000}}

\begin{figure}[H]
	\centering
	\caption{Three fiber crossing: $n=81$ gradient directions, $b=3000s/mm^2$ and $ratio=10$. From left to right: SH $l_{\max}=8$ representation of true FOD,  \texttt{SH-ridge},  \texttt{SCSD8}, \texttt{SCSD12}, \texttt{SN-lasso}; From top to bottom: separation angle $sep=90^{\circ}$, $sep=75^{\circ}$, $sep=60^{\circ}$. The lines indicate the true fiber directions, the opaque part in the plots corresponds to mean estimated FOD across $100$ replicates, and the  translucent  part in the plots corresponds to mean plus two standard deviations of the estimated FOD.
		\label{fig:3fib_b3_ratio10_lmax8_N81}    }
\end{figure}

\begin{table}[H]                                                                      
	\centering                    
	\caption{Three fiber crossing: $n=81$ gradient directions, $b=3000s/mm^2$ and $ratio=10$.\label{table:3fib_b3_ratio10_lmax8_N81}      }                                              
	{\scriptsize
	\begin{tabular}{c|ccc|ccc|ccc|cc}\hline\hline                               
		\multicolumn{12}{c}{Separation angle:  $sep=90^{\circ}$}\\                                                                      
		Estimator & Correct & Under & Over & Error1  & Error2 & Error3 & Mean & Mean & Mean & Mean  & SD  \\
		& &  &  &   &  &  & Sep1 & Sep2 & Sep3 & H-dist. & H-dist. \\
		\hline        
		SH-ridge & 0.99 & 0.00 & 0.01 & 1.97 & 3.94 & 4.58 & 87.73 & 87.25 & 87.51 & 0.56 & 0.0063 \\                                     
				\hline                                                                                                                      
				SCSD8 & 1.00 & 0.00 & 0.00 & 2.01 & 4.32 & 4.67 & 87.59 & 87.53 & 87.53 & 0.50 & 0.0079 \\                                 
				\hline                                                                                                                      
				SCSD12 & 0.98 & 0.00 & 0.02 & 1.92 & 4.34 & 4.88 & 87.52 & 87.34 & 87.28 & 0.42 & 0.0216 \\                                
				\hline                                                                                                                      
				SN-lasso & 0.94 & 0.00 & 0.06 & 2.59 & 5.02 & 4.91 & 86.68 & 86.35 & 86.70 & 0.40 & 0.0284 \\                                     
				\hline\hline
				
				\multicolumn{12}{c}{Separation angle:  $sep=75^{\circ}$}\\                                                                      
		 		Estimator & Correct & Under & Over & Error-1  & Error-2 & Error-3 & Sep-1. & Sep-2. & Sep-3. & H-dist. & SD H-dist. \\
		 		\hline 
				SH-ridge & 1.00 & 0.00 & 0.00 & 6.69 & 7.17 & 5.37 & 83.34 & 82.34 & 82.30 & 0.54 & 0.0107 \\                                     
				\hline                                                                                                                      
				SCSD8 & 1.00 & 0.00 & 0.00 & 5.06 & 5.69 & 4.16 & 72.80 & 73.62 & 72.57 & 0.45 & 0.0092 \\                                 
				\hline                                                                                                                      
				SCSD12 & 0.83 & 0.00 & 0.17 & 4.48 & 5.34 & 3.89 & 74.12 & 75.21 & 72.97 & 0.39 & 0.0481 \\                                
				\hline                                                                                                                      
				SN-lasso & 0.93 & 0.00 & 0.07 & 4.44 & 5.09 & 4.10 & 75.22 & 76.23 & 74.34 & 0.37 & 0.0292 \\                                     
				\hline\hline
				
				\multicolumn{12}{c}{Separation angle:  $sep=60^{\circ}$}\\                                                                      
		 		Estimator & Correct & Under & Over & Error-1  & Error-2 & Error-3 & Sep-1. & Sep-2. & Sep-3. & H-dist. & SD H-dist. \\
		 		\hline 
		 		SH-ridge & 0.25 & 0.75 & 0.00 & 8.73 & 8.70 & 8.09 & 67.79 & 68.77 & 67.66 & 0.59 & 0.0031 \\                                     
				\hline                                                                                                                      
				SCSD8 & 0.97 & 0.03 & 0.00 & 5.11 & 6.67 & 6.09 & 54.92 & 55.61 & 56.26 & 0.47 & 0.0187 \\                                 
				\hline                                                                                                                      
				SCSD12 & 0.41 & 0.00 & 0.59 & 5.55 & 6.88 & 5.52 & 57.31 & 58.70 & 58.00 & 0.47 & 0.0640 \\                                
				\hline                                                                                                                      
				SN-lasso & 0.88 & 0.00 & 0.12 & 3.85 & 5.77 & 5.50 & 58.54 & 60.13 & 59.95 & 0.39 & 0.0500 \\               
		\hline\hline
	\end{tabular}
}
\end{table}

\subsection{Real D-MRI Data Experiments \label{sec:realdatasupp}}

\subsubsection*{Single tensor model}
The single tensor model \citep{LeBihan1995,Basser2002,Mori2007} for the DWI signal along the gradient field direction $\mathbf{u}$ has: 
$$
S(\mathbf{u})=S_{0} \exp(-b \mathbf{u}^{T} \mathbf{D} \mathbf{u}), 
$$  
where 
$\mathbf{D}$ is the  diffusion tensor, a $3\times 3$ positive definition matrix.  Let $\lambda_1 
\leq \lambda_2 \leq \lambda_3$ be the three eigenvalues of $\mathbf{D}$, then \textit{fractional anisotropy (FA)} and  \textit{mean diffusivity (MD)} are defined as follows:

	$$
	FA = \sqrt{\frac{1}{2}}\frac{\sqrt{(\lambda_1-\lambda_2)^2+(\lambda_2-\lambda_3)^2+(\lambda_3-\lambda_1)^2}}{\sqrt{\lambda_1^2+\lambda_2^2+\lambda_3^2}}.
	$$

	$$
	MD = \frac{\lambda_1+\lambda_2+\lambda_3}{3}.
	$$
	
	Note that $FA \in [0,1]$ and large FA indicates anisotropic diffusion along the principal eigenvector of the diffusion tensor $\mathbf{D}$, whereas small FA indicates isotropic diffusion (if the single tensor model holds). On the other hand, MD captures the rate of water diffusion: Large MD means fast diffusion  and small MD indicates slow diffusion (e.g., due to tissue barrier).

	Figure \ref{fig:realdata_MD_FA} shows (estimated) $S_0, \sigma, SNR=S_0/\sigma, FA, MD$ and the scatter plot of $MD$ versus $FA$ for one ADNI data set used in this paper. Here, FA and MD are estimated under the single tensor model through nonlinear regression \citep{CarmichaelCPP2013} 
	and $S_0, \sigma$ are estimated based on the five $b_0$ images under the Rician noise model \citep{HahnPHH2006}.

	\begin{figure}[H]
	\centering
	\caption{ADNI data set: Preliminary analysis under the single tensor model. \label{fig:realdata_MD_FA}}
\end{figure}

\subsubsection*{Estimation of the response function}

We first
identify voxels with a single dominant fiber bundle characterized by $FA>0.8$ and the ratio between the two smaller eigenvalues $<1.5$. For each such voxel,  we set the minor eigenvalue as the average of the two smaller eigenvalues. We then take the median of the leading eigenvalue and the minor eigenvalue across such voxels and  denote these medians by $\bar{\lambda}_3$ and $\bar{\lambda}_1$, respectively. The response function is then set as 
$$
R(\cos(\theta))= S_0\exp^{-b (\bar{\lambda}_1 \sin^2\theta  + \bar{\lambda}_3 \cos^2\theta )}, ~~~ \theta \in  [0,\pi]
$$
according to the single tensor model (a.k.a. the Gaussian diffusion model).

\subsubsection*{Additional results}
\begin{figure}[H]
	\centering
	\caption{ADNI data sets. Fiber orientation maps of  z-slices (a) $z=40$; (b): $z=32$. ROI I,II are  indicated by the white boxes. \label{fig:realdata_FOM}}
\end{figure}

\begin{figure}[H]
	\centering
	\caption{ROI I - Subregion 3: FOD estimates on a $5 \times 5$ subregion (columns 12-16  and rows 8- 12). This subregion is  indicated by the white boxes on the colormap, FA map and MD map of ROI I. 
		\label{fig:realdata_regionb_sub3}}
	
			
			
		
\end{figure}



\end{document}